\documentclass[prb,twocolumn]{revtex4}
\usepackage{times}
\usepackage[dvips]{graphicx}
\usepackage{latexsym,amsmath,amssymb,bm,euscript}
\usepackage[dvips]{color}
\usepackage{multirow}

\def\etal{~\textit{et~al.}} % etal
\def\ra{\rangle} % bra
\def\la{\langle} % ket
\def\up{\uparrow}
\def\dn{\downarrow}
\def\Hc{{\rm H.c.}}
\def\ET{{$\kappa$-(ET)$_2$Cu$_2$(CN)$_3$}}
\begin{document}

\title{Spin Bose-Metal phase in a spin-1/2 model with ring exchange on a 
two-leg triangular strip}

\author{D. N. Sheng$^1$, Olexei I. Motrunich$^2$, and 
Matthew P. A. Fisher$^3$}
\affiliation{
$^1$Department of Physics and Astronomy, California State University, 
Northridge, California 91330\\
$^2$Department of Physics, California Institute of Technology,
Pasadena, California 91125\\
$^3$Microsoft Research, Station Q, University of California, 
Santa Barbara, California 93106\\
}

\date{\today}

\begin{abstract}
Recent experiments on triangular lattice organic Mott insulators 
have found evidence for a 2D spin liquid in close proximity to the 
metal-insulator transition.     
A Gutzwiller wavefunction study of the triangular lattice 
Heisenberg model with a four-spin ring exchange term appropriate in
this regime has found that the projected spinon Fermi sea state has a 
low variational energy.  This wavefunction, together with a 
slave particle gauge theory analysis,
suggests that this putative spin liquid possesses spin correlations 
that are singular along surfaces in momentum space, 
i.e.\ ``Bose surfaces''.  Signatures of this state, which we will 
refer to as a ``Spin Bose-Metal" (SBM), are expected to be manifest in 
quasi-1D ladder systems:  The discrete transverse momenta cut through 
the 2D Bose surface leading to a distinct pattern of 1D gapless modes.
Here, we search for a quasi-1D descendant of the triangular lattice 
SBM state by exploring the Heisenberg plus ring model on a two-leg 
triangular strip (zigzag chain).
Using DMRG supplemented by variational wavefunctions and a 
Bosonization analysis, we map out the full phase diagram.
In the absence of ring exchange the model is equivalent to the 
$J_1 - J_2$ Heisenberg chain, and we find the expected Bethe-chain 
and dimerized phases.  Remarkably, moderate ring exchange reveals a 
new gapless phase over a large swath of the phase diagram.
Spin and dimer correlations possess singular wavevectors at particular 
``Bose points'' (remnants of the 2D Bose surface) and allow us to 
identify this phase as the hoped for quasi-1D descendant of the 
triangular lattice SBM state.  
We use Bosonization to derive a low energy effective theory for the 
zigzag Spin Bose-Metal and find three gapless modes and one 
Luttinger parameter controlling all power law correlations.
Potential instabilities out of the zigzag SBM give rise to other 
interesting phases such as a period-3 Valence Bond Solid or 
a period-4 Chirality order, which we discover in the DMRG.
Another interesting instability is into a Spin Bose-Metal phase with 
partial ferromagnetism (spin polarization of one spinon band), 
which we also find numerically using the DMRG.
\end{abstract}

\maketitle

%%%%%%%%%%%%%%%%%%%%%%%%%%%%%%%%%%%%%%%%%%%%%%%%%%%%%%%%%%%%%%%%%%%%%
%%%%%%%%%%%%%%%%%%%%%%%%%%%%%%%%%%%%%%%%%%%%%%%%%%%%%%%%%%%%%%%%%%%%%
%%%%%%%%%%%%%%%%%%%%%%%%%%%%%%%%%%%%%%%%%%%%%%%%%%%%%%%%%%%%%%%%%%%%%
\section{Introduction}
\label{sec:intro}

A promising regime to search for elusive 2D spin liquids is in the
proximity of the Mott metal-insulator transition.  
In such ``weak Mott insulators" significant local charge fluctuations
induce multi-spin ring exchange processes which tend to suppress 
magnetic or other types of ordering.  
Indeed, recent experiments\cite{Shimizu03, Kurosaki}
on the triangular lattice based organic Mott insulator \ET\ reveal 
no indication of magnetic order or other symmetry breaking down to 
temperature several orders of magnitude smaller than the characteristic
exchange interaction energy $J \approx 250$~K.  
Under pressure the \ET\ undergoes a weak first order transition into a 
metallic state, while at ambient pressure it has a small charge gap of 
$200$~K, as expected in a weak Mott insulator.
Thermodynamic, transport, and spectroscopic experiments
\cite{Shimizu03, SYamashita, MYamashita}
all point to the presence of a plethora of low energy excitations in the 
\ET, indicative of a {\it gapless} spin liquid phase.
Several authors have proposed\cite{ringxch, SSLee, Senthil_Mott} 
that this putative spin liquid can be described in terms of a 
Gutzwiller-projected Fermi sea of spinons.

Quantum chemistry calculations suggest that a one-band triangular 
lattice Hubbard model at half filling is an appropriate theoretical 
starting point to describe \ET.\cite{McKenzie, Shimizu03}
Variational studies of the triangular lattice Hubbard model\cite{Morita}
find indications of a non-magnetic spin liquid phase just on the 
insulating side of the Mott transition.
Moreover, exact diagonalization studies of the triangular lattice
Heisenberg model show that the presence of a four-site ring exchange 
term appropriate near the Mott transition can readily destroy the 
120$^\circ$ antiferromagnetic order.\cite{LiMing} 
One of us\cite{ringxch} performed variational wavefunction studies 
on this spin model and found that the Gutzwiller-projected Fermi sea 
state\cite{SSLee} has the lowest energy for sufficiently strong 
four-site ring exchange interactions appropriate for the \ET.

Despite these encouraging hints, the theoretical evidence for a 
spin liquid phase in the triangular lattice Hubbard model or 
Heisenberg spin model with ring exchanges is at best suggestive.
Variational studies are biased by the choice of wavefunctions and 
can be notoriously misleading.
Exact diagonalization studies are restricted to very small sizes, 
which is especially problematic for gapless spin liquids. 
Quantum Monte Carlo fails due to the sign problem.
The density matrix renormalization group (DMRG) can reach the 
ground state of large 1D systems, but capturing the highly entangled and 
non-local character of a 2D gapless spin liquid state is a 
formidable challenge.
Thus, with new candidate spin liquid materials and increasingly refined 
experiments available, the gap between theory and experiment becomes 
ever more dire.  

Effective field theory approaches such as slave particle gauge theories
or vortex dualities, while unable to solve any particular Hamiltonian,
do indicate the possibility of stable gapless 2D spin liquid phases.
Such gapless 2D spin liquids generically exhibit spin correlations that 
decay as a power law in space, perhaps with anomalous exponents, 
and which can oscillate at particular wavevectors.  
The location of these dominant singularities in momentum space provides a
convenient characterization of gapless spin liquids.
In the  ``algebraic" or ``critical" spin liquids
\cite{WenPSG, Rantner02, Hermele_U1, LeeNagaosaWen} 
these wavevectors are limited to a finite discrete set, often at 
high symmetry points in the Brilloin zone, and their effective 
field theories can often exhibit a relativistic structure.   
But the singularities can also occur along {\it surfaces} in momentum 
space, as they do in the Gutzwiller-projected spinon Fermi sea state, 
the 2D Spin Bose-Metal (SBM) phase. 
It must be stressed that it is the {\it spin} (i.e., bosonic) 
correlation functions that possess such singular surfaces -- 
there are no fermions in the system -- and the low energy excitations 
cannot be described in terms of weakly interacting quasiparticles.    
It has been proposed recently\cite{DBL} that a 2D ``Boson-ring" model 
describing itinerant hard core bosons hopping on a square lattice with a 
frustrating four-site term can have an analogous liquid ground state 
which we called a d-wave Bose liquid (DBL).
The DBL is also a Bose-Metal phase, possessing a singular Bose surface 
in momentum space.

Recently we have suggested\cite{2legDBL, SBM_Solvay} that it should be 
possible to access such Bose-Metals by systematically approaching 2D 
from a sequence of quasi-1D ladder models.  On a ladder the quantized 
transverse momenta cut through the 2D surface, leading to a 
quasi-1D descendant state with a set of low-energy modes whose number 
grows with the number of legs and whose momenta are inherited from the 
2D Bose surfaces.  These quasi-1D descendant states can be accessed in a 
controlled fashion by analyzing the 1D ladder models using numerical 
and analytical approaches.
These multi-mode quasi-1D liquids constitute a new and previously 
unanticipated class of quantum states interesting in their own right.  
But more importantly they carry a distinctive quasi-1D ``fingerprint" 
of the parent 2D state.

The power of this approach was demonstrated in Ref.~\onlinecite{2legDBL} 
where we studied the new Boson-ring model on a two-leg ladder and 
mapped out the full phase diagram using the DMRG and ED, supported by 
variational wavefunction and gauge theory analysis.  
Remarkably, even for a ladder with only two legs, we found compelling 
evidence for the quasi-1D descendant of the 2D DBL phase.   
This new quasi-1D quantum state possessed all of the expected 
signatures reflecting the parent 2D Bose surface.

\begin{figure}[t]
\centerline{\includegraphics[width=\columnwidth]{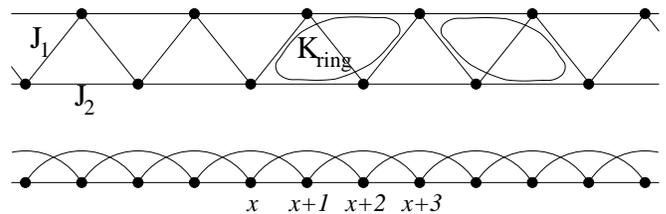}}
\caption{
Top: Heisenberg plus ring exchange model on a 2-leg triangular strip.
Bottom: Convenient representation of the model as a $J_1 - J_2$ chain 
with additional four-site terms; 
the Hamiltonian is written out in Eq.~(\ref{Hring}).
}
\label{fig:model}
\end{figure}

In this paper we turn our attention to the triangular lattice 
Heisenberg model with ring exchange appropriate for the \ET\ material.
In hopes of detecting the quasi-1D descendant of the triangular lattice 
Spin Bose-Metal (Gutzwiller-projected spinon Fermi sea state), 
we place this model on a triangular strip with only two legs shown in 
Fig.~\ref{fig:model}. 
The all-important ring exchange term acts around four-site plackets as 
illustrated; we also allow different Heisenberg exchange couplings
along and transverse to the ladder.

\begin{figure}[t]
\centerline{\includegraphics[width=\columnwidth]{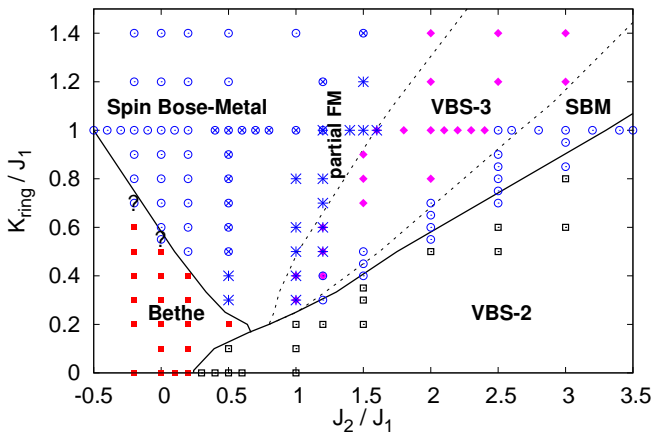}}
\caption{
(Color online)
Phase diagram of the ring model Eq.~(\ref{Hring}) determined in the DMRG 
using system sizes $L = 48 - 96$.
Filled squares (red) denote Bethe-chain phase.
Open squares (with black outlines) denote Valence Bond Solid with 
period 2.
Open circles (blue) denote Spin Bose-Metal.
Open circles with crosses denote where the DMRG has difficulties 
converging to singlet for the larger sizes, but where we still think 
this is spin-singlet SBM.
Star symbols denote points where the ground state appears to have true 
non-zero spin (for all points here, the magnetization is smaller than 
full polarization of the smaller Fermi sea in the SBM interpretation).
Filled diamonds (magenta) denote VBS with period 3. Our identifications
are ambiguous in the lower VBS-3 region approaching VBS-2.
Lines indicate phase boundaries determined in VMC using spin-singlet 
wavefunctions described in the text
(we also used appropriate dimerized wavefunctions for the VBS states).
A detailed study of a cut $K_{\rm ring} / J_1 = 1$ is presented in
Sec.~\ref{sec:HringDMRG} (cf.~Fig.~\ref{fig:qsing_J3_0p0}).
}
\label{fig:phased_J3_0p0}
\end{figure}

It is convenient to view the two-leg strip as a $J_1 -J_2$ chain 
(studied extensively before\cite{White_zigzag, Nersesyan}) 
with additional four-spin exchanges.  The Hamiltonian reads
\begin{eqnarray}
\hat{H} = \sum_x &\big[& 
  2 J_1 \vec{S}(x) \cdot \vec{S}(x+1) 
+ 2 J_2 \vec{S}(x) \cdot \vec{S}(x+2) \nonumber \\
&& + K_{\rm ring} \left( P_{x, x+2, x+3, x+1} + \Hc \right) 
\big] ~.
\label{Hring}
\end{eqnarray}
The four-spin operators act as
$P_{1234} : |\sigma_1, \sigma_2, \sigma_3, \sigma_4 \ra 
        \to |\sigma_4, \sigma_1, \sigma_2, \sigma_3 \ra$,
$P_{1234}^\dagger = (P_{1234})^{-1}$.
We attack this model using a combination of numerical and analytical
techniques - DMRG, exact diagonalization (ED), 
variational Monte Carlo (VMC), 
as well as employing bosonization to obtain a low energy effective 
field theory from the slave particle-gauge formulation (and/or from an 
interacting electron Hubbard-type model).  
Our key findings are summarized in Fig.~\ref{fig:phased_J3_0p0} which
shows the phase diagram for antiferromagnetic couplings 
$J_1$, $J_2$, and $K_{\rm ring}$.
For $K_{\rm ring} \to 0$, the $J_1 - J_2$ model has the familiar 
1D Bethe-chain phase for $J_2 \lesssim 0.24 J_1$ and period-2 
Valence Bond Solid (VBS-2) for larger $J_2$.
For $K_{\rm ring} \gtrsim 0.2 J_1$, new physics opens up.
In fact, Klironomos\etal\cite{Klironomos} considered such
$J_1 - J_2 - K_{\rm ring}$ model motivated by the study of Wigner 
crystals in a quantum wire.\cite{Meyer08}  
Using ED of systems up to $L = 24$, they found an unusual phase in this 
intermediate regime, 
called ``4P'' in Fig.~8 of Ref.~\onlinecite{Klironomos}, 
but it had proven difficult to clarify its nature. 
We identify this region as a descendant of the triangular lattice 
Spin Bose-Metal phase (or further derivatives of the descendant as 
discussed below).

\begin{figure}
\centerline{\includegraphics[width=\columnwidth]{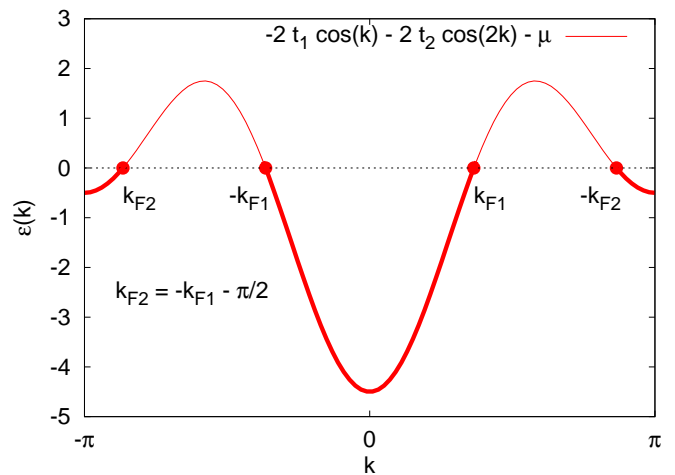}}
\caption{
Spinon dispersion for $t_2 > 0.5 t_1$ showing two occupied Fermi sea 
segments (here and throughout we use the 1D chain language, 
see bottom Fig.~\ref{fig:model}).
Gutzwiller projection of this is the zigzag SBM state at the focus
of this work.
}
\label{fig:fbands}
\end{figure}

A caricature of the zigzag Spin Bose-Metal is provided by considering
a Gutzwiller trial wavefunction construction on the two-leg strip.  
The 2D SBM is obtained by letting spinons hop on the triangular lattice 
with no fluxes and then Gutzwiller-projecting to get a trial 
spin wavefunction.
So here we also take spinons hopping on the ladder with no fluxes,
which is $t_1 - t_2$ hopping in the 1D chain language that we mainly use.
For $t_2 < 0.5 t_1$, the mean field state has one Fermi sea segment
spanning $[-\pi/2, \pi/2]$ (spinons are at half-filling), and the 
Gutzwiller projection of this is known to be an excellent state for the 
Bethe chain.  
On the other hand, for $t_2 > 0.5 t_1$, the spinon band has two 
Fermi seas as shown in Fig.~\ref{fig:fbands}.  
The Gutzwiller projection of this is a new phase that we identify 
as a quasi-1D descendant of the triangular lattice Spin Bose-Metal.
The wavefunction has one variational parameter $t_2/t_1$, 
or, equivalently, the ratio of the two Fermi sea volumes.
Using this restricted family of states, our VMC energetics study of the 
$J_1 - J_2 - K_{\rm ring}$ model finds three regimes broadly delineated 
by solid lines in Fig.~\ref{fig:fbands} for larger $K_{\rm ring}$: 
{\it i)} In the Bethe-chain regime the optimal state has one Fermi sea.
{\it ii)} For sufficiently large $K_{\rm ring}$ and upon 
increasing $J_2$, we enter a different regime where it is advantageous 
to start populating the second Fermi sea.  As we further increase $J_2$ 
moving away from the Bethe-chain phase, we gradually transfer more 
spinons from the first to the second Fermi sea.  This whole region is the
SBM.
{\it iii)} Finally, at still larger $J_2$, the volumes of the two 
Fermi seas become equal, which corresponds to $t_2/t_1 \to \infty$, 
i.e., decoupled-legs limit.

The DMRG is the crucial tool that allows us to answer how much of this 
trial state picture actually holds in the $J_1 - J_2 - K_{\rm ring}$ 
model.  Fig.~\ref{fig:phased_J3_0p0} shows all points that were studied
using the DMRG and their tentative phase identifications by looking at 
various ground state properties.  Remarkably, in a broad-brush sense, 
the three regimes found in VMC for $K_{\rm ring} > 0.2$
(one spinon Fermi sea, two generic Fermi seas, and decoupled legs)
match quite closely different qualitative behaviors found by the 
DMRG study and marked as Bethe-chain, SBM, and VBS-2 regions. 
Here we note that the decoupled-legs Gutzwiller wavefunction is gapless 
and does not have VBS-2 order, but it is likely unstable towards 
opening a spin gap;\cite{White_zigzag, Nersesyan} still, it is a good 
initial description for large $J_2$.  On the other hand, away from the 
decoupled-legs limit, we expect a stable gapless SBM phase.
The DMRG measures spin and dimer correlations, and we identify
the SBM by observing singularities at characteristic wavevectors that 
evolve continuously as we move through this phase -- these are the 
quasi-1D ``Bose points'' (remnants of a 2D Bose surface).
The singular wavevectors are reproduced well by the VMC,
although the Gutzwiller wavefunctions apparently cannot capture the 
amplitudes and power law exponents.

An effective low energy field theory for the zigzag SBM phase can be 
obtained by employing Bosonization to analyze either a spinon 
gauge theory formulation or an interacting model of {\it electrons}
hopping on the zigzag chain.  In the latter case we identify an 
Umklapp term which drives the two-band metal of interacting electrons 
through a Mott metal-insulator transition.  The low energy Bosonized 
description of the Mott insulating state thereby obtained is identical to
that obtained from the zigzag spinon gauge theory.  In the interacting 
electron case, there are physical electrons that exist above the 
charge gap.  On the other hand, in the gauge theory the ``spinons" are 
unphysical and linearly confined.  

The low energy fixed point theory for the zigzag Spin Bose-Metal phase
consists of three gapless (free Boson) modes, two in the spin sector and 
one in the singlet sector (the latter we identify with spin chirality 
fluctuations).  Because of the SU(2) spin invariance, there is only one 
Luttinger parameter in the theory, and we can characterize all power laws
using this single parameter.  The dominant correlations occur at 
wavevectors $2k_{F1}$ and $2k_{F2}$ connecting opposite Fermi points,
and the power law can vary between $x^{-3/2}$ and $x^{-1}$ depending 
on the value of the Luttinger parameter.  We understand well the 
stability of this phase.  We also understand why the Gutzwiller-projected
wavefunctions, while capturing the singular wavevectors, are 
not fully adequate -- our trial wavefunctions appear to be described by a
specific value of the Luttinger parameter that gives $x^{-3/2}$ 
power law at $2k_{F1}$ and $2k_{F2}$.  
The difference between the DMRG and VMC in the SBM phase is 
qualitatively captured by the low energy Bosonized theory.

The full DMRG phase diagram findings are in fact much richer.
Prominently present in Fig.~\ref{fig:phased_J3_0p0} is a new phase 
occurring inside the SBM and labeled VBS-3.  This has period-3 
Valence Bond Solid order ``dimerizing'' every third bond and also has 
coexisting effective Bethe-chain-like state formed by non-dimerized 
spins (see Sec.~\ref{subsec:DMRG:VBS3} and Fig.~\ref{fig:dimer3} for 
more explanations).
A careful look at the SBM theory reveals that at a special 
commensuration where the volume of the first Fermi sea is 
twice as large as that of the second Fermi sea, 
the SBM phase can be unstable gapping out the first Fermi sea 
and producing such VBS-3 state.

Another observation in Fig.~\ref{fig:phased_J3_0p0} is the 
possibility of developing a partial ferromagnetic (FM) moment in the 
SBM region labeled ``partial FM'' to the left of the VBS-3 phase.
We do not understand all details in this region.
In the SBM further to the left, we think the ground state is 
spin-singlet, which is what we find from the DMRG for smaller
system sizes up to $L=48$.  However, the DMRG already has 
difficulties converging to the spin-singlet state for the 
larger system sizes $L \sim 96$.  In the partial FM region, it seems 
that the ground state has a small magnetization.  Given our SBM picture,
it is conceivable that one or both spinon Fermi seas could develop 
some spin polarization.  The most likely scenario is for the 
polarization to first appear in the smaller Fermi sea since it is 
more narrow in energy (more flat-band-like).  
The total spin that we measure in the partial FM region in 
Fig.~\ref{fig:phased_J3_0p0} is smaller than what would be expected 
from a full polarization of the second Fermi sea, and it is hard for us 
to analyze such states.

To check our intuition, we have also considered a modified model with an 
additional third-neighbor coupling $J_3$ which can be either 
antiferromagnetic or ferromagnetic tailored to either suppress the 
ferromagnetic tendencies or to reveal them more fully.
We have studied this model at $K_{\rm ring} = J_1 = 1$, varying $J_2$.
With antiferromagnetic $J_3 = 0.5$, we have indeed increased the 
stability of spin-singlet states in the region to the left of the VBS-3.
Interestingly, this study, which is not polluted by the small moment
difficulties, also revealed a new spin-gapped phase near the VBS-3.  
The new phase has a particular period-4 order in the spin chirality,  
and we can understand the occurrence as an instability at another 
commensuration point hit by the singular wavevectors as they vary 
in the SBM phase (see Sec.~\ref{subsec:commens_other} for details).
Turning now to ferromagnetic $J_3 = -0.5$, we have found a more clear 
example of the partial ferromagnetism where the ground state is well 
described by Gutzwiller-projecting a state with a fully polarized second 
Fermi sea and an unpolarized first Fermi sea.

The paper is organized as follows.
To set the stage, in Sec.~\ref{sec:SBMtheory} we develop general
theory of the zigzag SBM phase.
In Sec.~\ref{sec:HringDMRG} we present the DMRG study of the ring
model that leads to the phase diagram Fig.~\ref{fig:phased_J3_0p0}.
We consider carefully the cut at $K_{\rm ring} = J_1$ and provide 
detailed characterization of the new SBM phase.
In Sec.~\ref{sec:SBMstability} we study analytically the stability of 
the SBM.  We also consider possible phases that can arise as some
instabilities of the SBM.  This is done in particular to address the 
DMRG findings of the VBS-3 and Chirality-4 states, which we present in 
Sec.~\ref{sec:DMRG_commens}.
To clarify the regime to the left of the VBS-3 where the DMRG runs into
convergence difficulties or small moment development, we also perturb the
model with antiferromagnetic (Sec.~\ref{subsec:DMRG:afJ3}) or 
ferromagnetic (Sec.~\ref{sec:ferroJ3}) third-neighbor interaction $J_3$ 
and discuss partially polarized SBM.
Finally, in Sec.~\ref{sec:concl} we briefly summarize and suggest 
some future directions one might explore.
In Appendix~\ref{app:SBMprops}, within our effective field theory 
analysis, we summarize the bosonization expressions for physical 
observables that are measured in the DMRG and VMC. 
In Appendix~\ref{app:Gutzw} we provide details of the Gutzwiller 
wavefunctions that are used throughout in the VMC analysis.
In Appendix~\ref{app:DMRG4nonSBM} we summarize the DMRG results for
the conventional Bethe-chain and VBS-2 phases.

%%%%%%%%%%%%%%%%%%%%%%%%%%%%%%%%%%%%%%%%%%%%%%%%%%%%%%%%%%%%%%%%%%%%%%%
%%%%%%%%%%%%%%%%%%%%%%%%%%%%%%%%%%%%%%%%%%%%%%%%%%%%%%%%%%%%%%%%%%%%%%%
%%%%%%%%%%%%%%%%%%%%%%%%%%%%%%%%%%%%%%%%%%%%%%%%%%%%%%%%%%%%%%%%%%%%%%%
\section{Spin Bose-Metal theory on the zigzag strip}
\label{sec:SBMtheory}

Since a wavefunction does not constitute a theory, and can at best
capture a caricature of the putative SBM phase, it is highly desirable
to have a field-theoretic approach.  The goal here is to obtain an 
{\it effective} low energy theory for the SBM on the zigzag chain.  
In 2D the usual approach is to decompose the spin operators in terms
of an SU(2) spinor -- the fermionic spinons:
\begin{equation}
\vec{S} = \frac{1}{2} 
f_\alpha^\dagger \vec{\sigma}_{\alpha\beta} f_\beta ~;
\quad\quad f_\alpha^\dagger f_\alpha = 1 ~.
\label{f}
\end{equation}
In the mean field one assumes that the spinons do not interact with 
one another and are hopping freely on the 2D lattice.  
For the present problem the mean field Hamiltonian would have 
the spinons hopping in zero magnetic field, and the ground state would 
correspond to filling up a spinon Fermi sea.
In doing this one has artificially enlarged the Hilbert space, 
since the spinon hopping Hamiltonian allows for unoccupied and 
doubly-occupied sites, which have no meaning in terms of the 
spin model of interest.  It is thus necessary to project back down
into the physical Hilbert space for the spin model, restricting the 
spinons to single occupancy.  If one is only interested in constructing 
a variational wavefunction, this can be readily achieved by the
Gutzwiller projection, where one simply drops all terms in the 
wavefunction with unoccupied or doubly-occupied sites.  
The alternate approach to implement the single occupancy constraint is 
by introducing a gauge field, a $U(1)$ gauge field in this instance, 
that is minimally coupled to the spinons in the hopping Hamiltonian.
This then becomes an intrinsically strongly-coupled lattice gauge field 
theory.  To proceed, it is necessary to resort to an approximation by 
assuming that the gauge field fluctuations are (in some sense) weak.  
In 2D one then analyzes the problem of a Fermi sea of spinons 
coupled to a weakly fluctuating gauge field.
This problem has a long history,\cite{IoffeLarkin, Holstein, Reizer, PALee, LeeNagaosa, Polchinski94, Altshuler94, YBKim94, LeeNagaosaWen} 
but all the authors have chosen to sum the same class of diagrams.  
Within this (uncontrolled) approximation one can then compute 
physical spin correlation functions, which are gauge invariant.  
It is unclear, however, whether this is theoretically legitimate, 
and even less clear whether or not the spin liquid phase thereby 
constructed captures correctly the universal properties of a 
physical spin liquid that can (or does) occur for some spin Hamiltonian.

Fortunately, on the zigzag chain we are in much better shape.  
Here it is possible to employ Bosonization to analyze the quasi-1D 
gauge theory, as we detail below.
While this still does not give an exact solution for the ground state 
of any spin Hamiltonian, with regard to capturing universal low energy 
properties it is controlled.
As we will see, the low energy effective theory for the SBM phase 
is a Gaussian field theory, and perturbations about this can be 
analyzed in a systematic fashion to check for stability of the SBM and 
possible instabilities into other phases.

As we will also briefly show, the low energy effective theory for the 
SBM can be obtained just as readily by starting with a model of 
interacting {\it electrons} hopping on the zigzag chain, i.e.\ a 
Hubbard-type Hamiltonian.
If one starts with interacting electrons, it is (in principle) possible 
to construct the gapped electron excitations in the SBM Mott insulator.
Within the gauge theory approach, the analogous gapped spinon 
excitations are unphysical, being confined together with a linear 
potential.  Moreover, within the electron formulation one can access the 
metallic phase, and also the Mott transition to the SBM insulator.

%%%%%%%%%%%%%%%%%%%%%%%%%%%%%%%%%%%%%%%%%%%%%%%%%%%%%%%%%%%%%%%%%%%%%%%
%%%%%%%%%%%%%%%%%%%%%%%%%%%%%%%%%%%%%%%%%%%%%%%%%%%%%%%%%%%%%%%%%%%%%%%
\subsection{SBM via Bosonization of gauge theory}
\label{subsec:SBMspinon}

We first start by using Bosonization\cite{Shankar_Acta, Lin98, Fjaerestad02} 
to analyze the gauge theory.\cite{KimLee, Hosotani_1, Mudry}
Motivated by the 2D triangular lattice with ring exchanges,
we assume a mean field state in which the spinons are hopping in 
zero flux.
Here the spinons are hopping on the zigzag strip with near-neighbor 
and second-neighbor hopping strengths denoted $t_1$ and $t_2$.
This is equivalent to a strictly 1D chain with likewise first- and 
second-neighbor hopping.  The dispersion is 
\begin{equation}
\xi(k) = - 2 t_1 \cos(k) - 2 t_2 \cos(2k) - \mu ~.
\label{xi}
\end{equation}
For $t_2 > 0.5 t_1$ there are two sets of Fermi crossings at wave 
vectors $\pm k_{F1}$ and $\pm k_{F2}$ as shown in Fig.~\ref{fig:fbands}.
Our convention is that fermions near $k_{F1}$ and $k_{F2}$ are moving
to the right; the corresponding group velocities are $v_1, v_2 > 0$.  
The spinons are at half-filling, which implies 
$k_{F1} + k_{F2} = -\pi/2 \mod 2\pi$.

The spinon operators are expanded in terms of continuum fields,
\begin{equation}
f_\alpha(x) =  \sum_{a, P} e^{i P k_{Fa} x} f_{Pa\alpha} ~,
\end{equation}
with $a=1,2$ denoting the two Fermi seas, $\alpha = \up,\dn$ denoting the
spin, and $P = R/L = \pm$ denoting the right and left moving fermions.
We now use Bosonization,\cite{Shankar_Acta, Lin98, Fjaerestad02}
re-expressing these low energy spinon operators with Bosonic fields,
\begin{equation}
f_{P a \alpha} = \eta_{a\alpha} 
e^{ i (\varphi_{a\alpha} + P \theta_{a\alpha}) } ~,
\label{fbosonize}
\end{equation}
with canonically conjugate boson fields:
\begin{eqnarray}
[\varphi_{a\alpha}(x) , \varphi_{b\beta}(x^\prime)] &=&
[\theta_{a\alpha}(x) , \theta_{b\beta}(x^\prime)] = 0 ~, \\ ~
[\varphi_{a\alpha}(x) , \theta_{b\beta}(x^\prime)] &=& 
i \pi \delta_{ab} \delta_{\alpha\beta} \, \Theta(x - x^\prime) ~,
\end{eqnarray}
where $\Theta(x)$ is the Heaviside step function.
Here, we have introduced Klein factors, the Majorana fermions
$\{ \eta_{a\alpha}, \eta_{b\beta} \} 
 = 2 \delta_{ab} \delta_{\alpha \beta}$,
which assure that the spinon fields with different flavors 
anti-commute with one another.
The (slowly varying) fermionic densities are simply
$f_{P a \alpha}^\dagger f_{P a \alpha} = 
\partial_x (P\varphi_{a\alpha} + \theta_{a\alpha})/(2\pi)$.

A faithful formulation of the physical system in this slave particle
approach Eq.~(\ref{f}) is a compact U(1) lattice gauge theory. 
In 1+1D continuum theory, we work in the gauge eliminating spatial 
components of the gauge field.
The imaginary-time bosonized Lagrangian density is then:
\begin{equation}
{\cal L} = \frac{1}{2\pi} \sum_{a\alpha} 
\left[ \frac{1}{v_a} (\partial_\tau \theta_{a \alpha})^2
       + v_a (\partial_x \theta_{a \alpha})^2 \right] 
+ {\cal L}_A  ~.
\label{Lbosonized}
\end{equation}
Here ${\cal L}_A$ encodes the coupling to the slowly varying 1D (scalar) 
potential field $A(x)$,
\begin{equation}
{\cal L}_A = \frac{1}{m} (\partial_x A/\pi)^2 + i \rho_A A ~,
\end{equation}
where $\rho_A$ denotes the total ``gauge charge'' density,
\begin{equation}
\rho_A =  \sum_{a\alpha} \partial_x \theta_{a\alpha} /\pi ~.
\end{equation}
It is useful to define ``charge" and ``spin" boson fields, 
\begin{equation}
\theta_{a \rho /\sigma} = \frac{1}{\sqrt{2}}
(\theta_{a \uparrow} \pm \theta_{a \downarrow}) ~,
\end{equation}
and ``even" and ``odd" flavor combinations,
\begin{equation}
\theta_{\mu \pm} = \frac{1}{\sqrt{2}} 
(\theta_{1\mu} \pm \theta_{2\mu}) ~, 
\end{equation}
with $\mu = \rho, \sigma$.  Similar definitions hold for the 
$\varphi$ fields.  The commutation relations for the new 
$\theta, \varphi$ fields are unchanged.

Integration over the gauge potential generates a mass term,
\begin{equation}
{\cal L}_A =  m ( \theta_{\rho +} - \theta_{\rho +}^{(0)} )^2 ~,
\label{m4rho+}
\end{equation}
for the field $\theta_{\rho +} = \sum_{a\alpha} \theta_{a \alpha}/2$.
In the gauge theory analysis, we cannot determine the mean value 
$\theta_{\rho +}^{(0)}$, which is important for detailed properties 
of the SBM in Appendix~\ref{app:SBMprops} as well as for the discussion 
of nearby phases in Secs.~\ref{subsec:g.gt.1}-\ref{subsec:commens_other}.
But if we start with an interacting electron model, one can readily 
argue that the correct value in the SBM phase satisfies
\begin{eqnarray}
4 \theta_{\rho +}^{(0)} = \pi \mod 2\pi ~.
\label{u8pos}
\end{eqnarray}

%%%%%%%%%%%%%%%%%%%%%%%%%%%%%%%%%%%%%%%%%%%%%%%%%%%%%%%%%%%%%%%%%%%%%%%
%%%%%%%%%%%%%%%%%%%%%%%%%%%%%%%%%%%%%%%%%%%%%%%%%%%%%%%%%%%%%%%%%%%%%%%
\subsection{SBM by Bosonizing interacting electrons}
\label{subsec:SBMelectron}

Consider then a model of {\it electrons} hopping on the zigzag strip.
We assume that the electron hopping Hamiltonian is identical to the 
spinon mean field Hamiltonian, with first and second neighbor hopping
strengths, $t_1, t_2$;
\begin{eqnarray}
H \!&=&\! -\sum_x [t_1 c^\dagger_\alpha(x) c_\alpha(x+1) 
                   + t_2 c^\dagger_\alpha(x) c_\alpha(x+2) + \Hc] 
\nonumber \\
&& + H_{int} ~.
\end{eqnarray}
The electrons are taken to be at half-filling.
The interactions between the electrons could be taken as a Hubbard 
repulsion, perhaps augmented with further neighbor interactions,
but we do not need to specify the precise form for what follows.

For $t_2 < 0.5 t_1$, the electron Fermi sea has only one segment
spanning $[-\pi/2, \pi/2]$, and at low energy the model is essentially 
the same as the 1D Hubbard model.  We know that in this case even an 
arbitrary weak repulsive interaction will induce an allowed 
four-fermion Umklapp term that will be marginally relevant 
driving the system into a 1D Mott insulator.
The residual spin sector will be described in terms of the Heisenberg 
chain, and is expected to be in the gapless Bethe-chain phase.

On the other hand, for $t_2 > 0.5 t_1$, the electron band has two 
Fermi seas as shown in Fig.~\ref{fig:fbands}.  This is the case of 
primary interest to us.  As in the one-band case, Umklapp terms are 
required to drive the system into a Mott insulator.
But in this two-band case there are no allowed four-fermion
Umklapp terms.  While it is possible to study perturbatively the effects
of the momentum conserving four-fermion interactions and address 
whether or not the two-band metal is stable for some particular form 
of the lattice Coulomb repulsion, we do not pursue this here.  
Rather, we focus on the allowed eight-Fermion Umklapp term which takes 
the form, 
\begin{equation}
H_8 = v_8
(c_{R1\up}^\dagger c_{R1\dn}^\dagger c_{R2\up}^\dagger c_{R2\dn}^\dagger
 c_{L1\up} c_{L1\dn} c_{L2\up} c_{L2\dn} + \Hc ) ~,
\end{equation}
where we have introduced slowly varying electron fields for the 
two bands, at the right and left Fermi points.
For repulsive electron interactions we have $v_8 > 0$.
This Umklapp term is strongly irrelevant at weak coupling since its 
scaling dimension is $\Delta_8 = 4$ (each electron field has 
scaling dimension $1/2$),
much larger than the space-time dimension $D = 2$.

To make progress we can Bosonize the electrons, just as we did for the 
spinons,
$c_{P a \alpha} \sim e^{ i (\varphi_{a\alpha} + P \theta_{a\alpha}) }$.
The eight-Fermion Umklapp term becomes,
\begin{equation}
H_8 = 2 v_8 \cos(4 \theta_{\rho+}) ~,
\label{H8bosonized}
\end{equation}
where as before $\theta_{\rho +} = \sum_{a \alpha} \theta_{a \alpha} /2$
and $\rho_e(x) = 2 \partial_x \theta_{\rho +}/\pi$ is now the 
{\it physical} slowly varying {\it electron} density.
The Bosonized form of the non-interacting electron Hamiltonian 
is precisely the first part of Eq.~(\ref{Lbosonized}), and one can 
readily confirm that $\Delta_8 = 4$.
But now imagine adding a {\it strong} density-density repulsion between 
the electrons.  The slowly varying contributions, on scales somewhat 
larger than the lattice spacing, will take the simple form, 
$H_\rho \sim V_\rho \rho_e^2(x) 
\sim V_\rho (\partial_x \theta_{\rho +})^2$.
These forward scattering interactions will ``stiffen" the 
$\theta_{\rho +}$ field and will reduce the scaling dimension $\Delta_8$.
If $\Delta_8$ drops below $2$ then the Umklapp term becomes relevant and
will grow at long scales.  This destabilizes the two-band metallic state,
driving a Mott metal-insulator transition.
The $\theta_{\rho +}$ field gets pinned in the minima of the $H_8$
potential, which gives Eq.~(\ref{u8pos}).
Expanding to quadratic order about the minimum gives a mass term of the 
form Eq.~(\ref{m4rho+}).  For the low energy spin physics of primary 
interest this shows the equivalence between the direct Bosonization of 
the electron model and the spinon gauge theory approach.

The difference between the spinon gauge theory and the interacting 
electron theory are manifest in the charge sector.  In the latter case 
the electron excitations $c^\dagger$ above the gap will correspond to 
instantons connecting adjacent minima of the cosine potential 
Eq.~(\ref{H8bosonized}).  In the spinon gauge theory there are 
no such fermionic excitations $f^\dagger$,
and the spinon excitations are linearly confined.
This is appropriate for the spin model which has no ``charge sector",
and no notion of spinons.  In the weak Mott insulating phase of the 
electron model, the Fermi wavevectors $k_{F1}, k_{F2}$ denote the 
momenta of the minimum energy gapped electron excitations.
What is the meaning, then, of the spinon Fermi wavevectors if the 
spinon excitations are unphysical?   
Within the spinon gauge theory the only gauge invariant (i.e.\ physical) 
momenta are the sums and differences of $k_{F1},k_{F2}$, which 
correspond to momenta of the (low energy) spin excitations.
In the electron model, the spin excitations below the charge 
localization length of the Mott insulator will be similar to that of 
electrons in the metal.  On longer scales, the spin sector remains 
gapless, and this is the regime described below by the low energy 
effective theory of the SBM Mott insulator.
It is these physical longer length scale spin excitations which are
correctly captured by both the spinon gauge theory and interacting 
electron approaches.

%%%%%%%%%%%%%%%%%%%%%%%%%%%%%%%%%%%%%%%%%%%%%%%%%%%%%%%%%%%%%%%%%%%%%%%
%%%%%%%%%%%%%%%%%%%%%%%%%%%%%%%%%%%%%%%%%%%%%%%%%%%%%%%%%%%%%%%%%%%%%%%
\subsection{Fixed point theory of the SBM phase}
\label{subsec:SBMfp}

The low energy spin physics in either formulation can be obtained by
integrating out the massive $\theta_{\rho +}$ field, as we now 
demonstrate.  Performing this Gaussian integration leads to the 
effective fixed-point (quadratic) Lagrangian for the SBM spin liquid:
\begin{equation}
{\cal L}_0^{\rm SBM} =  {\cal L}_0^\rho + {\cal L}_0^\sigma ~,
\label{LSBM0}
\end{equation}
with the ``charge" sector contribution,
\begin{equation}
{\cal L}_0^\rho =  \frac{1}{2\pi g_0} 
\left[ \frac{1}{v_0} (\partial_\tau \theta_{\rho -})^2
       + v_0 (\partial_x \theta_{\rho -})^2 \right] ~,
\label{L0rho}
\end{equation}
and the spin sector contribution,
\begin{equation}
{\cal L}_0^\sigma = \frac{1}{2\pi} \sum_{a} 
\left[ \frac{1}{v_a} (\partial_\tau \theta_{a \sigma})^2
       + v_a (\partial_x \theta_{a \sigma})^2 \right] ~.
\label{L0sigma}
\end{equation}
The velocity $v_0$ in the ``charge" sector depends on the 
product of the flavor velocities, $v_0 = \sqrt{v_1 v_2}$, while the 
dimensionless ``conductance" depends on their ratio:
\begin{equation}
g_0 = \frac{2}{\sqrt{v_1/v_2} + \sqrt{v_2/v_1}} ~.
\label{g0}
\end{equation}
Notice that $g_0 \le 1$, with $g_0 \to 0$ upon approaching the 
limit of a single Fermi surface ($v_1 \neq 0$, $v_2 \to 0$),
and $g_0 \to 1$ in the limit of two equally-sized Fermi surfaces
($v_2/v_1 \to 1$) that occurs when the two legs of the 
triangular strip decouple.

In Sec.~\ref{subsec:stability}, we also consider all symmetry allowed 
residual short-range interactions between the low energy degrees of 
freedom and conclude that the above fixed-point theory can indeed 
describe a stable phase, with the only modification that $g_0 \to g$ is 
now a general Luttinger parameter.  Stability requires $g < 1$.
There are also three marginal interactions that need to have 
appropriate signs to be marginally irrelevant.

The gapless excitations in the SBM lead to power law correlations in 
various physical quantities at wavevectors connecting the Fermi points.
Here and in the numerical study Sec.~\ref{sec:HringDMRG}, 
we focus on the following observables: 
spin $\vec{S}(x)$, bond energy ${\cal B}(x)$ (i.e., VBS order parameter),
and spin chirality $\chi(x)$:
\begin{eqnarray}
{\cal B}(x) &=& \vec{S}(x) \cdot \vec{S}(x+1) ~, 
\label{B_def} \\ 
\chi(x) &=& \vec{S}(x-1) \cdot [\vec{S}(x) \times \vec{S}(x+1)] ~.
\label{chi_def}
\end{eqnarray}
In Appendix~\ref{app:SBMprops}, we give detailed expressions in the
continuum theory.  
The most straightforward contributions are obtained by writing out,
e.g., $\vec{S}(x) \sim f^\dagger(x) \vec{\sigma} f(x)$ in terms of the 
continuum fermion fields and then bosonizing 
[see also Eqs.~(\ref{B_via_f}) and (\ref{chi_via_f}) for ${\cal B}(x)$ 
and $\chi(x)$].
We expect dominant power laws at wavevectors $\pm 2 k_{Fa}$ and 
$\pm \pi/2 = \mp (k_{F1}+k_{F2})$, originating from fermion bilinears 
composed of a particle and a hole moving in opposite directions.
Such bilinears become enhanced upon projecting down into the spin sector 
(i.e.\ upon integrating out the massive $\theta_{\rho +}$ in the 
Bosonized field theory), and it is possible to compute the scaling 
dimension of any operator in terms of the single Luttinger parameter, 
$g$.  It is also important to consider more general contributions, e.g., 
containing four fermion fields; this is best done using symmetry 
arguments and the corresponding expressions can be found in 
Appendix~\ref{app:SBMprops}.

Table~\ref{tab:SBMprops} summarizes such analysis of the observables
by listing scaling dimensions at various wavevectors.
We describe power law correlation of a given operator $A$ at a 
wavevector $Q$ by specifying the scaling dimension $\Delta_{A_Q}$
defined from the real-space decay
\begin{equation}
\la A(x) A(0) \ra \sim \sum_Q \frac{e^{i Q x}}{|x|^{2\Delta_{A_Q}}} ~.
\end{equation}
The corresponding static structure factor (i.e.\ Fourier transform) 
has momentum-space singularity $\sim |q-Q|^{2\Delta_{A_Q} - 1}$.

\begin{table}
\begin{tabular}{|c|c|c|c|c|c|c|}
\hline
\multicolumn{1}{|c|}{} &
\multicolumn{1}{|c|}{\multirow{2}{*}{$Q=\, 0$}} &
\multicolumn{1}{|c|}{$\pm 2k_{F1}$;} &
\multicolumn{1}{|c|}{\multirow{2}{*}{$\pm \pi/2$}} &
\multicolumn{1}{|c|}{$\pm (k_{F2}-k_{F1})$;} &
\multicolumn{1}{|c|}{\multirow{2}{*}{$\pi$}} &
\multicolumn{1}{|c|}{\multirow{2}{*}{$\pm 4k_{F1}$}} 
\\
\multicolumn{1}{|c|}{} &
\multicolumn{1}{|c|}{\multirow{2}{*}{}} &
\multicolumn{1}{|c|}{$\pm 2k_{F2}$} &
\multicolumn{1}{|c|}{\multirow{2}{*}{}} &
\multicolumn{1}{|c|}{$\pm (3k_{F1}+k_{F2})$} &
\multicolumn{1}{|c|}{\multirow{2}{*}{}} &
\multicolumn{1}{|c|}{\multirow{2}{*}{}}
\\ \hline
\multicolumn{1}{|c|}{$\vec{S}$} &
\multicolumn{1}{|c|}{\multirow{3}{*}{$1$}} &
\multicolumn{1}{|c|}{\multirow{2}{*}{$\frac{1}{2} + \frac{g}{4}$}} &
\multicolumn{1}{|c|}{\multirow{3}{*}{$\frac{1}{2} + \frac{1}{4g}$}} &
\multicolumn{1}{|c|}{\multirow{3}{*}{$\frac{1}{2} + \frac{1}{4g} + \frac{g}{4}$}} & 
\multicolumn{1}{|c|}{1} &
\multicolumn{1}{|c|}{subd.}
\\ \cline{1-1} \cline{6-7}
\multicolumn{1}{|c|}{$\cal B$} &
\multicolumn{1}{|c|}{\multirow{3}{*}{}} &
\multicolumn{1}{|c|}{\multirow{2}{*}{}} &
\multicolumn{1}{|c|}{\multirow{3}{*}{}} &
\multicolumn{1}{|c|}{\multirow{3}{*}{}} &
\multicolumn{1}{|c|}{1} &
\multicolumn{1}{|c|}{$g$} 
\\ \cline{1-1} \cline{3-3} \cline{6-7}
\multicolumn{1}{|c|}{$\chi$} &
\multicolumn{1}{|c|}{\multirow{3}{*}{}} &
\multicolumn{1}{|c|}{subd.} &
\multicolumn{1}{|c|}{\multirow{3}{*}{}} &
\multicolumn{1}{|c|}{\multirow{3}{*}{}} &
\multicolumn{1}{|c|}{$1/g$} &
\multicolumn{1}{|c|}{subd.} 
\\ \hline
\end{tabular}
\caption{Spin Bose-Metal fixed-point theory:
Scaling dimensions of the spin $\vec{S}$, bond energy $\cal B$, and 
chirality $\chi$ observables at various wavevectors $Q$ in the top row.
Entries with subdominant power laws are listed as ``subd.''.
}
\label{tab:SBMprops}
\end{table}

The $Q=0$ entries in Table~\ref{tab:SBMprops} come from simple 
identifications
\begin{eqnarray}
S^z_{Q=0} &\sim& \partial_x \theta_{1\sigma} + \partial_x \theta_{2\sigma} ~, 
\label{SzQ0} \\
{\cal B}_{Q=0} &\sim& \partial_x \theta_{\rho-} ~, 
\label{BQ0} \\
\chi_{Q=0} &\sim& \partial_x \varphi_{\rho-} ~.
\label{chiQ0}
\end{eqnarray}
In particular, the last line provides physical meaning to the ``$\rho-$''
sector -- this spin-singlet sector encodes low energy fluctuations of the
chirality.
A direct way to observe the propagating $\rho-$ mode would be to measure
the spectral function of the chirality, while in the present DMRG study 
we detect it by a $|q|$ (i.e., V-shaped) behavior in the static 
structure factor at small wavevector $q$.

%%%%%%%%%%%%%%%%%%%%%%%%%%%%%%%%%%%%%%%%%%%%%%%%%%%%%%%%%%%%%%%%%%%%%
%%%%%%%%%%%%%%%%%%%%%%%%%%%%%%%%%%%%%%%%%%%%%%%%%%%%%%%%%%%%%%%%%%%%%
%%%%%%%%%%%%%%%%%%%%%%%%%%%%%%%%%%%%%%%%%%%%%%%%%%%%%%%%%%%%%%%%%%%%%

\section{DMRG study of the Spin Bose-Metal in the 
$J_1 - J_2 - K_{\rm ring}$ model on the zigzag chain}
\label{sec:HringDMRG}

We study the ring model Eq.~(\ref{Hring}) on the two-leg triangular
strip shown in Fig.~\ref{fig:model}.  We use the 1D chain picture and 
take site labels $x=1, \dots, L$ where $L$ is the length of the system. 
We use exact diagonalization (ED) and density matrix renormalization 
group (DMRG)\cite{White_dmrg1, White_dmrg2, Schollwock} methods 
supplemented with variational Monte Carlo (VMC)\cite{Ceperley77, Gros89} 
to determine the nature of the ground state of the Hamiltonian 
Eq.~(\ref{Hring}).

\subsection{Measurement details}
\label{subsec:measurmnts}
We first describe numerical measurements.
All calculations use periodic boundary conditions in the $\hat{x}$ 
direction.  In the ED, we can characterize states by a momentum quantum 
number $k$.  On the other hand, our DMRG calculations are done with 
real-valued wavefunctions.  This gives no ambiguity when the 
ground state carries momentum $0$ or $\pi$ and is unique.  
However, if the ground state carries nontrivial momentum $k \neq 0, \pi$,
then its time-reversed partner carries $-k$, and the DMRG state is some 
combination of these.
While the real space measurements depend on the specific combination 
in the finite system, the momentum space measurements described below 
do not depend on this and are unique. 
Most of the calculations are done in the sector with $S^z_{\rm tot} = 0$,
which contains any ground state of the SU(2)-invariant system.

The DMRG calculations keep more than $m=3200$ states per block
\cite{White_dmrg1, White_dmrg2, Schollwock} to ensure accurate results, 
and the density matrix truncation error is of the order of $10^{-6}$.  
Typical relative error for the ground-state energy is 
of the order of $10^{-4}$ or smaller for the systems we have studied.
Using ED, we have confirmed that all DMRG results are numerically exact 
when the system size is $L=24$.
The DMRG convergence depends strongly on the phase being studied, 
the system size, the type of the correlations, and the distance 
between operators.  
In the Bethe-chain and VBS-2 phases there is still good
convergence for size $L=192$, while in the SBM we are limited to 
$L = 96 - 144$ systems.
The entanglement entropy calculations are done with up to $m=6000$
states in each block, which is necessary for capturing the long range 
entanglement in the SBM states where we find an effective 
``central charge'' $c \simeq 3$.

We have already specified the main observables in Sec.~\ref{subsec:SBMfp}
[cf.~Eqs.~(\ref{B_def}-\ref{chi_def})].
We measure spin correlations, bond energy (dimer) correlations, 
and chirality correlations defined as follows:
\begin{eqnarray}
C(x, x') &=& \la \vec{S}(x) \cdot \vec{S}(x') \ra ~,  \\
D(x, x') &=& \la {\cal B}(x) {\cal B}(x') \ra  - {\la {\cal B} \ra}^2 ~,
\\
X(x, x') &=& \la \chi(x) \chi(x') \ra ~.
\end{eqnarray}
For simplicity, we set $D(x, x') = 0$ if bonds ${\cal B}(x)$ and
${\cal B}(x')$ have common sites, and similarly for $X(x, x')$.
Structure factors $C(q)$, $D(q)$, and $X(q)$ are obtained through
Fourier transformation:
\begin{eqnarray}
O(q) = \frac{1}{L} \sum_{x, x'} O(x, x') e^{-i q (x-x')} ~,
\label{Oq}
\end{eqnarray}
where $O = C, D, X$.  We have $O(q) = O(-q)$ and usually show only 
$0 \leq q \leq \pi$.
The spin structure factor at $q=0$ gives the total spin in the 
ground state:
\begin{eqnarray}
C(q=0) = \frac{\la \vec{S}_{\rm tot}^2 \ra}{L}
= \frac{S_{\rm tot} (S_{\rm tot} + 1)}{L} ~.
\label{Stot}
\end{eqnarray}
In all figures, we loosely use ``$\la {\cal B}_q {\cal B}_{-q} \ra$''
and ``$\la \chi_q \chi_{-q} \ra$'' to denote $D(q)$ and $X(q)$
respectively.

Turning to the VMC calculations, the trial wavefunctions are described
in broad terms in Sec.~\ref{sec:intro} and in more detail in 
Appendix~\ref{app:Gutzw}.  The states are labeled by occupation numbers 
of the two Fermi seas, $(N_1, N_2)$.  Since $N_1 + N_2 = L/2$, 
there is only one variational parameter.
There are three distinct regimes: 
{\it i)} $N_1 = L/2$, $N_2 = 0$, i.e., a single Fermi sea, which is 
appropriate for the Bethe-chain phase; 
{\it ii)} $N_1 \neq N_2 \neq 0$ appropriate for the SBM;
and {\it iii)} $N_1 = N_2 = L/4$, i.e., decoupled legs, which is a 
reasonable starting point for the large $J_2$ limit.

In Appendix~\ref{app:Gutzw}, we describe correlations in the 
generic SBM wavefunctions and identify characteristic wavevectors 
$2k_{Fa} = 2\pi N_a/L$, $a=1,2$, and also $\pi/2$
(see Sec.~\ref{sec:SBMtheory} and Table~\ref{tab:SBMprops}). 
One observation is that such wavefunctions correspond to a special 
case $g=1$ in the SBM theory and thus cannot capture general exponents.
Despite this shortcoming, the wavefunctions capture the locations of 
the singular wavevectors observed in the DMRG.
We also try to improve the wavefunctions by using a
``gapless superconductor'' modification described in 
Appendix~\ref{subapp:gaplessSC} and designed to preserve the 
singular wavevectors while allowing more variational parameters.  
This indeed improves the trial energy and provides better match with the 
DMRG correlations at short scales, even if the long distance properties 
are still not captured fully.
When presenting the DMRG structure factors, we also show the 
corresponding VMC results for wavefunctions determined by minimizing 
the trial energy over the described family of states.

Using the DMRG, we find four distinct quantum phases in the
$J_2/J_1 - K_{\rm ring}/J_1$ plane as illustrated in 
Fig.~\ref{fig:phased_J3_0p0}.
In the small $K_{\rm ring}$ region, we have the conventional 
Bethe-chain phase at small $J_2$ and Valence Bond Solid
state with period 2 (VBS-2) at larger $J_2$.
The SBM phase emerges in the regime $K_{\rm ring} > 0.2 J_1$
and dominates the intermediate parameter space.
Inside this region, we discover a new VBS state with period 3 (VBS-3).
To fully understand the VBS-3 (in particular, its relationship to the 
flanking Spin Bose-Metals) we will need the stability analysis of the 
SBM in Sec.~\ref{sec:SBMstability}, while the DMRG results are discussed 
afterwards in Sec.~\ref{sec:DMRG_commens}.

We explore more finely a cut $K_{\rm ring} = J_1$ through the phase
diagram Fig.~\ref{fig:phased_J3_0p0}, and our presentation points are 
taken from this cut.  The Bethe-chain and VBS-2 phases are fairly 
conventional (for this $K_{\rm ring}$, the VBS-2 is close to the 
decoupled-legs state at large $J_2$ values).  Nevertheless, it is 
useful to see measurements in these phases for comparisons.  
Such examples are given in Appendix~\ref{app:DMRG4nonSBM}, while here 
we focus on the Spin Bose-Metal point deep in the phase.  We will 
discuss more difficult parts of the phase diagram 
Fig.~\ref{fig:phased_J3_0p0} once we have the overall picture of the SBM.

%%%%%%%%%%%%%%%%%%%%%%%%%%%%%%%%%%%%%%%%%%%%%%%%%%%%%%%%%%%%%%%%%%%%%%
%%%%%%%%%%%%%%%%%%%%%%%%%%%%%%%%%%%%%%%%%%%%%%%%%%%%%%%%%%%%%%%%%%%%%%
\subsection{Representative Spin Bose-Metal points}
\label{subsec:DMRG:SBM}

\begin{figure}
\centerline{\includegraphics[width=\columnwidth]{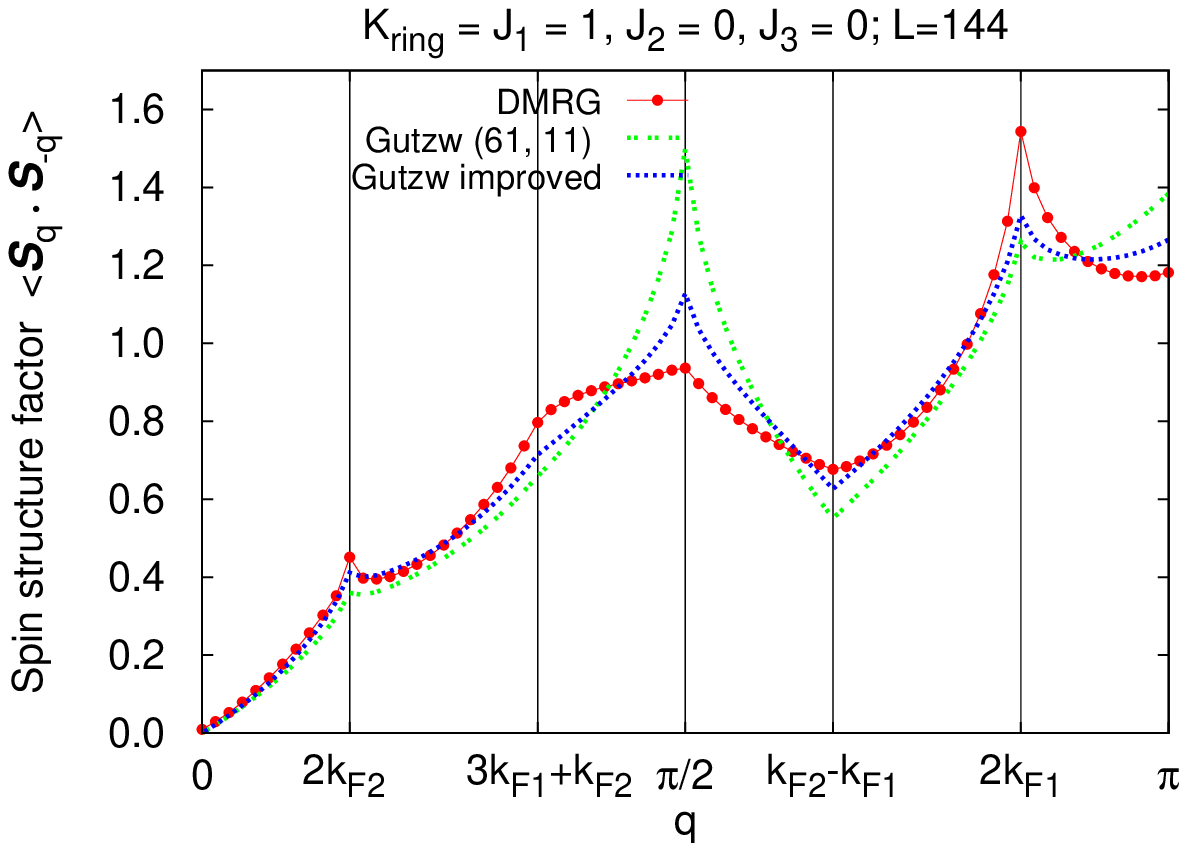}}
\centerline{\includegraphics[width=\columnwidth]{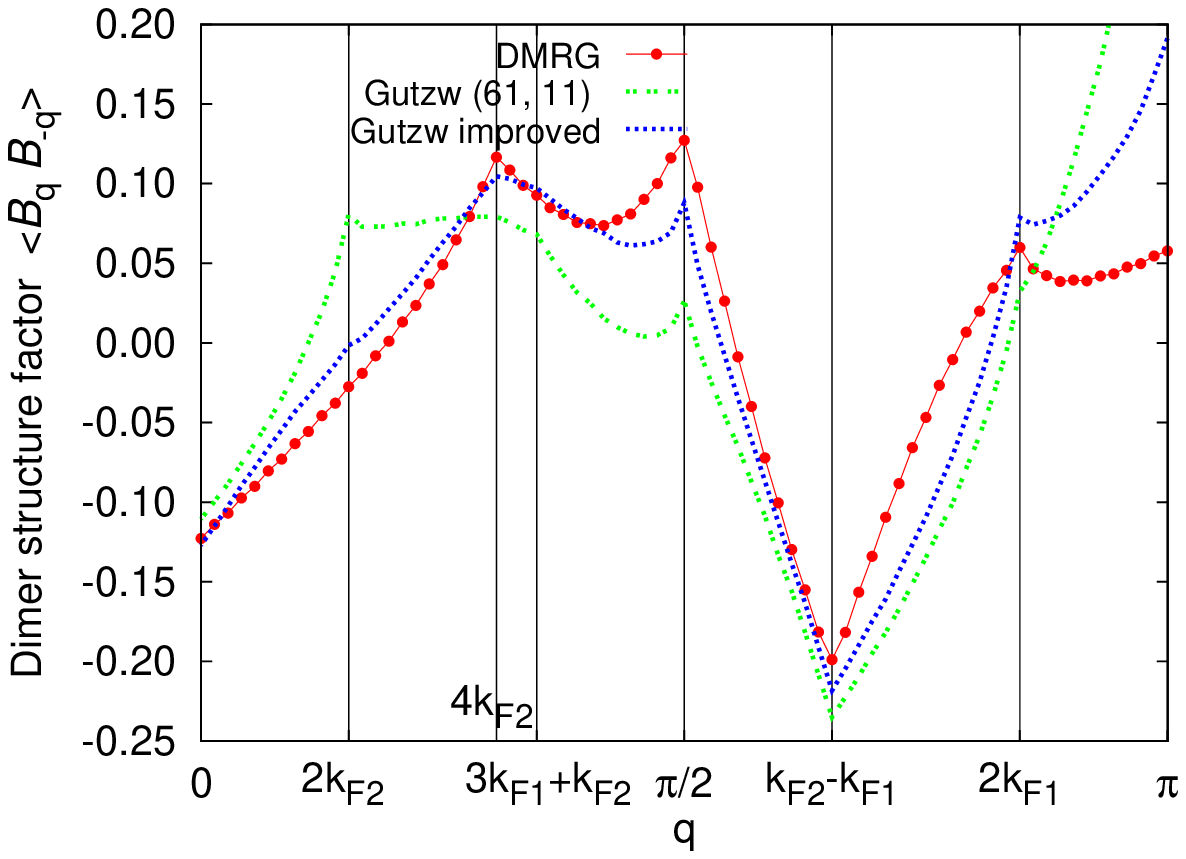}}
\centerline{\includegraphics[width=\columnwidth]{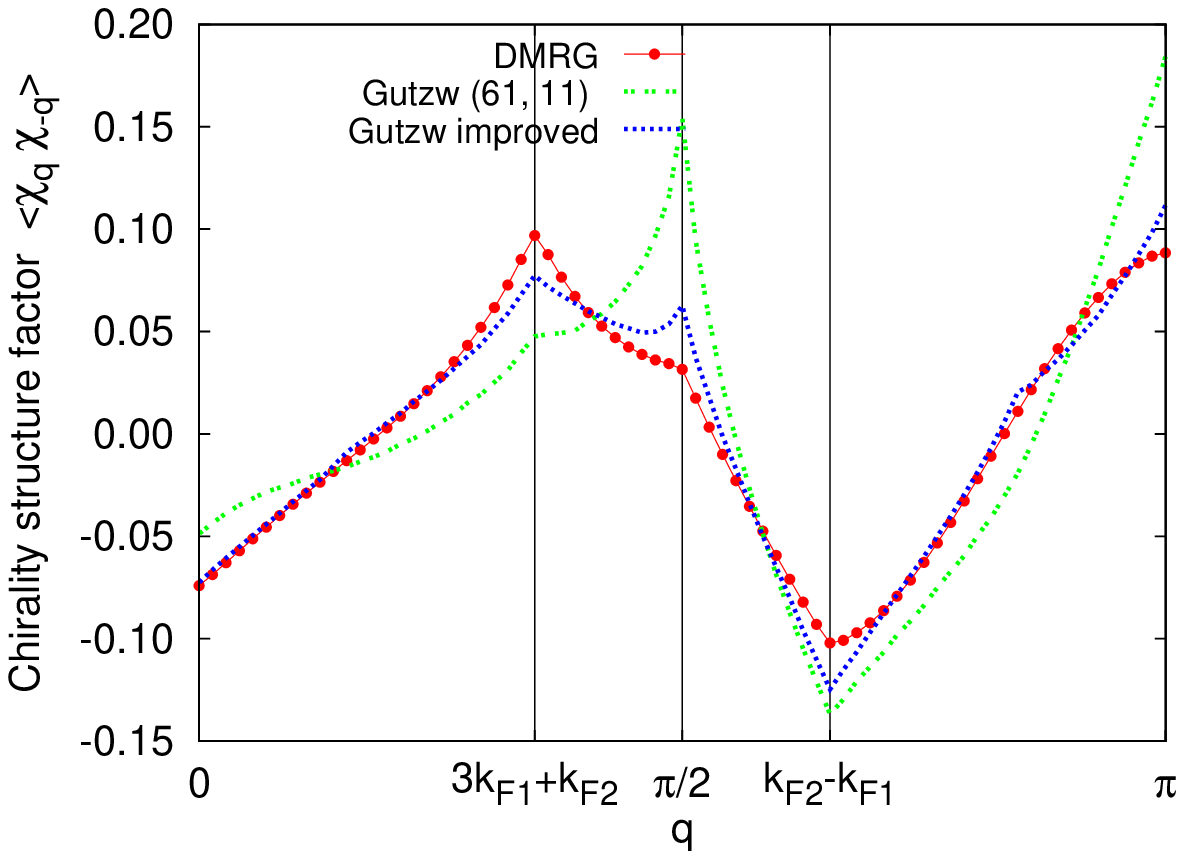}}
\caption{
(Color online)
Spin, dimer, and chirality structure factors at a
representative point in the Spin Bose-Metal phase, 
$K_{\rm ring} = J_1 = 1$, $J_2 = 0$ (close to the Bethe-chain phase), 
measured in the DMRG for system size $L = 144$. 
We also show the structure factors in the Gutzwiller projection of 
two Fermi seas $(N_1, N_2) = (61, 11)$, and in the improved Gutzwiller 
wavefunction with parameters $f_0 = 1, f_1 = 0.65, f_2 = -0.5$ 
(see Appendix~\ref{app:Gutzw}; the parameters are determined by 
optimizing the trial energy within the given family of states).
Vertical lines label important wavevectors expected in the SBM theory 
for such spinon Fermi sea volumes.
We discuss the comparison of the DMRG, VMC, and analytical theory
in the text.
}
\label{fig:J2_0p0_J3_0p0}
\end{figure}

\begin{figure}
\centerline{\includegraphics[width=\columnwidth]{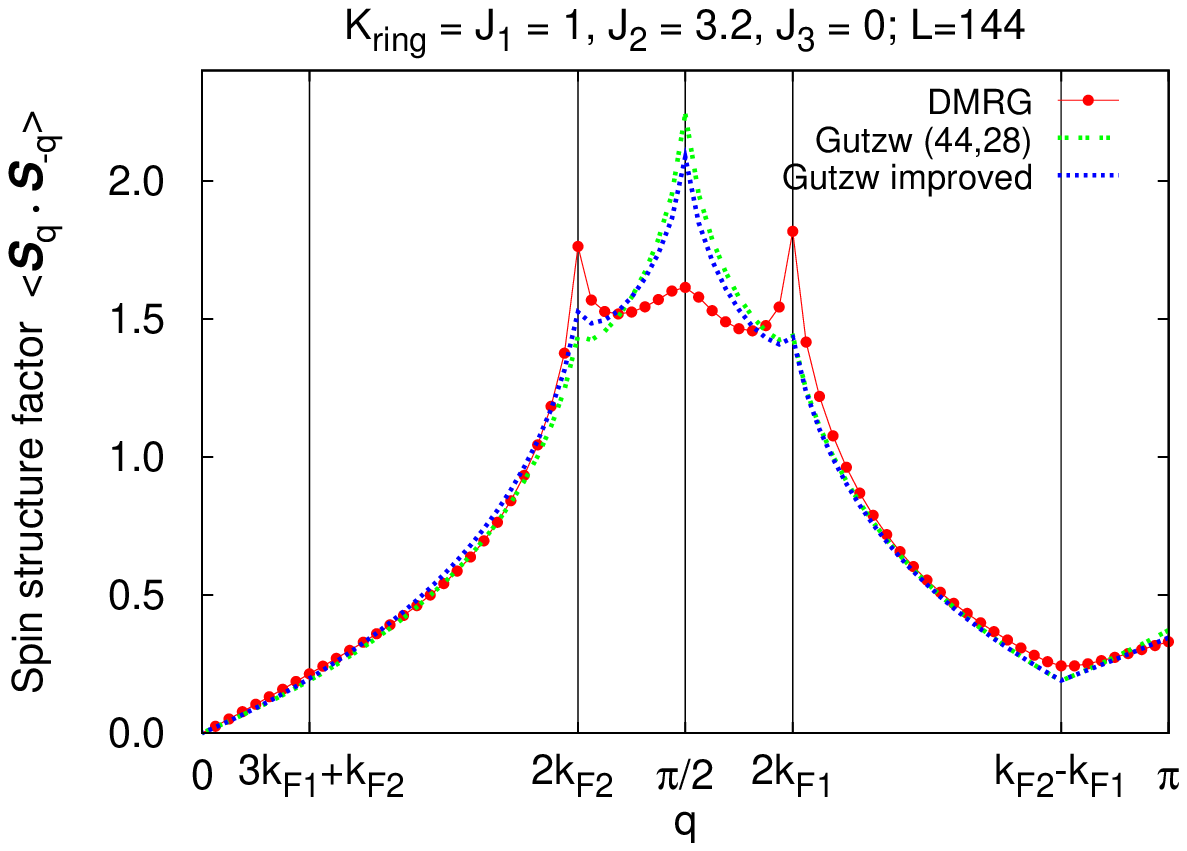}}
\centerline{\includegraphics[width=\columnwidth]{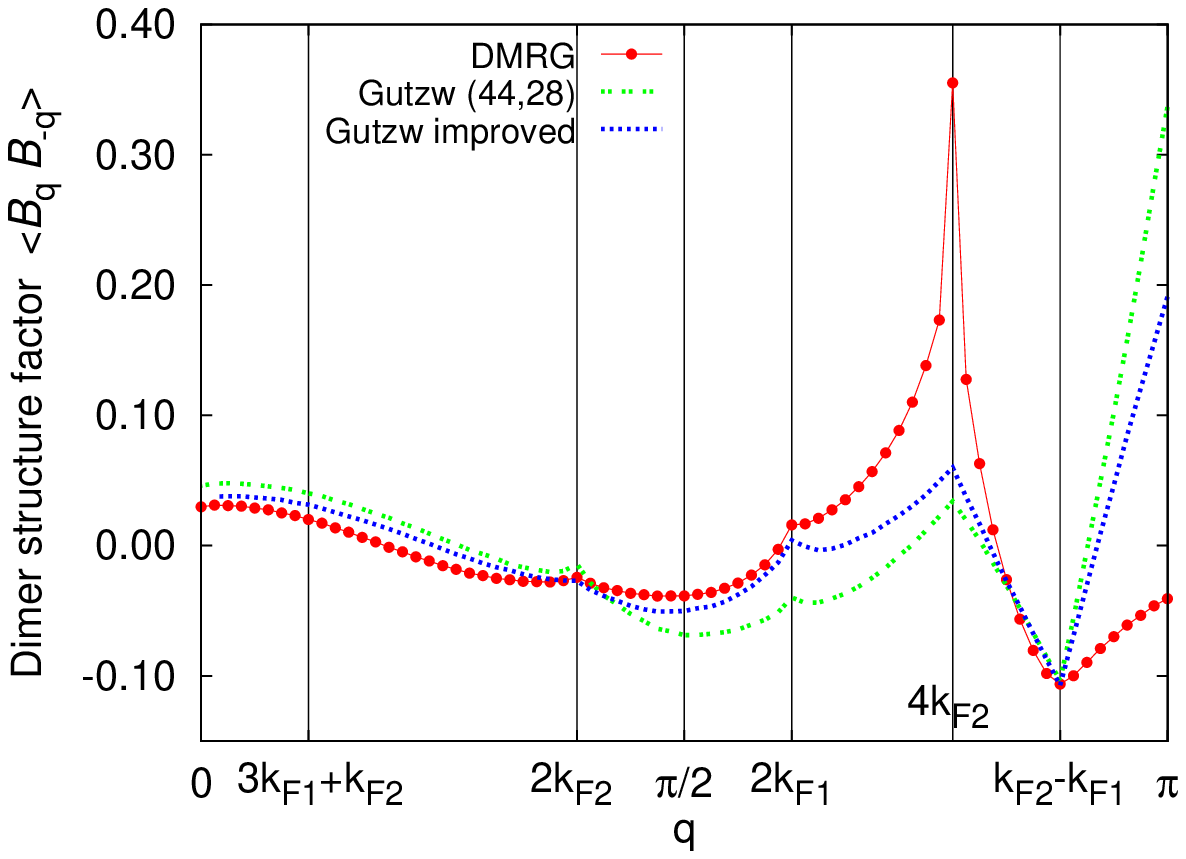}}
\centerline{\includegraphics[width=\columnwidth]{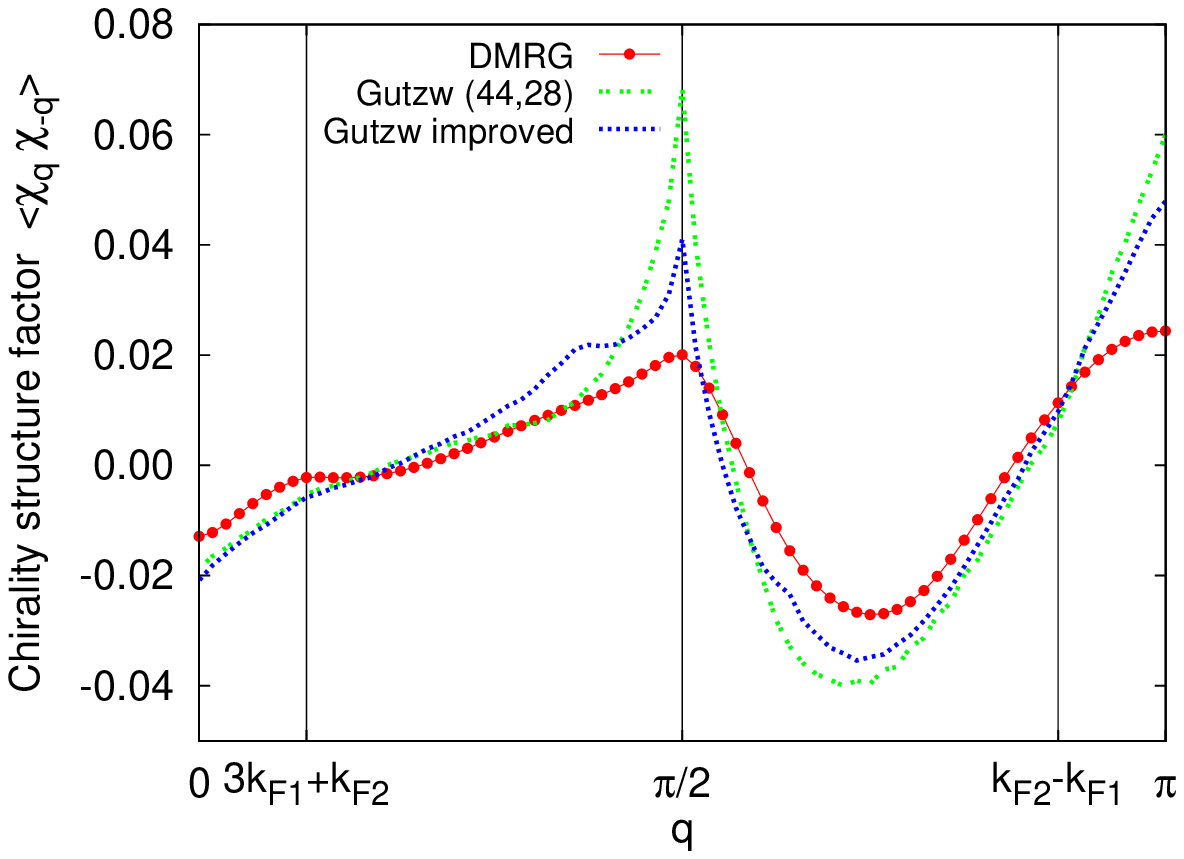}}
\caption{
(Color online)
Spin, dimer, and chirality structure factors at a
representative point in the Spin Bose-Metal phase, 
$K_{\rm ring} = J_1 = 1$, $J_2 = 3.2$ (between the VBS-3 and 
VBS-2 phases) measured in the DMRG for system size $L = 144$. 
We also show results in the Gutzwiller projection of two Fermi seas 
$(N_1, N_2) = (44, 28)$, and in the improved Gutzwiller wavefunction with
parameters $f_0 = 1, f_1 = 0.75, f_2 = -0.1$
(see Appendix~\ref{app:Gutzw} for details).
Vertical lines label important wavevectors expected in the SBM theory.
We discuss the comparison of the DMRG, VMC, and analytical theory
in the text.
}
\label{fig:J2_3p2_J3_0p0}
\end{figure}

Proceeding along the $K_{\rm ring} = J_1 = 1$ cut through 
Fig.~\ref{fig:phased_J3_0p0}, we start in the Bethe-chain phase
at large negative $J_2$ (a representative point is discussed in 
Appendix~\ref{subapp:DMRG:Bethe}).  As we change $J_2$ towards positive 
value, the system undergoes a transition at $J_2 = -0.6$.  
The new phase has characteristic spin correlations that are markedly 
different from the Bethe-chain phase.
Figure~\ref{fig:J2_0p0_J3_0p0} shows a representative point $J_2 = 0$.
The DMRG calculations are more difficult to converge and are done for 
smaller size $L = 144$ than in the Bethe-chain phase example 
(see also the entanglement entropy discussion below).

Comparing with the Bethe-chain state (e.g., Fig.~\ref{fig:J2_m1p0_J3_0p0}
in Appendix~\ref{subapp:DMRG:Bethe}), there is no $q=\pi$ dominance 
in the spin structure factor.  
Instead, we see three singular wavevectors located at 
$11 \times 2\pi/144$, $\pi/2$, and $61 \times 2\pi/144$.
Our Gutzwiller wavefunctions determined from the energetics have
$(N_1, N_2) = (61, 11)$, and the corresponding 
$2k_{F2}$, $-k_{F2}-k_{F1} = \pi/2 \mod 2\pi$, and $2k_{F1}$
match precisely the DMRG singular wavevectors.
The improved Gutzwiller wavefunction reproduces crude short-distance 
features better, but it has the same long-distance properties as the 
bare Gutzwiller.  
As discussed in Appendix~\ref{app:Gutzw}, our Gutzwiller wavefunctions 
do not capture all power laws predicted in the general analytical theory.
The wavefunctions appear to have equal exponents for spin correlations 
at these three wavevectors, while the theory summarized in 
Table~\ref{tab:SBMprops} gives stronger singularities at 
$2k_{F1}, 2k_{F2}$ and a weaker singularity at $\pi/2$.  
Very encouragingly, these theoretical expectations are consistent with 
what we find in the DMRG, where we can visibly see the difference in the 
behaviors at these wavevectors, particularly when we reference against 
the VMC results.

Similar discussion applies to the bond energy (dimer) correlations 
shown in the middle panel of Fig.~\ref{fig:J2_0p0_J3_0p0}.
The dominant features are at $2k_{F1}$ and $\pi/2$, and we also
see a peak at a wavevector identified as $4k_{F2}$, which is indeed
expected from the SBM theory, cf.~Table~\ref{tab:SBMprops}.
The theory predicts similar singularities at $2k_{F1}$ and $2k_{F2}$,
but for some reason we do not see the $2k_{F2}$ in the DMRG data,
even though it is clearly present in the bare Gutzwiller.
We suspect that this is caused by the narrowness in energy of the 
second Fermi sea when its population is low, so the amplitude of
the $2k_{F2}$ bond energy feature may be much smaller.
The $2k_{F2}$ can still show up in the bare Gutzwiller since, as we 
discuss in Appendix~\ref{app:Gutzw}, this wavefunction knows only about 
the Fermi sea sizes and not about the spinon energy scales like 
bandwidths or Fermi velocities.  The improved Gutzwiller clearly 
tries to remedy this, although within its limitations.  
The $4k_{F2}$ feature is not associated solely with the second 
Fermi sea and is less affected by this argument; 
indeed, $4k_{F2} = -4k_{F1}$, and both Fermi seas ``participate''
in producing this feature as can be seen from Eq.~(\ref{e4kF}).

Turning to the chirality structure factor in the bottom panel
of Fig.~\ref{fig:J2_0p0_J3_0p0},
we see a feature at $\pi/2$, which in the SBM theory is expected 
to have the same singularity as the spin and dimer at this wavevector.
We also see features at wavevectors $k_{F2} - k_{F1}$ and 
$3k_{F1} + k_{F2}$ in all our observables; these features are expected 
to be $\sim |\delta q|$ (i.e., V-shaped) in the Gutzwiller wavefunctions 
but have weaker singularity in the SBM theory, which is reasonably 
consistent with the DMRG measurements.
Very notable in the chirality structure factor is a $\sim |q|$ shape
at small $q$.  This can be viewed as a direct evidence for the
gapless ``$\rho-$'' mode in the SBM, cf.~Eq.~(\ref{chiQ0}).
On the other hand, a feature at $\pi$ is weaker than V-shaped, 
in contrast with the Gutzwiller wavefunctions but in agreement 
with the SBM theory expectations in Table~\ref{tab:SBMprops}.

To summarize, the correlations in the SBM phase are dramatically
different from the Bethe-chain phase, and we can match all the 
characteristic wavevectors using the Gutzwiller wavefunctions projecting 
two Fermi seas.  We also understand the failure of the wavefunctions 
to reproduce the nature of the singularities and the amplitudes, 
while the Bosonization theory of the SBM is consistent with all 
DMRG observations even when the wavefunctions fail.

With further increase of $J_2$, continuing along the 
$K_{\rm ring} = J_1 = 1$ cut through Fig.~\ref{fig:phased_J3_0p0},
the SBM phase becomes prominently unstable towards the VBS-3 phase 
occupying the range $1.5 < J_2 < 2.5$ and described in 
Sec.~\ref{subsec:DMRG:VBS3}.
Interestingly, as we increase $J_2$ still further, the SBM phase 
reappears with its characteristic correlations shown in
Fig.~\ref{fig:J2_3p2_J3_0p0} for a representative point $J_2 = 3.2$.
Much of the SBM physics discussion that we have just done at $J_2 = 0$
carries over here, with the appropriately shifted locations of the 
singular wavevectors.
The singularities at wavevectors $q_{\rm low} = 2k_{F2}$ and
$q_{\rm high} = 2k_{F1}$ are now more equally developed and are detected 
in the spin as well as dimer structure factors.  
The two wavevectors are closer to $\pi/2$ and are located symmetrically 
in accordance with the general ``sum rule'' $2k_{F1} + 2k_{F2} = \pi$,
while the comparable strengths reflect the more similar energy bandwidths
of the two Fermi seas.  The apparent lack of the wavevector $\pi/2$ in 
the DMRG dimer structure factor is similar to that in the Gutzwiller 
wavefunction and is a matrix element effect for the first-neighbor bond 
when the sizes of the two Fermi seas approach each other.  
On the other hand, the strength of the $4k_{F2}$ dimer feature is very 
notable here; it can indeed be dominant in the SBM theory for 
sufficiently small Luttinger parameter $g$, cf.~Table~\ref{tab:SBMprops}.
The improved Gutzwiller wavefunction modifies the structure factors
in the right direction but clearly does not succeed reproducing
them accurately -- as noted before, our wavefunctions do not contain 
the full physics expected in the Bosonized theory.

One technical remark that we want to make here is that the DMRG 
ground state at this point $J_2 = 3.2$ and size $L = 144$ appears to 
have non-trivial momentum quantum number $k \neq 0,\pi$.  
We deduce this by observing that the measured correlations $O(x,x')$ 
depend not just on $x-x'$ but on both $x$ and $x'$, and by seeing 
characteristic beatings as a function of $x$ and $x'$
(while the $q$-space structure factors are well-converged).
On the other hand, the VMC wavefunction shown in 
Fig.~\ref{fig:J2_3p2_J3_0p0} has momentum zero 
(see Appendix~\ref{app:Gutzw}) and all measurements depend on 
$x-x'$ only.  If we assume that the beatings originate from the DMRG 
state being a superposition of $|k \ra$ and $|-k \ra$, 
we can extract $2k$ and find it to be consistent with the state $|k \ra$ 
constructed from the VMC by moving one spinon across one of the 
Fermi seas ($2k = 4k_{F2} = -4k_{F1}$).
It is plausible that such state happens to have a slightly lower energy 
in the given finite system (e.g., at the same $J_2 = 3.2$, we find
trivial $k$ for $L=72$ but non-trivial $k$ for $L=96$, likely reflecting
finite-size effects on the filling of the last few spinon orbitals).
We have not attempted to construct a trial spin-singlet state with the 
right momentum quantum number for the present $L$.  Still, we expect that
the structure factors are not very sensitive to such rearrangements of 
few spinons in the large system limit.  Indeed, we find that
the structure factors have the same features for different system
lengths $L=72$, $96$, and $144$.
It is also worth repeating that our structure factor measurements using
Eq.~(\ref{Oq}) do not depend on which specific combination of $|k \ra$ 
and $|-k \ra$ is found by the DMRG procedure.

%%%%%%%%%%%%%%%%%%%%%%%%%%%%%%%%%%%%%%%%%%%%%%%%%%%%%%%%%%%%%%%%%%%%%%
%%%%%%%%%%%%%%%%%%%%%%%%%%%%%%%%%%%%%%%%%%%%%%%%%%%%%%%%%%%%%%%%%%%%%%
\subsection{Evolution of the singular wavevectors in the SBM}
\label{subsec:qsing}

\begin{figure}[th]
\centerline{\includegraphics[width=\columnwidth]{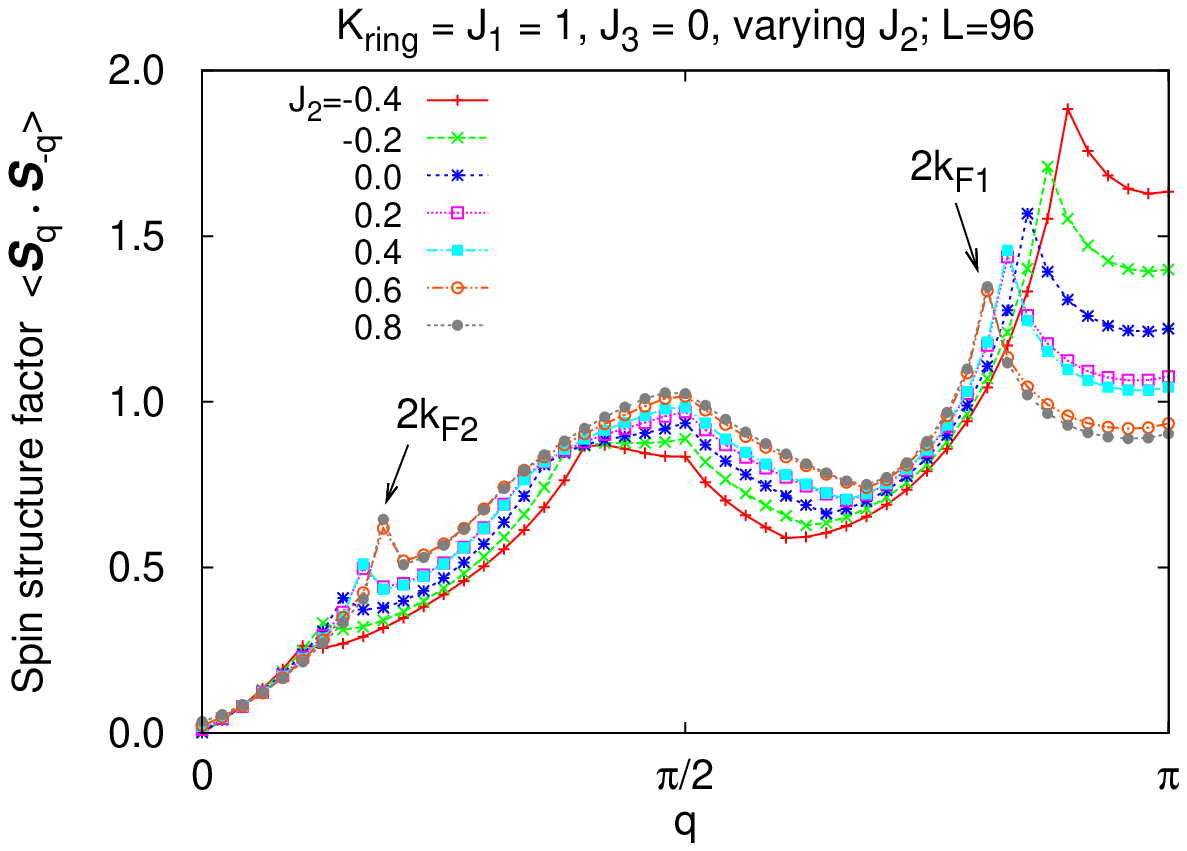}}
\centerline{\includegraphics[width=\columnwidth]{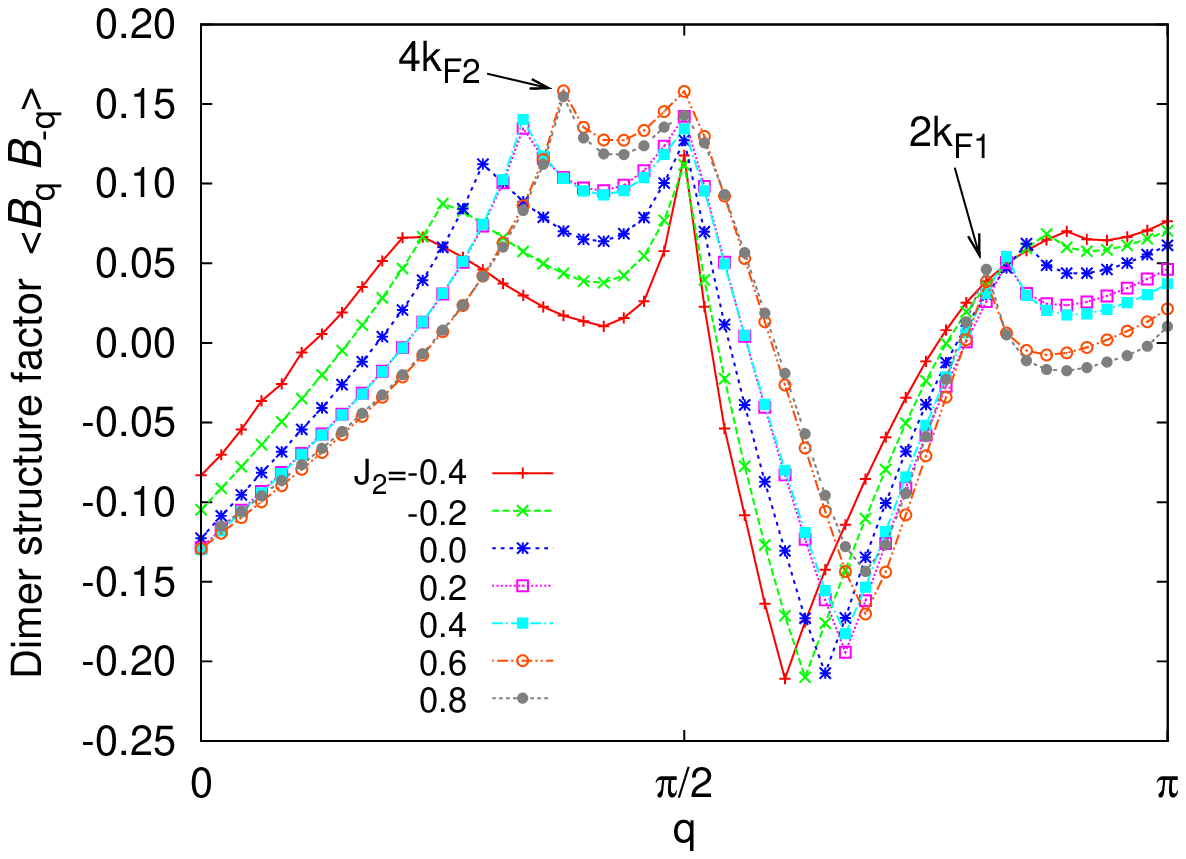}}
\caption{
(Color online)
Evolution of the structure factors in the Spin Bose-Metal phase 
between the Bethe-chain and VBS-3 (or partial FM), measured by the 
DMRG for system size $L = 96$.
We can track singular wavevectors (``Bose surfaces'') as spinons are 
transfered from the first to the second Fermi sea upon increasing $J_2$.
In the spin structure factor, the wavevector $q_{\rm high}$ that starts 
near $\pi$ is identified as $2k_{F1}$ and the wavevector $q_{\rm low}$ 
that starts near $0$ is identified as $2k_{F2}$ 
(see Fig.~\ref{fig:J2_0p0_J3_0p0} and text for details).
The $q_{\rm high}$ and $q_{\rm low}$ are summarized in 
Fig.~\ref{fig:qsing_J3_0p0}.
} \label{fig:evolution1}
\end{figure}

\begin{figure}[th]
\centerline{\includegraphics[width=\columnwidth]{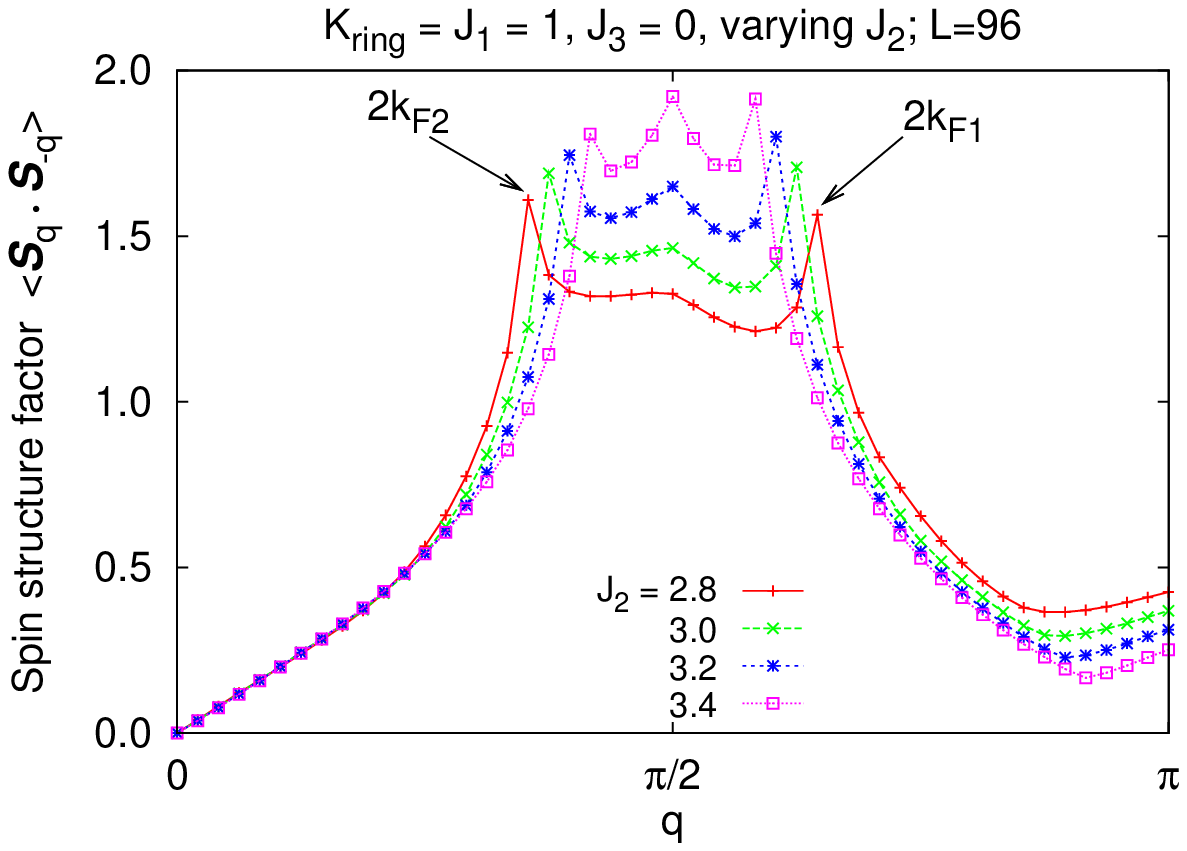}}
\centerline{\includegraphics[width=\columnwidth]{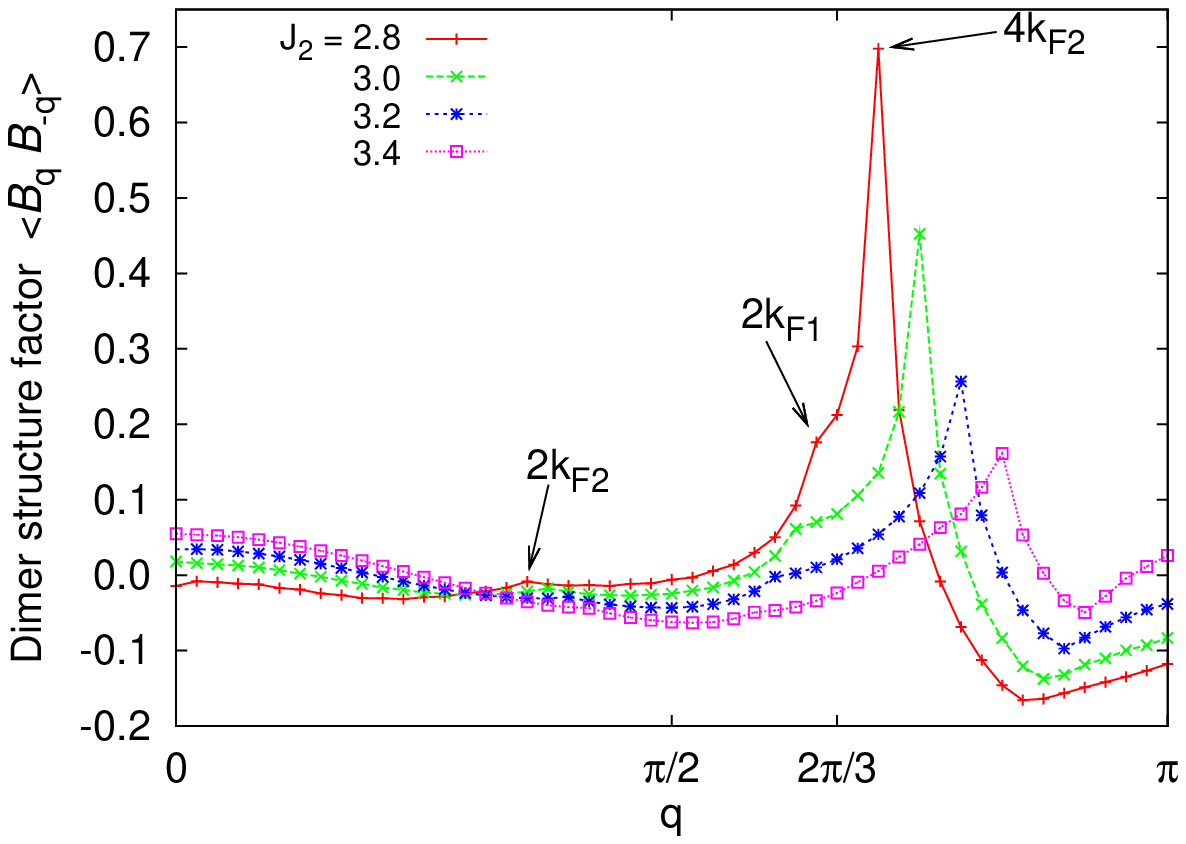}}
\caption{
(Color online)
Evolution of the structure factors in the Spin Bose-Metal phase 
between VBS-3 and VBS-2, measured by the DMRG for system size $L = 96$.
The $q_{\rm high} = 2k_{F1}$ and $q_{\rm low} = 2k_{F2}$ bracket the 
$\pi/2$ and approach each other with increasing $J_2$
(they are summarized in Fig.~\ref{fig:qsing_J3_0p0}).
Very notable here is the strong $4k_{F2}$ peak in the dimer structure 
factors that evolves out of the Bragg peak present in the VBS-3 phase 
at $2\pi/3$ 
(see Sec.~\ref{subsec:DMRG:VBS3} and Fig.~\ref{fig:J2_2p0_J3_0p0}); 
the $2k_{F1}$ moves away from the $2\pi/3$ in the opposite direction.
} 
\label{fig:evolution2}
\end{figure}

We further illustrate the Spin Bose-Metal by showing evolution of the 
DMRG structure factors and singular wavevectors as we move inside the 
phase.  The spin and dimer structure factors are shown in
Fig.~\ref{fig:evolution1} for $L=96$ and varying parameter $J_2/J_1$ 
inside the SBM phase adjacent to the Bethe-chain phase.  
With increasing $J_2$, the singular wavevector $q_{\rm high}$ 
(identified as $2k_{F1}$ in the VMC) 
is moving away from $\pi$ towards smaller values, while the 
singular wavevector $q_{\rm low}$ (identified as $2k_{F2}$) is 
moving to larger values; this corresponds to spinons being transfered 
from the first to the second Fermi sea as found in the VMC energetics.
The spin and dimer correlations show similar behavior at $2k_{F1}$,
and both have features also at the wavevector $\pi/2$.
The lack of visible dimer feature at $2k_{F2}$ was discussed 
for the point $J_2 = 0$ earlier, and it is likely that something similar 
is at play here.  On one hand, the second band is narrow when we just 
enter the SBM since the second Fermi sea starts small.  
On the other hand, in the region labeled ``partial FM''close to the 
VBS-3 phase in Fig.~\ref{fig:phased_J3_0p0}, the DMRG finds nonzero
magnetization in the ground state.  We think that this occurs in the 
second Fermi sea and indicates an effective narrowness of this band 
near the VBS-3 phase as well, so the $2k_{F2}$ energy feature may indeed 
be weak in the whole SBM phase between the Bethe-chain and VBS-3.

We think that the same physics also starts causing convergence 
difficulties in the DMRG for the $L=96$ systems shown in 
Fig.~\ref{fig:evolution1}.  
Specifically, we can use Eq.~(\ref{Stot}) to extract the ground state 
spin $S_{\rm tot}$ and find non-integer values of order 1, e.g.,
$S_{\rm tot} (S_{\rm tot} + 1) = 0.3$, $1.0$, $1.9$, $2.1$, $3.4$ for
the points $J_2 = 0$, $0.2$, $0.4$, $0.6$, $0.8$ in 
Fig.~\ref{fig:evolution1}.
This can only happen if the DMRG is not converged and is mixing
several states that are close in energy but have different spin.
Correspondingly, we observe a difference between 
$\la S^z_q S^z_{-q} \ra$ and $\la S^x_q S^x_{-q} \ra$ structure factors 
(recall that we are working in the sector $S^z_{\rm tot} = 0$).  
We do not show these graphs, but the difference is localized near
$2k_{F2}$, where the $\la S^z_q S^z_{-q} \ra$ has much sharper feature 
while the $\la S^x_q S^x_{-q} \ra$ has weaker feature.  
On the other hand, there is essentially no difference near $2k_{F1}$.
This strongly suggests that the origin of the 
convergence difficulties lies with the second Fermi sea.
The $\la \vec{S}_q \cdot \vec{S}_{-q} \ra$ structure factor that is 
shown in Fig.~\ref{fig:evolution1} does not mix different $S_{\rm tot}$ 
states but only sums the corresponding structure factors and is less 
sensitive to these convergence issues.
The fact that the peaks in the top panel of Fig.~\ref{fig:evolution1} are
located symmetrically with respect to $\pi/2$ suggests that this region
is still spin-singlet SBM (but is on the verge of some magnetism in the 
second band).  Finally, the $S_{\rm tot}$ values for the same parameters 
but system size $L=48$ are indeed well converged to zero, however with 
increasing number of states needed in the DMRG blocks for larger $J_2$.
We thus conclude that the points in Fig.~\ref{fig:evolution1} are 
spin-singlet SBM.  The DMRG convergence difficulties for the 
larger $L$ are in accord with the presence of many low-energy 
excitations (see also the discussion of the entanglement entropy below).
We will further test our intuition that this region is close to some 
weak subband ferromagnetism in Secs.~\ref{subsec:DMRG:afJ3} and 
\ref{sec:ferroJ3} by adding antiferromagnetic or ferromagnetic $J_3$ to 
suppress or enhance the FM tendencies.

Consider now Fig.~\ref{fig:evolution2} that shows evolution of the 
structure factors in the SBM phase between the VBS-3 and VBS-2. 
The $q_{\rm high} = 2k_{F1}$ and $q_{\rm low} = 2k_{F2}$ continue moving
towards each other with increasing $J_2$, and the spin structure factors 
become nearly symmetric with respect to $\pi/2$.  
When the $q_{\rm high}$ and $q_{\rm low}$ peaks merge at $\pi/2$,
which in the VMC would correspond to decoupled legs, 
one expects\cite{White_zigzag, Nersesyan} that a new instability will 
likely emerge resulting in a VBS-2 state 
(we discuss a representative point in Appendix~\ref{subapp:DMRG:VBS2}).

A very notable feature in the SBM dimer structure factor is the strong 
$4k_{F2}$ peak.  
Foretelling a bit, this peak can be traced as evolving out of the 
dimer Bragg peak at $2\pi/3$ of the VBS-3 phase to be discussed in 
Sec.~\ref{subsec:DMRG:VBS3}.
Turning this around and approaching the VBS-3 phase by decreasing $J_2$, 
we can view the VBS-3 as an SBM instability when the dimer $4k_{F2}$ peak
merges with the $2k_{F1}$ singularity, $4k_{F2} = 2k_{F1} = 2\pi/3$.

Finally, in this SBM region the DMRG converges well to spin-singlet 
states.  The remark we made when discussing the point $J_2 = 3.2$
applies to all points shown in Fig.~\ref{fig:evolution2} with $L=96$: 
they show correlations $O(x,x')$ beating in both $x$ and $x'$, 
which can be interpreted similarly to the earlier $J_2 = 3.2$ case by 
assuming superposition of degenerate ground states with opposite 
momentum quantum numbers.

\begin{figure}[t]
\centerline{\includegraphics[width=\columnwidth]{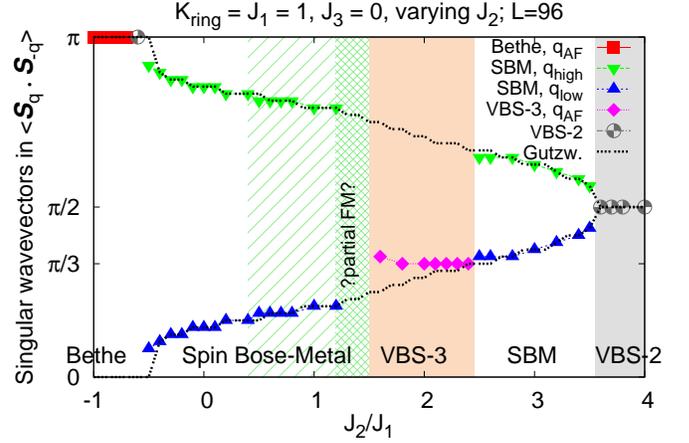}}
\caption{
(Color online)
Cut through the phase diagram Fig.~\ref{fig:phased_J3_0p0} at 
$K_{\rm ring}/J_1 = 1$, showing evolution of the most prominent 
wavevectors in the spin structure factor. 
In the Bethe-chain phase we have singular antiferromagnetic 
$q_{\rm AF} = \pi$, cf.~Fig.~\ref{fig:J2_m1p0_J3_0p0}.
In the Spin Bose-Metal we have singular $q_{\rm high} = 2k_{F1}$ and 
$q_{\rm low} = 2k_{F2}$ located symmetrically about $\pi/2$, 
cf.~Figs.~\ref{fig:evolution1} and \ref{fig:evolution2}.
In the VBS-3 we have singular $q_{\rm AF} = \pi/3$, 
cf.~Fig.~\ref{fig:J2_2p0_J3_0p0}.
In the VBS-2 region for large $J_2$, cf.~Fig.~\ref{fig:J2_4p0_J3_0p0},
we have dominant correlations at $\pi/2$ corresponding to the 
decoupled legs fixed point, which is likely to be unstable towards 
opening a spin gap.\cite{White_zigzag, Nersesyan}
The dotted lines show results for the improved Gutzwiller wavefunction.
}
\label{fig:qsing_J3_0p0}
\end{figure}

Fig.~\ref{fig:qsing_J3_0p0} summarizes the singular spin wavevectors 
extracted from plots like Figs.~\ref{fig:evolution1} and 
\ref{fig:evolution2}, superimposed on the phase diagram of the 
model along the cut $K_{\rm ring} = J_1$.
Remarkably, the singular wavevectors throughout the entire SBM phase are 
well captured by the improved Gutzwiller wavefunctions, as we have 
illustrated in Figs.~\ref{fig:J2_0p0_J3_0p0} and \ref{fig:J2_3p2_J3_0p0}.
These singular wavevectors are intimately connected to the 
sign structure of the ground state wavefunction, indicating a striking 
coincidence between the exact DMRG ground state wavefunction and the 
Gutzwiller projected VMC wavefunction.
Besides the SBM regions, Fig.~\ref{fig:qsing_J3_0p0} also shows the 
Bethe-chain (cf.~Appendix~\ref{subapp:DMRG:Bethe}),
VBS-2 (Appendix~\ref{subapp:DMRG:VBS2}), and 
VBS-3 (Sec.~\ref{subsec:DMRG:VBS3}) phases.

We now mention more difficult points in the overall phase diagram.  
The lightly hatched SBM region in Fig.~\ref{fig:qsing_J3_0p0} indicates 
the discussed rising DMRG difficulties of not converging to an 
exact singlet for $L = 96$.
Such DMRG states are shown as open circles with crosses in 
Fig.~\ref{fig:phased_J3_0p0}.
As we have already mentioned, the estimated $S_{\rm tot}$ values 
are not converged and are of order $1$ for $L = 96$ (but are converged
to zero for $L = 48$), while $q_{\rm low}$ and $q_{\rm high}$ are 
located symmetrically around $\pi/2$; all this suggests that the 
phase is spin-singlet SBM.
On the other hand, at points $J_2 = 1.2 - 1.5$ not marked in 
Fig.~\ref{fig:qsing_J3_0p0} but shown as star symbols in 
Fig.~\ref{fig:phased_J3_0p0}, the estimated $S_{\rm tot}$ values are
larger and the apparent dominant wavevectors are no longer located 
symmetrically.  Here we suspect a modification of the ground state, 
likely towards partial polarization of the second Fermi sea;
this ``partial FM'' region is also indicated by cross-hatching in
Fig.~\ref{fig:qsing_J3_0p0}.

%%%%%%%%%%%%%%%%%%%%%%%%%%%%%%%%%%%%%%%%%%%%%%%%%%%%%%%%%%%%%%%%%%%%%%
%%%%%%%%%%%%%%%%%%%%%%%%%%%%%%%%%%%%%%%%%%%%%%%%%%%%%%%%%%%%%%%%%%%%%%
\subsection{Entanglement entropy and effective central charge of the SBM}
\label{subsec:entropy}

\begin{figure}
\centerline{\includegraphics[width=\columnwidth]{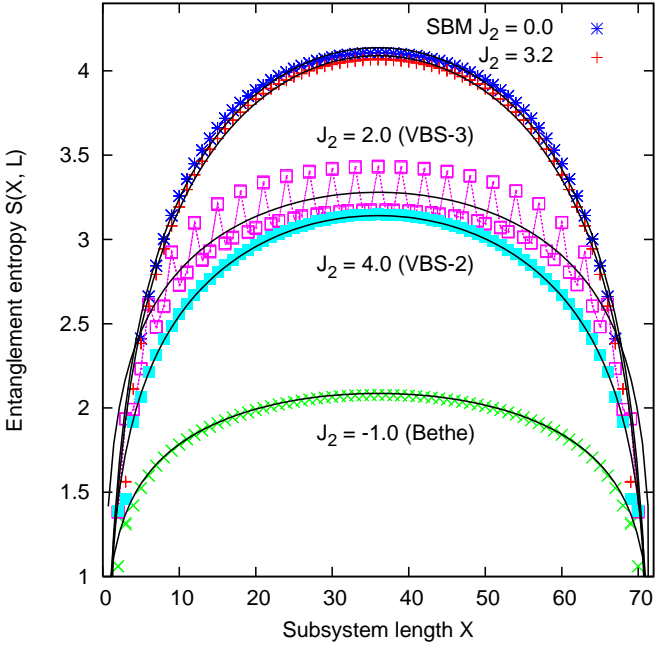}}
\caption{
(Color online)
Entanglement entropy at representative points in the 
Bethe-chain ($J_2 = -1$), SBM ($J_2 = 0$ and $J_2 = 3.2$), 
VBS-3 ($J_2 = 2$), and VBS-2 ($J_2 = 4$) phases, taken from the 
cut $K_{\rm ring} = J_1 = 1$, measured in the DMRG for system 
size $L=72$ with periodic boundary conditions.
We use Eq.~(\ref{eq:entropy}) to fit data over the range 
$6 \leq X \leq 66$, which gives the best estimates as 
$c = 1.0$ (Bethe-chain), 
$c = 3.1$ (SBM at $J_2 = 0$), 
$c = 3.2$ (SBM at $J_2 = 3.2$), 
$c = 1.6$ (VBS-3), and 
$c = 2.1$ (VBS-2).
}
\label{fig:entropy}
\end{figure}

We explore properties of the SBM phase that can further distinguish 
it from the Bethe-chain and VBS states.  Earlier we have noted that we 
need to keep more states per block to achieve similar convergence 
for the SBM phase in comparison with the Bethe-chain and VBS phases, 
indicating stronger entanglement between subsystems in the SBM.  
Bosonization analysis finds that the SBM fixed point theory 
has three free Boson modes.  One can associate a central charge 
$1$ with each mode.  Despite the fact that they have different 
velocities (so the full system is not conformally invariant), 
we expect that the total entanglement entropy should have a
universal behavior described by a combined central charge $c = 3$.

In general, for a one-dimensional gapless state with conformally 
invariant correlation functions in space-time, the entanglement entropy 
for a finite subsystem of length $X$ inside a system of length $L$ 
with periodic boundary conditions varies as\cite{CalabreseCardy}
\begin{eqnarray}
S(X, L) = \frac{c}{3}
\log \left( \frac{L}{\pi} \sin\frac{\pi X}{L} \right) + A ~,
\label{eq:entropy}
\end{eqnarray}
where $A$ is a constant (independent of the subsystem length) 
and $c$ is the effective central charge.
The virtue of the entanglement entropy is that it does not depend on the 
mode velocities and in principle measures the number of gapless modes 
directly from the ground state wavefunction.

Figure~\ref{fig:entropy} shows the entanglement entropy $S(X, L)$ 
as a function of $X$ for different quantum phases
for a finite system length $L = 72$.
The results are obtained from the DMRG for representative points taken
from the same cut $K_{\rm ring} = J_1 = 1$ discussed earlier.
The entropy can be well fitted by the ansatz 
Eq.~(\ref{eq:entropy}) with different $c$ values.

The Bethe-chain state (at $J_2 = -1$) gives central charge $c = 1.0$ 
consistent with one gapless mode.  
On the other hand, the entanglement entropy for either of the two SBM 
examples $J_2 = 0$ and $J_2 = 3.2$ is much larger and can be fitted 
by close values $c = 3.1$ and $c = 3.2$, respectively. 
The closeness of the central charges in these two different SBM states 
(cf.~Figs.~\ref{fig:J2_0p0_J3_0p0} and \ref{fig:J2_3p2_J3_0p0})
indicates the universal behavior of the entanglement which is 
independent of the details like the relative sizes of the 
spinon Fermi seas.  

Interestingly, the VBS-2 point at $J_2 = 4$ is fitted by $c = 2.1$, 
which is related to the fact that the wavefunction is close to the 
decoupled-legs limit (see Appendix~\ref{subapp:DMRG:VBS2} and 
Fig.~\ref{fig:J2_4p0_J3_0p0}), 
and this $L = 72$ system ``does not know'' about the eventual spin gap 
and very small dimerization.

Finally, the fitted effective central charge for the VBS-3 example is 
around $c = 1.6$.  The oscillatory behavior of $S(X)$ reflects 
translational symmetry breaking in the DMRG state.  
Note that the entropy values are larger here compared with the 
Bethe-chain or VBS-2 cases, which is probably due to a mix of degenerate 
states in the DMRG wavefunction. 
However, the overall $X$-dependence is clearly weaker than in the VBS-2 
and is approaching the Bethe-chain behavior for large $X$.

\begin{figure}
\centerline{\includegraphics[width=\columnwidth]{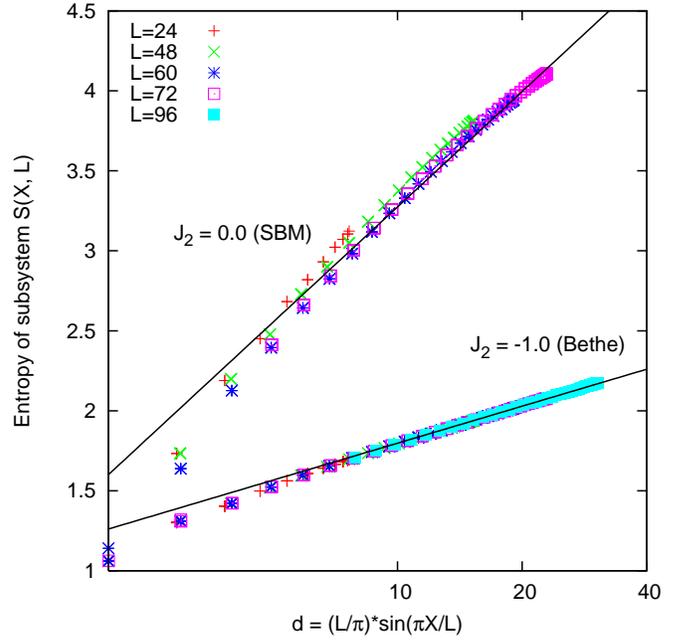}}
\caption{
(Color online)
Entanglement entropy for various system sizes for the 
Bethe-chain ($J_2 = -1$) and SBM ($J_2 = 0$) points,
cf.~Fig.~\ref{fig:entropy}, plotted versus scaling variable 
$d =\frac{L}{\pi} \sin \frac{\pi X}{L}$.
We also show fits to Eq.~(\ref{eq:entropy}) done for the larger sizes.
The Bethe-chain data collapses well and is fitted with $c=1$.
The SBM data is fitted with $c=3.1$ (we show the same fit as in 
Fig.~\ref{fig:entropy} for this SBM point).  The collapse of different $L$
is less good, which is likely due to imprecise scaling of the 
discrete shell filling numbers with $L$ (see text for details).
}
\label{fig:entropy_scaling}
\end{figure}

To better understand finite size effects, we focus on the SBM and 
Bethe-chain phases and discuss the universal dependence of the entropy
on the scaling variable $d = \frac{L}{\pi} \sin\frac{\pi X}{L}$.
Figure~\ref{fig:entropy_scaling} shows $S(X, L)$ as a function of $d$ 
for several system sizes for the SBM point $J_2 = 0$ and the Bethe-chain
point $J_2 = -1$.

At the SBM point, the data for the two larger sizes $L = 60$ and $72$ 
collapse onto one curve, which can be reasonably fitted by 
$S(X, L) = (3.1/3) \log(d) + 0.88$, strongly suggesting the 
effective central charge $c \simeq 3$.  The smaller sizes $L=24$ and 
$48$ have somewhat shifted entropy values compared with the $L=60$ and 
$72$ collapse, but show roughly similar slope for the largest $d$.  
The differences are likely due to finite-size shell filling effects.
Indeed, we can measure the structure factors and characterize the
presumed SBM states by the spinon occupation numbers of the two 
Fermi seas; we find $(N_1, N_2) = (10,2)$, $(20,4)$, $(26,4)$, and 
$(31,5)$ for $L=24$, $48$, $60$, and $72$ respectively, 
and these numbers in each Fermi sea do not precisely ``scale'' with $L$.

On the other hand, the Bethe-chain case does not have such effects 
and data for all sizes collapse.
The results can be well fitted by $S(X, L) = (1/3) \log(d) + 0.99$ 
as shown in the same figure for system sizes $L = 24$ to $96$.
We note that while the entropy for the Bethe-chain phase for $L=96$ is 
fully converged by keeping up to $m = 4200$ states in the DMRG block, 
the entropy for the SBM for $L=72$ is still increasing slowly with the
number of states kept, and we estimate that the error in $S(X=L/2, L)$ 
is around a few percent when $m=6000$ states are kept per block
(comparing to an extrapolation to $m=\infty$).
Indeed, the SBM data for $L=72$ is bending down slightly from the 
fitted line at the larger $d$ corresponding to $X \sim L/2$, 
as can be seen in Fig~\ref{fig:entropy_scaling},
which is probably because the data is less converged.

To summarize, the entanglement entropy calculations establish the SBM 
as a critical phase with three gapless modes and clearly distinguish
it from the Bethe-chain and VBS-2 phases (or the decoupled-legs limit).  
We also note that the structure factor measurements and detection of all 
features as in Figs.~\ref{fig:evolution1}~and~\ref{fig:evolution2}
did not require as much effort and was done for larger sizes than the
entropy; however, to characterize the long distance power laws accurately
one would probably need to capture all entanglement, 
which we have not attempted.

%%%%%%%%%%%%%%%%%%%%%%%%%%%%%%%%%%%%%%%%%%%%%%%%%%%%%%%%%%%%%%%%%%%%%%
%%%%%%%%%%%%%%%%%%%%%%%%%%%%%%%%%%%%%%%%%%%%%%%%%%%%%%%%%%%%%%%%%%%%%%
%%%%%%%%%%%%%%%%%%%%%%%%%%%%%%%%%%%%%%%%%%%%%%%%%%%%%%%%%%%%%%%%%%%%%%
\section{Stability of the Spin Bose-Metal phase;
nearby phases out of the SBM}
\label{sec:SBMstability}

\subsection{Residual interactions and stability of the SBM}
\label{subsec:stability}

We account for residual interactions between low-energy degrees of
freedom in the SBM theory, Sec.~\ref{sec:SBMtheory}, 
by considering all allowed short-range interactions of the spinons.  
The four-fermion interactions can be conveniently expressed in terms of 
chiral currents:
\begin{equation}
J_{Pab} = f^\dagger_{P a \alpha} f_{P b \alpha} ~; \quad
\vec{J}_{Pab} = \frac{1}{2} 
f^\dagger_{P a \alpha} \vec{\sigma}_{\alpha\beta} f_{P b \beta} ~.
\label{JP}
\end{equation}
We assume that interactions that are chiral, say involving 
only right movers, can be neglected apart from velocity renormalizations.
The most general four-fermion interactions which mix right and left 
movers can be succinctly written as,
\begin{eqnarray}
{\cal L}^\rho_1 &=& \sum_{a,b}
\left( w^\rho_{ab} J_{R a b} J_{L a b} 
       + \lambda^\rho_{ab} J_{R a a} J_{L b b} 
\right)~, \\
{\cal L}^\sigma_1 &=& -\sum_{a,b}
\left( w^\sigma_{ab} \vec{J}_{R a b} \cdot \vec{J}_{L a b} 
       + \lambda^\sigma_{ab} \vec{J}_{R a a} \cdot \vec{J}_{L b b}
\right)~,
\end{eqnarray}
with $w_{11} = w_{22} = 0$ (convention), 
$w_{12} = w_{21}$ (from Hermiticity), and
$\lambda_{12} = \lambda_{21}$ (from $R \leftrightarrow L$ symmetry).  
There are 8 independent couplings:
$w_{12}^{\rho/\sigma}$, $\lambda_{11}^{\rho/\sigma}$, 
$\lambda_{22}^{\rho/\sigma}$, and $\lambda_{12}^{\rho/\sigma}$.

We treat these interactions perturbatively as follows.
First we bosonize the interactions and obtain terms quadratic in 
$\partial_x \theta_{a\alpha}$ and $\partial_x \varphi_{a\alpha}$
as well as terms involving products of four exponentials
$e^{\pm i \theta_{a\alpha}}$ and $e^{\pm i \varphi_{a\alpha}}$.
We next impose the condition that $\theta_{\rho +}$ is pinned
and compute the scaling dimensions of the exponential operators.   

The $w_{12}^{\rho/\sigma}$ terms give
\begin{eqnarray}
W \equiv (w_{12}^\rho J_{R12} J_{L12} 
- w_{12}^\sigma \vec{J}_{R12} \cdot \vec{J}_{L12}) + \Hc = ~~~~~~~~ 
\label{w12} \\
= \cos(2\varphi_{\rho-}) \Bigg\{
4 w_{12}^\rho \Big[\cos(2\varphi_{\sigma-}) 
                   - \hat\Gamma \cos(2\theta_{\sigma-}) \Big] ~~~~~~~ 
\label{w12rho} \\
~ - w_{12}^\sigma \Big[\cos(2\varphi_{\sigma-}) 
                       + \hat\Gamma \cos(2\theta_{\sigma-}) 
                       + 2 \hat\Gamma \cos(2\theta_{\sigma+}) \Big]
\Bigg\},
\label{w12sigma}
\end{eqnarray}
where
\begin{eqnarray}
\hat\Gamma = \eta_{1\up} \eta_{1\dn} \eta_{2\up} \eta_{2\dn} ~.
\label{Gamma}
\end{eqnarray} 
The $w_{12}^{\rho/\sigma}$ terms have scaling dimension 
$1 + g_0^{-1} \ge 2$ and are irrelevant in the bare theory,
Eq.~(\ref{LSBM0}), and henceforth dropped.  
The detailed $W$ expression will be used later when we analyze 
phases neighboring the SBM.

The remaining exponentials only depend on the fields $\theta_{a\sigma}$ 
and $\varphi_{a\sigma}$ so that the charge and spin sectors decouple.  
Since the rest of ${\cal L}^\rho_1$ is quadratic,
${\cal L}^{\rho} = {\cal L}^\rho_0 + {\cal L}^\rho_1$
takes the same form as ${\cal L}^\rho_0$ in Eq.~(\ref{L0rho}) except with
$g_0, v_0 \to g, v$, where
\begin{eqnarray}
g^2 &=& 4 
\frac{(v_1 - \lambda_{11}^\rho/\pi)(v_2 - \lambda_{22}^\rho/\pi) 
      - (\lambda_{12}^\rho/\pi)^2}
     {(v_1 + v_2 - 2\lambda_{12}^\rho/\pi)^2 
      - (\lambda_{11}^\rho/\pi + \lambda_{22}^\rho/\pi)^2} ~, \\
v &=& \frac{g}{2} 
(v_1 + v_2 + \lambda_{11}^\rho/\pi + \lambda_{22}^\rho/\pi
 - 2\lambda_{12}^\rho/\pi) ~.
\end{eqnarray}

In the spin sector, the remaining interactions are given by,
\begin{equation}
\tilde{\cal L}^\sigma_1 = 
- \sum_a \lambda^\sigma_{aa} \vec{J}_{Raa} \cdot \vec{J}_{Laa} 
- \lambda^\sigma_{12} (\vec{J}_{R11} \cdot \vec{J}_{L22} 
                       + \vec{J}_{L11} \cdot \vec{J}_{R22}) ~.
\label{vJRvJL}
\end{equation}
When we write this in the bosonization, the $J^z_R J^z_L$ pieces 
contribute to the harmonic part of the action
\begin{eqnarray}
V_z &=& \sum_a \frac{\lambda_{aa}^\sigma}{8\pi^2}
\left[(\partial_x \varphi_{a\sigma})^2 - (\partial_x \theta_{a\sigma})^2
\right] \\
&+& \frac{\lambda_{12}^\sigma}{4\pi^2} 
\left[(\partial_x \varphi_{1\sigma}) (\partial_x \varphi_{2\sigma}) 
      - (\partial_x \theta_{1\sigma}) (\partial_x \theta_{2\sigma})
\right] ~,
\end{eqnarray}
while the $J^+_R J^-_L + J^-_R J^+_L$ produce nonlinear potential
\begin{eqnarray}
V_\perp &=& \sum_a \lambda_{aa}^\sigma \cos(2\sqrt{2}\theta_{a\sigma}) \\
&+& 2\lambda_{12}^\sigma \hat\Gamma \cos(2\theta_{\sigma+}) 
\cos(2\varphi_{\sigma-}) ~.
\label{Vperp}
\end{eqnarray}
A 1-loop RG analysis gives the following flow equations,
\begin{equation}
\frac{d \lambda^\sigma_{aa}}{d \ell} = 
-\frac{(\lambda^\sigma_{aa})^2}{2\pi v_a} ~, \quad
\frac{d \lambda^\sigma_{12}}{d \ell} = 
-\frac{(\lambda^\sigma_{12})^2}{\pi (v_1 + v_2)} ~.
\label{vJRvJL_flows}
\end{equation}
When $\lambda^\sigma_{11}, \lambda^\sigma_{22}$, and 
$\lambda^\sigma_{12}$ are positive, they scale to zero and the 
quadratic SBM Lagrangian ${\cal L}^{\rm SBM}_0$, Eq.~(\ref{LSBM0}), 
is stable.
We also require that the renormalized $g$ is smaller than $1$ 
so that the $w_{12}^{\rho/\sigma}$ terms in Eq.~(\ref{w12}) 
remain irrelevant.
In Sec.~\ref{subsec:g.gt.1}, we consider what happens when $g>1$ 
or when some of the $\lambda^\sigma_{11}, \lambda^\sigma_{22}$, and 
$\lambda^\sigma_{12}$ change sign and become marginally relevant.

The above stability considerations are complete for generic 
incommensurate Fermi wavevectors.  
At special commensurations, new interactions may be allowed and
can potentially destabilize the SBM. 
Such situations need to be analyzed separately, and in 
Secs.~\ref{subsec:commens_VBS3}~and~\ref{subsec:commens_other} 
we consider cases relevant for the VBS-3 and Chirality-4
phases found by the DMRG in the ring model, 
Secs.~\ref{subsec:DMRG:VBS3}~and~\ref{subsec:DMRG:afJ3}.

We want to make one remark about the allowed interactions, 
which will be useful later.
Let us ignore for a moment the pinning of $\theta_{\rho+}$;
for example, let us think about the electron interactions in the 
approach of Sec.~\ref{subsec:SBMelectron}.
Three of the eight $\theta$ and $\varphi$ fields, namely 
$\varphi_{\rho+}$, $\varphi_{\sigma+}$, and $\theta_{\rho-}$,
do not appear as arguments of the cosines, and the action has 
continuous symmetries under independent shifts of these.
The first two symmetries correspond to microscopic conservation laws
for the total charge $Q \sim \int \partial_x \theta_{\rho+}$
and the total spin $S^z \sim \int \partial_x \theta_{\sigma+}$.
On the other hand, the invariance under the shifts of 
$\theta_{\rho-}$ corresponds to the conservation of
$X = N_{R1} - N_{L1} - N_{R2} + N_{L2} 
\sim \int \partial_x \varphi_{\rho-}$,
where $N_{Pa}$ denotes the total number of fermions near 
Fermi point $P k_{Fa}$.
This is not a microscopic symmetry, but emerges in the continuum theory 
for generic $k_{F1}, k_{F2}$.  Indeed, writing the total momentum
$P = (N_{R1} - N_{L1}) k_{F1} + (N_{R2} - N_{L2}) k_{F2}
= X k_{F1} - (N_{R2} - N_{L2}) \pi/2$, we see that any attempt to change 
$X$ violates the momentum conservation except for special
commensurate $k_{F1}$.
We thus conclude that $\varphi_{\rho+}$ and $\varphi_{\sigma+}$ 
can never be pinned by the interactions, while $\theta_{\rho+}$ 
can not be pinned generically except at special commensurate points.

%%%%%%%%%%%%%%%%%%%%%%%%%%%%%%%%%%%%%%%%%%%%%%%%%%%%%%%%%%%%%%%%%%%%%
%%%%%%%%%%%%%%%%%%%%%%%%%%%%%%%%%%%%%%%%%%%%%%%%%%%%%%%%%%%%%%%%%%%%%
\subsection{Gapped paramagnets when $g > 1$}
\label{subsec:g.gt.1}

We now consider phases that can emerge as some instabilities of the 
Spin Bose-Metal.
We use heavily the Bosonization expressions for various observables 
given in Appendix~\ref{app:SBMprops}.
As we have already mentioned, when $g > 1$ the interactions in
Eq.~(\ref{w12}) are relevant and the SBM phase is unstable.
We can safely expect that as a result of the runaway flows, 
the variables $\varphi_{\rho-}$ and $\theta_{\sigma+}$ will be pinned.
The situation is less clear with the remaining parts of the potential
since we cannot pin simultaneously $\theta_{\sigma-}$ and 
$\varphi_{\sigma-}$.  Still, it is possible that the situation is 
resolved by pinning one variable or the other.
For example, depending on whether $w_{12}^\sigma$ and $w_{12}^\rho$ have 
the same or opposite signs, it is advantageous to pin $\theta_{\sigma-}$ 
or $\varphi_{\sigma-}$.  If either pinning scenario happens, 
there remain no gapless modes in the system.  
It is readily established in the cases below that all spin correlations 
are short-ranged, i.e., we have fully-gapped paramagnets;
also, $\la {\cal B}_\pi \ra \neq 0$ in all cases, so the translational
symmetry is necessarily broken.

%%%%%%%%%%%%%%%%%%%%%%%%%%%%%%%%%%%%%%%%%%%%%%%%%%%%%%%%%%%%%%%%%%%%%%
\subsubsection{$w_{12}^\sigma w_{12}^\rho > 0$ and pinned
$\varphi_{\rho-}, \theta_{\sigma+}, \theta_{\sigma-}$: 
period-2 VBS}
\label{subsubsec:period2}
\begin{figure}
\centerline{\includegraphics[width=\columnwidth]{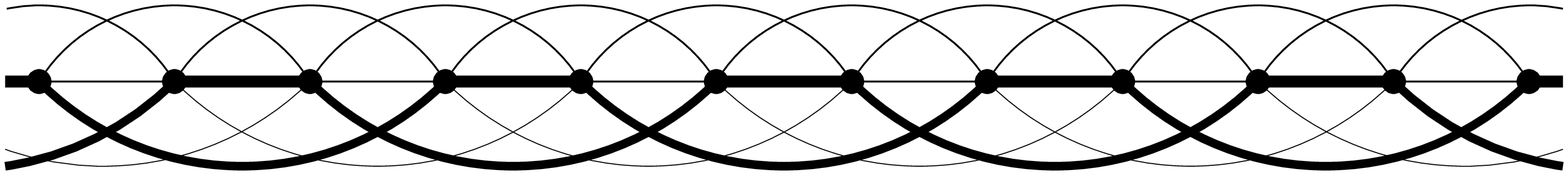}}
\caption{
Valence Bond Solid with period 2, where thicker lines indicate 
stronger bonds.  To emphasize the symmetries of the state,
we also show second- and third-neighbor bond energies, but
details can be different in different regimes. 
For example, the VBS-2 state in the $K_{\rm ring} = 0$ case has
dominant first-neighbor dimerization.  On the other hand, 
in the model with $K_{\rm ring} = J_1 = 1$, $J_3 = 0.5$,
the putative VBS-2 region between the Bethe-chain and SBM phases in 
Fig.~\ref{fig:qsing_J3_0p5} has significant third-neighbor modulation
but only very small first-neighbor one.
}
\label{fig:dimer2}
\end{figure}

Consider the case when the $\theta_{\sigma-}$ is pinned.
Using Appendix~\ref{app:SBMprops}, we see that ${\cal B}_\pi$ obtains an 
expectation value, while ${\cal B}_{\pi/2}$ and $\chi_{\pi/2}$ are 
short-ranged.
It is natural to identify this phase as a period-2 Valence Bond Solid
shown in Fig.~\ref{fig:dimer2}.
The pinning values and therefore some details of the state will differ 
depending on the sign of the coupling $w_{12}^\sigma$,
but in either case the ground state is two-fold degenerate.
[Here and below, when we find pinning values of appropriate $\varphi$-s 
and $\theta$-s minimizing a given potential, we determine which solutions
are physically distinct by checking if they produce distinct phases 
modulo $2\pi$ in the bosonization Eq.~(\ref{fbosonize}).
More practically, following Ref.~\onlinecite{Lin98}~Sec.~IV.E.1,
the chiral fermion fields remain unchanged under
$\varphi_{a\alpha} \to \varphi_{a\alpha} 
+ \pi (\ell_{Ra\alpha} + \ell_{La\alpha})$,
$\theta_{a\alpha} \to \theta_{a\alpha} 
+ \pi (\ell_{Ra\alpha} - \ell_{La\alpha})$,
where $\ell_{Pa\alpha}$ can be arbitrary integers.
This gives redundancy transformations for the $\rho\pm$, $\sigma\pm$ 
fields that we use to check if the minimizing solutions are 
physically distinct.]

%%%%%%%%%%%%%%%%%%%%%%%%%%%%%%%%%%%%%%%%%%%%%%%%%%%%%%%%%%%%%%%%%%%%%%
\subsubsection{$w_{12}^\sigma w_{12}^\rho < 0$ and pinned
$\varphi_{\rho-}, \theta_{\sigma+}, \varphi_{\sigma-}$: 
period-4 structures}
\label{subsubsec:period4}

Consider now the case when the $\varphi_{\sigma-}$ is pinned.  
We find that either 
$\epsilon_{\pi/2}$ in Eq.~(\ref{epsilon_halfpi_fixGamma}) or 
$\chi_{\pi/2}$ in Eq.~(\ref{chi_halfpi_fixGamma}), but not both, 
obtains an expectation value. 
Thus we either have a period-4 VBS or a period-4 structure in the 
chiralities.  Which one is realized depends on details of the pinning.

\begin{figure}
\centerline{\includegraphics[width=\columnwidth]{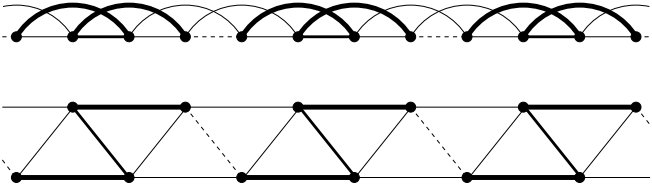}}
\caption{
Top: Valence Bond Solid with period 4 suggested as one of the 
instabilities out of the Spin Bose-Metal phase,
Sec.~\ref{subsubsec:period4}a.
Thick lines indicate stronger bonds.
Bottom: In the two-leg triangular ladder drawing, we see roughly 
independent spontaneous dimerization in each leg.
}
\label{fig:dimer4}
\end{figure}

As described in Appendix~\ref{app:SBMprops}, we work with the 
$+1$ eigenstate of the operator $\hat\Gamma$ 
[our Eqs.~(\ref{epsilon_halfpi_fixGamma}-\ref{chi_halfpi_fixGamma}) 
already assume this]. 
With this choice, to minimize the potential in Eq.~(\ref{w12}) we require
\begin{equation}
\cos(2\varphi_{\sigma-}) = \cos(2\theta_{\sigma+}) = \pm 1 ~.
\label{pin4lambd}
\end{equation}
Depending on the sign of $w_{12}^\rho$, we have:
\begin{eqnarray}
&\text{a)}& w_{12}^\rho>0:
\quad \cos(2\varphi_{\rho-}) = -\cos(2\theta_{\sigma+})~,\\
&\text{b)}& w_{12}^\rho<0: 
\quad \cos(2 \varphi_{\rho-}) = \cos(2\theta_{\sigma+})~.
\end{eqnarray}

a) In this case, 
$\la \epsilon_{\pi/2} \ra \neq 0, \la \chi_{\pi/2} \ra = 0$,
i.e., we find period-4 valence bond order.
Note that $\la \epsilon_{\pi/2} \ra$ can take four independent values 
$\la \epsilon_{\pi/2} \ra = e^{i\alpha} = 
e^{\pm i\pi/4}, e^{\pm i 3\pi/4}$, where we have assumed that 
$\theta_{\rho+}$ is fixed by Eq.~(\ref{u8pos}).
To visualize the state, we examine the corresponding contributions to
the first- and second-neighbor bond energies:
\begin{eqnarray}
\delta {\cal B}_{x, x+1} &\sim& 
\cos\left(\frac{\pi}{2} x + \frac{\pi}{4} + \alpha \right) =
\{+, 0, -, 0, \dots \} , ~~~~~~ \\
\delta {\cal B}_{x, x+2} &\sim& 
\cos\left(\frac{\pi}{2} x + \frac{\pi}{2} + \alpha \right) =
\{+, -, -, +, \dots \} .
\label{Bdimer4}
\end{eqnarray}
One can either use symmetry arguments or write out the microscopic 
hopping energies explicitly to fix the phases as above
[see Eq.~(\ref{BQ}), which generalizes to $n$-th neighbor bond as 
${\cal B}^{(n)}_Q \sim e^{i n Q/2} \epsilon_Q$ for $Q\neq \pi$].
Each line also shows schematically the sequence of bonds starting at 
$x=0$ for $\alpha=-\pi/4$.  
The four independent values of $\alpha$ correspond to four 
translations of the same VBS state along $x$.
The pattern of bonds is shown in Fig.~\ref{fig:dimer4}, where the 
more negative energy is associated with the stronger dimerization.
When viewed on the two-leg ladder, this state can be connected to
a state with independent spontaneous dimerization in each leg.

\begin{figure}
\centerline{\includegraphics[width=\columnwidth]{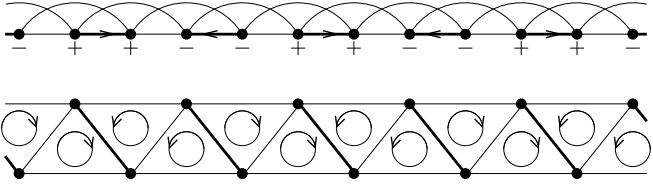}}
\caption{
Top: Chirality order with period 4 suggested as one of the 
instabilities out of the SBM phase, Sec.~\ref{subsubsec:period4}b.
In the 1D chain picture, the chirality pattern is given by 
Eq.~(\ref{chirality4}); $\chi(x)$ is associated with the $[x-1, x, x+1]$ 
loop and the arrows on the links show one ``gauge choice'' to 
produce such ``fluxes'' in the spinon hopping.
Bottom: In the two-leg triangular ladder drawing, we see alternating 
chiralities on the up-triangles along the strip and alternating 
chiralities on the down-triangles.
}
\label{fig:chirality4}
\end{figure}

b) Here we have 
$\la \epsilon_{\pi/2} \ra = 0, \la \chi_{\pi/2} \ra \neq 0$,
i.e., period-4 structure in the chirality $\chi$.
The pattern is
\begin{eqnarray}
\chi(x) \sim \cos\left(\frac{\pi}{2} x + \alpha \right) 
= \{+, -, -, +, \dots \} ~.
\label{chirality4}
\end{eqnarray}
There are four independent values of 
$\la \chi_{\pi/2} \ra = e^{i\alpha} = e^{\pm i\pi/4}, e^{\pm i 3\pi/4}$,
corresponding to four possible ways to register this pattern 
on the chain.
The state is illustrated in Fig.~\ref{fig:chirality4}.
When drawn on the two-leg ladder, chiralities on the upwards pointing 
triangles alternate along the strip, and so do chiralities on the 
downwards pointing triangles.

%%%%%%%%%%%%%%%%%%%%%%%%%%%%%%%%%%%%%%%%%%%%%%%%%%%%%%%%%%%%%%%%%%%%%%
\subsubsection{Gapped phases in the spinon language}
\label{subsubsec:gapspinons}

With an eye towards what might happen in the 2D spin liquid,
it is instructive to discuss the above phases in terms
of the spinons.  To this end, we can rewrite the $W$ term,
Eq.~(\ref{w12}), as follows:
\begin{eqnarray}
W &=& (w_{12}^\rho + w_{12}^\sigma/4) [P_1^\dagger P_2 + \Hc] \\
&-& (w_{12}^\rho - w_{12}^\sigma/4) 
[(f_{R1}^\dagger f_{L2}) (f_{L1}^\dagger f_{R2}) + \Hc] ~.
\end{eqnarray}
Here $P_a^\dagger = f_{Ra\up}^\dagger f_{La\dn}^\dagger -
f_{Ra\dn}^\dagger f_{La\up}^\dagger$ creates a ``Cooper pair'' in 
band $a$.
The preceding two sections can be then viewed as follows.
When $w_{12}^\rho w_{12}^\sigma > 0$, we minimize the first line 
by ``pairing and condensing'' the spinons; once everything is done,
we get the period-2 VBS state.
On the other hand, when $w_{12}^\rho w_{12}^\sigma < 0$, 
we minimize the second line by developing expectation values in the 
particle-hole channel.  Using
\begin{equation}
(f_{R1}^\dagger f_{L2}) (f_{L1}^\dagger f_{R2}) + \Hc
= 2 [ \epsilon_{\pi/2}^\dagger \epsilon_{\pi/2} 
      - \chi_{\pi/2}^\dagger \chi_{\pi/2} ] ~,
\end{equation}
we get either the period-4 dimer or period-4 chirality order 
depending on the sign of $w_{12}^\rho$.

%%%%%%%%%%%%%%%%%%%%%%%%%%%%%%%%%%%%%%%%%%%%%%%%%%%%%%%%%%%%%%%%%%%%%
%%%%%%%%%%%%%%%%%%%%%%%%%%%%%%%%%%%%%%%%%%%%%%%%%%%%%%%%%%%%%%%%%%%%%
\subsection{Nearby phases obtained when some of the 
$\vec{J}_R \cdot \vec{J}_L$ interactions, Eq.~(\ref{vJRvJL}), 
become marginally relevant}
\label{subsec:margrel}

Let us now assume $g<1$, so the singlet ``$\rho-$'' sector is not
a priori gapped.  We consider what happens when some of the couplings 
$\lambda_{11}^\sigma, \lambda_{22}^\sigma, \lambda_{12}^\sigma$ 
in Eq.~(\ref{vJRvJL}) change sign and become marginally relevant.  
We analyze this as follows.  
Consider the potential $V_\perp$, Eq.~(\ref{Vperp}),
again working with the $+1$ eigenstate of the operator $\hat\Gamma$.
If one (or several) of the couplings becomes negative, we have runaway 
flows Eq.~(\ref{vJRvJL_flows}) to still more negative values.
We then consider pinned field configurations that minimize the relevant 
part of the $V_\perp$ --- this is what happens in the spin sector.
Next we need to include the interactions Eq.~(\ref{w12}) with the 
singlet ``$\rho-$'' sector, since they can become relevant once 
some of the ``$\sigma$'' fields are pinned.
We now consider different possibilities.

%%%%%%%%%%%%%%%%%%%%%%%%%%%%%%%%%%%%%%%%%%%%%%%%%%%%%%%%%%%%%%%%%%%%
\subsubsection{$\lambda_{11}^\sigma > 0$, $\lambda_{22}^\sigma > 0$, 
$\lambda_{12}^\sigma < 0$}
In this case, only the $\lambda_{12}^\sigma$ is relevant and flows to 
large negative values.  We therefore pin the fields $\theta_{\sigma+}$ 
and $\varphi_{\sigma-}$.  To minimize $V_\perp$, the pinned values need
to satisfy Eq.~(\ref{pin4lambd}).
The spin sector is gapped and all spin correlations decay exponentially;
we also have $\la {\cal B}_\pi \ra \neq 0$, so the translational symmetry
is broken.

Next we include the interactions Eq.~(\ref{w12}).
Using Eq.~(\ref{pin4lambd}), the important part is
\begin{equation}
W = (4 w_{12}^\rho - 3 w_{12}^\sigma) 
\cos(2\theta_{\sigma+}) \cos(2\varphi_{\rho-}) ~.
\end{equation}
The $\varphi_{\rho-}$ is dynamical at this stage, but the 
$\theta_{\sigma+}$ is pinned and the $W$ now has scaling dimension $1/g$.
The possibilities are:

a) $g<1/2$: The $W$ term is irrelevant and the singlet sector remains 
gapless.  One manifestation of the gaplessness is that 
${\cal B}_{\pm \pi/2}$ and $\chi_{\pm \pi/2}$ have power law 
correlations characterized by scaling dimension $1/(4g)$. 
Thus we have a coexistence of the static period-2 VBS order and
power law VBS and chirality correlations at the wavevectors
$\pm \pi/2$.

b) $g>1/2$: The $W$ term is relevant and pins the field $\varphi_{\rho-}$
leaving no gapless modes in the system.  Such fully-gapped situation
has already been discussed in Sec.~\ref{subsubsec:period4}.
This gives either the period-4 VBS or period-4 chirality phase.

%%%%%%%%%%%%%%%%%%%%%%%%%%%%%%%%%%%%%%%%%%%%%%%%%%%%%%%%%%%%%%%%%%%%
\subsubsection{$\lambda_{11}^\sigma < 0$, $\lambda_{22}^\sigma < 0$, 
$\lambda_{12}^\sigma > 0$}
In this case, the $\lambda_{11}^\sigma$ and $\lambda_{22}^\sigma$ are 
relevant and flow to large negative values while $\lambda_{12}^\sigma$ 
is irrelevant. Then both $\theta_{1\sigma}$ and $\theta_{2\sigma}$ are 
pinned and satisfy 
$\cos(2\sqrt{2}\theta_{1\sigma}) = \cos(2\sqrt{2}\theta_{2\sigma}) = 1$.
The spin sector is gapped and all spin correlations are short-ranged.
All correlations at $\pi/2$ are also short-ranged.
The translational symmetry is broken since $\la {\cal B}_\pi \ra \neq 0$.
Including the interactions with the singlet sector as in the previous 
section, we have:

a) If $g<1/2$, the ``$\rho-$'' sector remains gapless and
${\cal B}_{2k_{F1}}$ and ${\cal B}_{2k_{F2}}$ have power
law correlations with scaling dimension $g/4$.
These coexist with the static period-2 VBS order.

b) If $g>1/2$, we also pin $\varphi_{\rho-}$ and the situation
is essentially the same as in Sec.~\ref{subsubsec:period2}.
This gives the fully-gapped period-2 VBS phase.

%%%%%%%%%%%%%%%%%%%%%%%%%%%%%%%%%%%%%%%%%%%%%%%%%%%%%%%%%%%%%%%%%%%%
\subsubsection{$\lambda_{11}^\sigma < 0$, $\lambda_{22}^\sigma > 0$, 
$\lambda_{12}^\sigma > 0$}
In this case, only the $\lambda_{11}^\sigma$ is relevant and pins 
$\theta_{1\sigma}$.
Spin correlations at $2k_{F1}$ and all correlations at $\pi/2$
are short-ranged.

We now include the interactions Eq.~(\ref{w12}); the important
part is
\begin{equation}
W = -(4 w_{12}^\rho + 3 w_{12}^\sigma) \cos(\sqrt{2}\theta_{1\sigma}) 
\cos(\sqrt{2}\theta_{2\sigma}) \cos(2\varphi_{\rho-}) ~.
\end{equation}
Both the ``$2\sigma$'' and ``$\rho-$'' modes are dynamical at this stage,
and the $W$ has scaling dimension $1/2 + 1/g$.   

a) $g<2/3$:  The $W$ term is irrelevant and we have two gapless
modes in this phase.  $(\vec{S}, {\cal B})_{2k_{F2}}$ have the
same scaling dimension $1/2 + g/4$ as in the SBM phase, 
while ${\cal B}_{2k_{F1}}$ has scaling dimension $g/4$.
Furthermore, $(\vec{S}, {\cal B})_{\pi}$ have scaling dimension $1/2$.

b) $g>2/3$:  The $W$ term is relevant pinning both $\theta_{2\sigma}$ 
and $\varphi_{\rho-}$.  This is the already encountered fully gapped 
period-2 VBS state.  

The case with $\lambda_{11}^\sigma > 0$, $\lambda_{22}^\sigma < 0$, 
$\lambda_{12}^\sigma > 0$ is considered similarly.

Finally, in the case $\lambda_{12}^\sigma < 0$ and either 
$\lambda_{11}^\sigma < 0$ or $\lambda_{22}^\sigma < 0$,
we can not easily minimize the potential Eq.~(\ref{Vperp}) since we have
non-commuting variables under the relevant cosines.  We do not know
what happens here, although one guess would be that one of the relevant 
terms wins over the others and the situation is reduced to the
already considered cases.

To summarize, we have found several phases that can be obtained
out of the Spin Bose-Metal:
1) fully gapped period-2 VBS; 
2),3) fully gapped period-4 phases, one with bond energy pattern and 
the other with chirality pattern;
4) period-2 VBS coexisting with one gapless mode in the singlet 
(``$\rho-$'') sector and power law correlations in 
${\cal B}_{\pi/2}, \chi_{\pi/2}$;
5) period-2 VBS coexisting with one gapless mode in the singlet sector
and power law correlations in ${\cal B}_{2k_{F1}}, {\cal B}_{2k_{F2}}$;
6) phase with two gapless modes, one in the spin sector and
one in the singlet sector.
It is possible that some of the gapless phases will be further
unstable to effects not considered here.

The above essentially covers all natural possibilities of gapping out 
some or all of the low-energy modes of the generic SBM phase.
Thus, as discussed at the end of Sec.~\ref{subsec:stability},
we cannot pin $\varphi_{\sigma+}$ because of the spin rotation 
invariance.  The SU(2) spin invariance also imposes restrictions 
on the values of the variables that are pinned; these conditions are 
automatically satisfied in the above cases since our starting 
interactions are SU(2)-invariant.
Furthermore, we cannot pin $\theta_{\rho-}$ because of the 
emergent conservation of $\int \partial_x \varphi_{\rho-}$.
One exception is when the Fermi wavevectors take special commensurate 
values; we discuss this next.

%%%%%%%%%%%%%%%%%%%%%%%%%%%%%%%%%%%%%%%%%%%%%%%%%%%%%%%%%%%%%%%%%%%%%
%%%%%%%%%%%%%%%%%%%%%%%%%%%%%%%%%%%%%%%%%%%%%%%%%%%%%%%%%%%%%%%%%%%%%
\subsection{Period-3 VBS state as a possible instability of
the SBM in the commensurate case with $k_{F1} = \pi/3$}
\label{subsec:commens_VBS3}

In the ring model Eq.~(\ref{Hring}), the DMRG observes translational 
symmetry breaking with period 3 in the intermediate parameter range 
flanked by the Spin Bose-Metal on both sides.
Motivated by this, we revisit the spinon-gauge theory in the special
case with $k_{F1} = \pi/3$ (then $k_{F2} = -5\pi/6$). 
Compared to fermion interactions present for generic incommensurate 
Fermi wavevectors, we find one new allowed term
\begin{eqnarray}
V_6 &=& u_6
(f_{R2\up}^\dagger f_{R2\dn}^\dagger f_{L1\alpha}^\dagger
 f_{L2\up} f_{L2\dn} f_{R1\alpha} + \Hc) \\
&=& -4 u_6 \cos(\sqrt{2}\theta_{1\sigma}) 
\sin(3\theta_{\rho-} - \theta_{\rho+}) ~.
\label{V6}
\end{eqnarray}
The pinned $\theta_{\rho+}$ value is kept general at this stage. 
The scaling dimension is $\Delta[V_6] = 1/2 + 9g/4$.
Let us study what happens when $g<2/3$ and $V_6$ becomes relevant,
so $u_6$ flows to large values.  
Then $\theta_{\rho-}$ and $\theta_{1\sigma}$ are pinned while the 
conjugate fields $\varphi_{\rho-}$ and $\varphi_{1\sigma}$ fluctuate 
wildly.  There remains one gapless mode $\theta_{2\sigma}$ 
that is still described by Eq.~(\ref{L0sigma}).

\begin{figure}
\centerline{\includegraphics[width=\columnwidth]{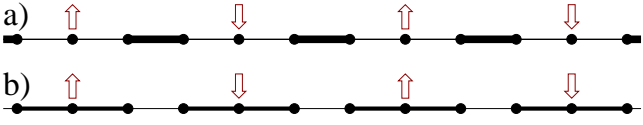}}
\caption{Valence Bond Solid states with period 3.
Thick lines indicate stronger bonds.
Remaining effective spin-1/2 degrees of freedom are also shown.
Coexisting with the translational symmetry breaking, 
we have $1/x$ power law spin correlations with the 
antiferromagnetic (dynamic) pattern as shown.
The two cases have slightly different microscopics, but are 
qualitatively similar on long length scales.
}
\label{fig:dimer3}
\end{figure}

We can use the bond energy and spin operators to characterize the 
resulting state.
First of all, ${\cal B}_{2k_{F1}}$ develops long-range order.
Since $2k_{F1} = 2\pi/3$, we thus have a Valence Bond Solid
with period 3.  Using Eqs.~(\ref{BQ}) and (\ref{epsilon_2kF}),
the microscopic bond energy is
\begin{eqnarray}
\delta {\cal B}(x) \sim \cos(\sqrt{2} \theta_{1\sigma}) 
\sin(2k_{F1} x + k_{F1} + \theta_{\rho-} + \theta_{\rho+}) ~.
\end{eqnarray}
In Eq.~(\ref{V6}), we write $3\theta_{\rho-} - \theta_{\rho+} = 
3(\theta_{\rho-} + \theta_{\rho+}) - 4\theta_{\rho+}$
and use the pinning condition on $4\theta_{\rho+}$,
Eq.~(\ref{H8bosonized}).
There are two cases:

a) $u_6 \cos(4\theta_{\rho+}) < 0$:
In this case, $V_6$ is minimized by inequivalent pinning values 
$\sqrt{2} \theta_{1\sigma} = \pi$;
$\theta_{\rho-} + \theta_{\rho+} = \pi/6, \pi/6+2\pi/3, \pi/6+4\pi/3$.
For $\theta_{\rho-} + \theta_{\rho+} = \pi/6$, the period-3 
pattern of bonds is
\begin{eqnarray}
\delta {\cal B}(x) = \{ \dots, -1, \frac{1}{2}, \frac{1}{2}, \dots \} ~,
\end{eqnarray}
while the other two inequivalent pinning values give translations of 
this pattern along the chain.  A lower bond energy is interpreted as a 
stronger antiferromagnetic correlation on the bond.  Then the above 
pattern corresponds to ``dimerizing'' every third bond as shown in 
Fig.~\ref{fig:dimer3}a).

b) $u_6 \cos(4\theta_{\rho+}) > 0$: 
In this case, $V_6$ is minimized by $\sqrt{2} \theta_{1\sigma} = 0$;
$\theta_{\rho-} + \theta_{\rho+} = \pi/6, \pi/6+2\pi/3, \pi/6+4\pi/3$.
For $\theta_{\rho-} + \theta_{\rho+} = \pi/6$, the period-3 
pattern of bonds is
\begin{eqnarray}
\delta {\cal B}(x) = \{ \dots, 1, -\frac{1}{2}, -\frac{1}{2}, \dots \} ~,
\end{eqnarray}
while the other two inequivalent pinning values give translations of 
this along the chain.  This pattern corresponds to every third bond 
being weaker as shown in Fig.~\ref{fig:dimer3}b).

Continuing with the characterization, we note that $\vec{S}_{2k_{F1}}$
and all operators at $\pi/2$ have exponentially decaying correlations.
On the other hand, $\vec{S}_{2k_{F2}}$ and ${\cal B}_{2k_{F2}}$
have $1/x$ power law correlations because of the remaining 
gapless $\theta_{2\sigma}$ mode.
Since $2k_{F2} = \pi/3$, we have period-6 spin correlations on the 
original 1D chain.

The physical interpretation is simple.
Consider first Fig.~\ref{fig:dimer3}a) where every third bond is 
stronger.  A caricature of this state is that spins in the strong 
bonds form singlets and are effectively frozen out.
The remaining ``free'' spins are separated by three lattice spacings 
and are weakly antiferromagnetically coupled forming a new effective 
1D chain.  
Thus we naturally have Bethe-chain-like staggered spin and bond energy 
correlations in this subsystem, which coexist with the static period-3
VBS order in the whole system.
The situation in Fig.~\ref{fig:dimer3}b) where every third bond is 
weaker is qualitatively similar.  Here we can associate an effective 
spin-1/2 with each three-site cluster formed by strong bonds. 
These effective spins are again separated by three lattice spacings
and form a new weakly coupled Bethe chain.
Note that while the theory analysis has the Fermi wavevectors tuned to 
the commensuration, the resulting state is a stable phase that can occupy
a finite region in the parameter space, as found by the DMRG in the 
ring model (see Sec.~\ref{subsec:DMRG:VBS3}).

We can construct trial wavefunctions using spinons as follows.  
In the mean field, we start with the band parameters 
$t_1$ and $t_2$ such that $k_{F1} = \pi/3$ and then add period-3 
modulation of the hoppings.  
The $\pm k_{F1}$ Fermi points are connected by the modulation
wavevector and are gapped out.  
The $\pm k_{F2}$ Fermi points remain gapless; just as in the 
Bethe chain case, the corresponding Bosonized field theory provides an 
adequate description of the long-wavelength physics, 
predicting $1/x$ decay of staggered spin and bond energy correlations.

The above wavefunction construction and theoretical analysis are 
implicitly in the regime where the residual spin correlations are 
antiferromagnetic.  In a given physical system forming such a 
period-3 VBS, one can also imagine ferromagnetic residual interactions 
between the non-dimerized spins.  Indeed, the DMRG finds some weak 
ferromagnetic tendencies in the ring model near the transition to this 
VBS state.  This is not covered by our spin-singlet SBM theory but 
could possibly be covered starting with a partially polarized SBM state.

%%%%%%%%%%%%%%%%%%%%%%%%%%%%%%%%%%%%%%%%%%%%%%%%%%%%%%%%%%%%%%%%%%%
%%%%%%%%%%%%%%%%%%%%%%%%%%%%%%%%%%%%%%%%%%%%%%%%%%%%%%%%%%%%%%%%%%%
\subsection{Other possible commensurate points}
\label{subsec:commens_other}

Alerted by the period-3 VBS case, we look for and find one additional 
commensurate case with an allowed new interaction that can destabilize 
the Spin Bose-Metal.  When $k_{F1} = 3\pi/8$, we find a new quartic term
\begin{eqnarray}
V_4 &\!=\!& u_4 
\Big[ f_{R1\up}^\dagger f_{R1\dn}^\dagger 
      \epsilon_{\alpha\beta} f_{L1\alpha} f_{R2\beta} 
     + (R \leftrightarrow L) + \Hc \Big] \\
&\!\sim\!&
-i \eta_{1\up} \eta_{2\up} 
\sin(2\theta_{\rho-} + \theta_{\rho+} - \theta_{\sigma+}) 
\sin(\varphi_{\rho-} + \varphi_{\sigma-}) \nonumber \\
&& - i \eta_{1\dn} \eta_{2\dn} 
\sin(2\theta_{\rho-} + \theta_{\rho+} + \theta_{\sigma+})
\sin(\varphi_{\rho-} - \varphi_{\sigma-}) \nonumber ~.
\end{eqnarray}
(For the schematic writing here, we have ignored the commutations
of the fields when separating the $\varphi$'s and $\theta$'s.)
The scaling dimension is $\Delta[V_4] = 1/2 + g + 1/(4g)$.
This is smaller than 2 for $g \in (0.191, 1.309)$ and the interaction
is relevant in this range.
Since we have conjugate variables $\theta_{\rho-}$ and $\varphi_{\rho-}$ 
both present in the above potential, we can not easily determine the 
ultimate outcome of the runaway flow.
It seems safe to assume that $\theta_{\sigma+}$ and $\varphi_{\sigma-}$
will be both pinned, which implies at least some period-2 translational 
symmetry breaking.
One possibility, perhaps aided by the interactions Eq.~(\ref{w12}), 
is that the $\varphi_{\rho-}$ is pinned;
in this case, the situation is essentially the same as in
Sec.~\ref{subsubsec:period4} and we get some period-4 structure.
Another possibility is that the $\theta_{\rho-}$ is pinned;
in this case $\la \epsilon_{4k_{F1}} \ra \neq 0$, and since 
$4k_{F1}=-\pi/2$, we get period-4 bond pattern.

To conclude, we note that the commensurate cases in this section and
in Sec.~\ref{subsec:commens_VBS3} can be understood phenomenologically 
by monitoring the wavevectors of the energy modes ${\cal B}_Q$.
The dominant wavevectors are $\pm 2k_{Fa}$, $\pm \pi/2$, and 
$\pm 4k_{F2} = \mp 4k_{F1}$.
When $4k_{F2}$ matches with $\pi/2$, we get the $k_{F1}=3\pi/8$
commensuration of this section (here also $2k_{F1}$ matches with 
$k_{F2}-k_{F1}$, while $2k_{F2}$ matches with $3k_{F1} + k_{F2}$).
When $4k_{F2}$ matches with $2k_{F1}$, we get the $k_{F1}=\pi/3$
commensuration of the previous section.
Tracking such singular wavevectors in the DMRG is then very helpful to 
alert us to possible commensuration instabilities, and both cases are 
realized in the ring model with additional antiferromagnetic 
$J_3 = 0.5 J_1$ discussed in Sec.~\ref{subsec:DMRG:afJ3},
cf.~Fig.~\ref{fig:qsing_J3_0p5}.

%%%%%%%%%%%%%%%%%%%%%%%%%%%%%%%%%%%%%%%%%%%%%%%%%%%%%%%%%%%%%%%%%%%%%
%%%%%%%%%%%%%%%%%%%%%%%%%%%%%%%%%%%%%%%%%%%%%%%%%%%%%%%%%%%%%%%%%%%%%
%%%%%%%%%%%%%%%%%%%%%%%%%%%%%%%%%%%%%%%%%%%%%%%%%%%%%%%%%%%%%%%%%%%%%
\section{DMRG study of commensurate instabilities inside the SBM:
VBS-3 and Chirality-4}
\label{sec:DMRG_commens}

%%%%%%%%%%%%%%%%%%%%%%%%%%%%%%%%%%%%%%%%%%%%%%%%%%%%%%%%%%%%%%%%%%%%%%
%%%%%%%%%%%%%%%%%%%%%%%%%%%%%%%%%%%%%%%%%%%%%%%%%%%%%%%%%%%%%%%%%%%%%%
\subsection{Valence Bond Solid with period 3}
\label{subsec:DMRG:VBS3}

\begin{figure}
\centerline{\includegraphics[width=\columnwidth]{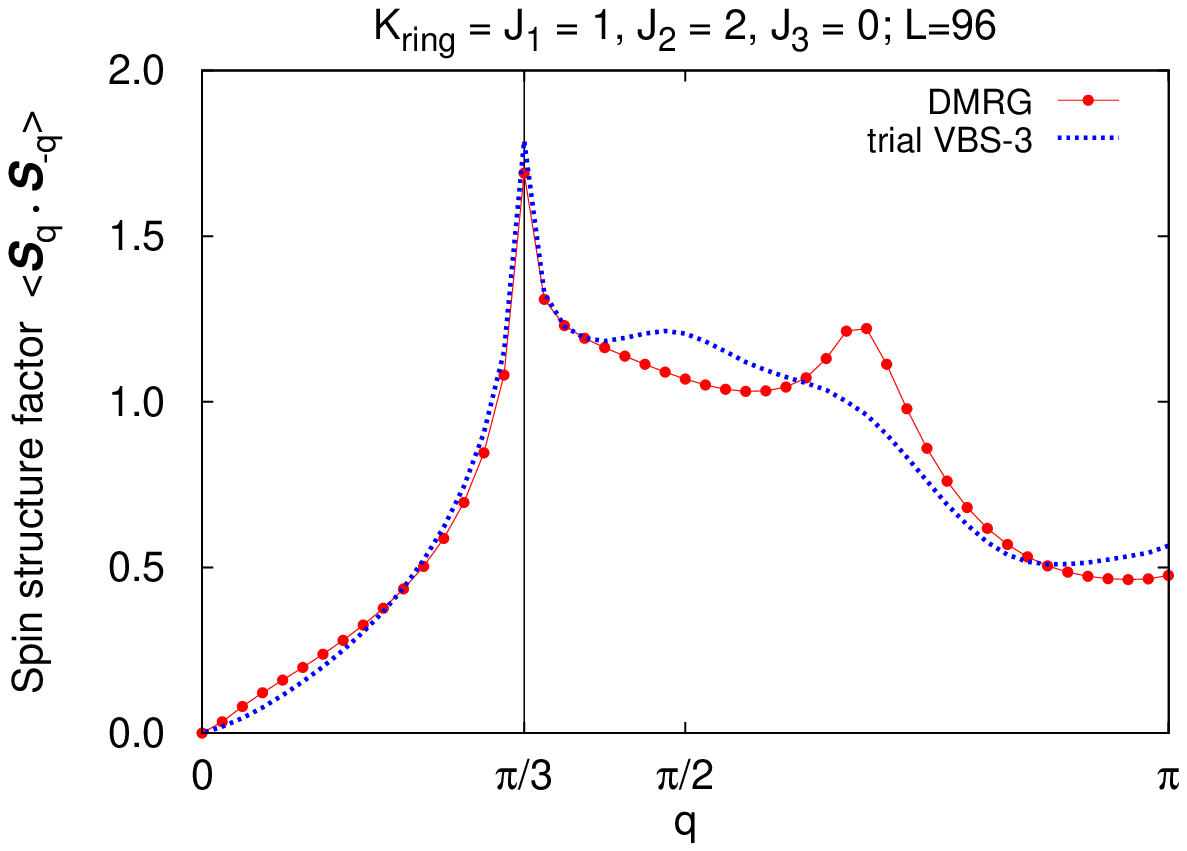}}
\centerline{\includegraphics[width=\columnwidth]{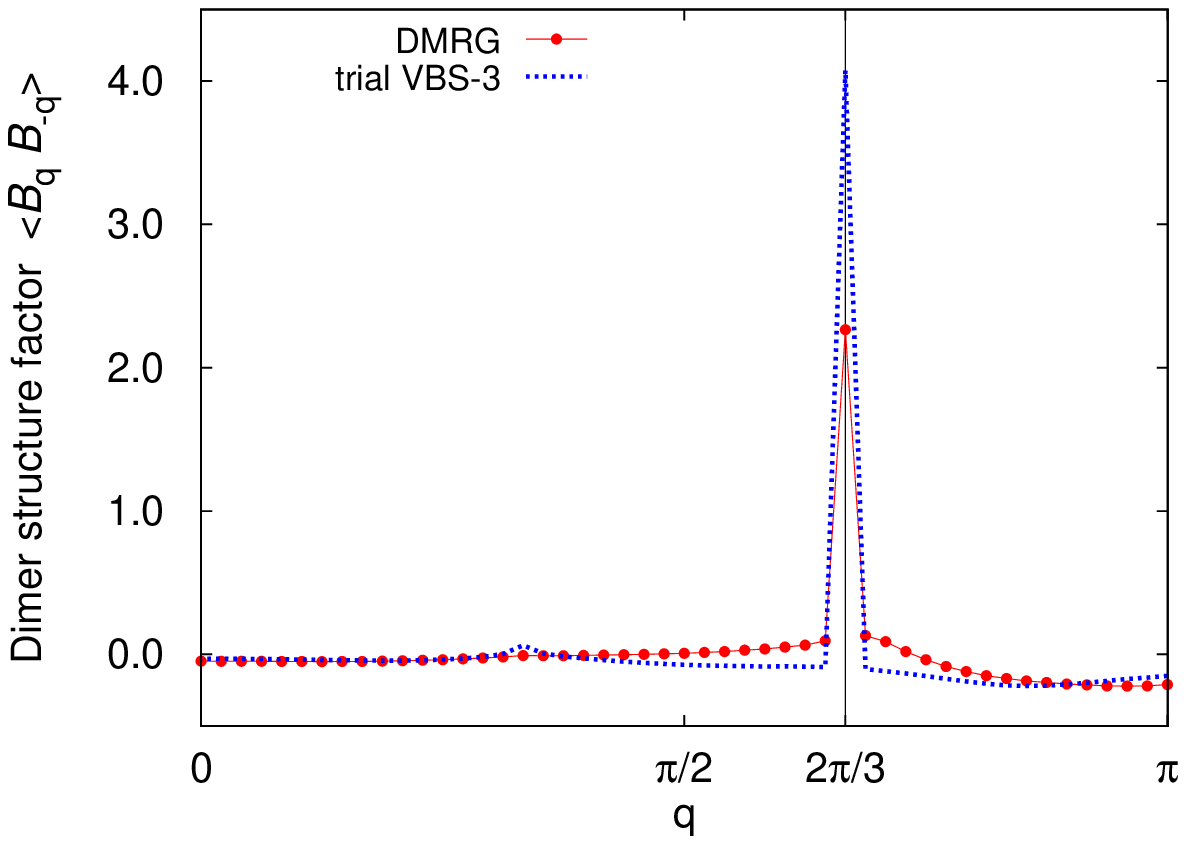}}
\caption{
(Color online)
Spin and dimer structure factors at a representative point in the 
VBS-3 phase, $K_{\rm ring} = J_1 = 1$, $J_2 = 2$, 
measured in the DMRG for system size $L=96$
(we do not show the chirality as it is not informative.)
The most notable features are the dimer Bragg peak at $2\pi/3$
corresponding to the static VBS order and also the spin singularity at 
$\pi/3$ corresponding to the effective spin-1/2 chain formed by the 
non-dimerized spins, see Fig.~\ref{fig:dimer3}.
The trial VBS-3 wavefunction is constructed as described in the text.
} 
\label{fig:J2_2p0_J3_0p0}
\end{figure}

As already mentioned in Sec.~\ref{sec:HringDMRG}, we find a range of 
parameters where the SBM phase is unstable towards a Valence Bond Solid 
with a period of 3 lattice spacings (VBS-3).
In the model with $K_{\rm ring} = J_1 = 1$, this occurs for 
$1.5 < J_2 < 2.5$, cf.~Fig.~\ref{fig:qsing_J3_0p0}.
The characteristic correlations are shown in Fig.~\ref{fig:J2_2p0_J3_0p0}
at a point $J_2 = 2$.  The dimer structure factor shows a Bragg peak at a
wavevector $2\pi/3$ corresponding to the period-3 VBS order.
The spin structure factor has a singularity at a wavevector $\pi/3$ 
corresponding to staggered correlations in the effective spin-1/2 chain 
formed by the non-dimerized spins, see Fig.~\ref{fig:dimer3}.
If we zoom in closer, the dimer structure factor also has a feature at 
$\pi/3$ that can be associated with this effective chain.

To construct a trial VBS-3 wavefunction, we start with the spinon 
hopping problem that would produce $k_{F1} = \pi/3$, so the first Fermi 
sea would be twice as large as the second.  
We then multiply every third first-neighbor hopping by $1 + \delta$
and Gutzwiller-project; for the point in Fig.~\ref{fig:J2_2p0_J3_0p0}
we find optimal $\delta = 1$.
This gaps out the larger Fermi sea but leaves the smaller Fermi sea 
gapless.  Our wavefunction is crude and shows a stronger dimer Bragg peak
than the DMRG and somewhat different spin correlations at short scales, 
but otherwise captures the qualitative features as can be seen
in Fig.~\ref{fig:J2_2p0_J3_0p0}.

The origin of the VBS-3 phase can be traced to the instability of the 
SBM at special commensuration, 
cf.~Secs.~\ref{subsec:commens_VBS3}-\ref{subsec:commens_other}.
Indeed, in Fig.~\ref{fig:evolution2} we can follow the evolution of
the singular wavevectors in the SBM phase between the VBS-2 and VBS-3.
As we decrease $J_2$ moving towards the VBS-3, the $4k_{F2}$ and 
$2k_{F1}$ singular wavevectors in the bond energy approach each other 
and coincide at $2\pi/3$.  
When this happens, there is a new umklapp term that can destabilize the 
SBM and produce the VBS-3 state as analyzed in Sec.~\ref{subsec:commens_VBS3}.
The instability requires $g < 2/3$ for the SBM Luttinger parameter. 
In this case according to Table~\ref{tab:SBMprops} the $4k_{F2}$ 
singularity in the dimer is stronger than the $2k_{F1,2}$, which is 
in agreement with what we see in the neighboring SBM in 
Fig.~\ref{fig:evolution2}.  
The re-emergence of the $4k_{F2}$ and $2k_{F1}$ at the other end of the 
VBS-3 phase is obscured here by the weak ferromagnetic tendency
(but is present in a model where this tendency is suppressed, see 
Fig.~\ref{fig:qsing_J3_0p5}).

%%%%%%%%%%%%%%%%%%%%%%%%%%%%%%%%%%%%%%%%%%%%%%%%%%%%%%%%%%%%%%%%%%%%%%
%%%%%%%%%%%%%%%%%%%%%%%%%%%%%%%%%%%%%%%%%%%%%%%%%%%%%%%%%%%%%%%%%%%%%%
\subsection{Enhancement of the spin-singlet SBM by antiferromagnetic
third-neighbor coupling $J_3 = 0.5$ and a new phase with
chirality order with period 4.}
\label{subsec:DMRG:afJ3}

As discussed in Sec.~\ref{subsec:qsing}, in the original 
$J_1 - J_2 - K_{\rm ring}$ model, states in the SBM region near the 
left boundary of the VBS-3 tend to develop a small magnetic moment.  
We conjecture that this occurs in the second spinon 
Fermi sea and suggest that an antiferromagnetic $J_3$ will stabilize the 
SBM phase with spin-singlet ground state.  
One motivation comes from the picture of the neighboring VBS-3,
where the non-dimerized spins are loosely associated with the 
second Fermi sea.  These spins are three lattice spacings apart,
so adding antiferromagnetic $J_3$ should lead to stronger 
antiferromagnetic tendencies among them and also in the physics 
associated with the second Fermi sea.

\begin{figure}[t]
\centerline{\includegraphics[width=\columnwidth]{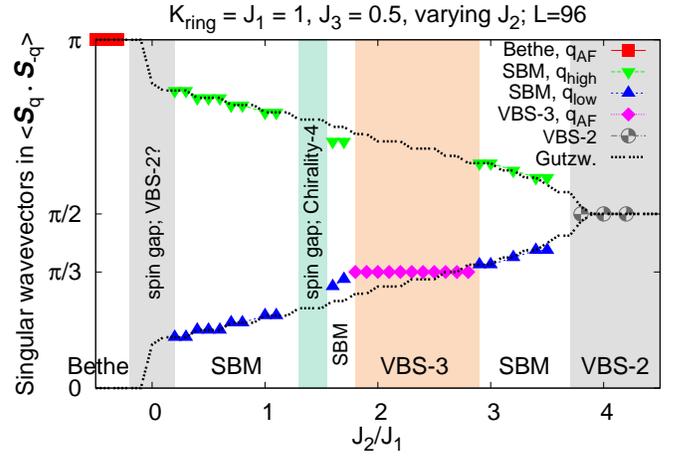}}
\caption{
(Color online)
Phase diagram of the $J_1 - J_2 - K_{\rm ring}$ model with additional
antiferromagnetic third-neighbor coupling $J_3 = 0.5 J_1$ 
introduced to stabilize spin-singlet states. 
The study is along the same cut $K_{\rm ring} = J_1$ as in 
Fig.~\ref{fig:qsing_J3_0p0}, and the overall features are similar,
with the following differences.
The ground state is singlet throughout eliminating the difficult
``partial FM'' region to the left of the VBS-3 
(and the VBS-3 phase is somewhat wider).
There is a sizable spin-gapped region between the Bethe-chain and 
SBM phases (see text for more details).
A new spin-gapped phase with period-4 chirality order appears inside the 
SBM to the left of the VBS-3.
}
\label{fig:qsing_J3_0p5}
\end{figure}

We have performed a detailed study adding a modest $J_3 = 0.5$
to the original model Eq.~(\ref{Hring}) along the same cut 
$K_{\rm ring} = J_1 = 1$.
Our motivating expectations are indeed borne out.
Figure~\ref{fig:qsing_J3_0p5} shows the phase diagram together with
the evolution of the singular wavevectors as a function of $J_2 / J_1$.
While the overall features are similar to the phase diagram in the 
$J_3 = 0$ case, Fig.~\ref{fig:qsing_J3_0p0}, a few points are worth 
mentioning.

First, the partial spin polarization is absent in the whole SBM phase
between the Bethe-chain and VBS-3 phases.  The DMRG converges confidently
to spin-singlet ground state for $L = 96$.  All properties are 
similar to those in Figs.~\ref{fig:J2_0p0_J3_0p0} and 
\ref{fig:evolution1}, providing further support for the singlet 
SBM phase in the original $J_3 = 0$ model.
The VBS-3 phase and the SBM phase between the VBS-3 and VBS-2 are 
qualitatively very similar in the two cases $J_3 = 0$ and $J_3 = 0.5$
and are not discussed further here.

An interesting feature in the model with $J_3 = 0.5$ is the
presence of a sizable phase with spin gap intervening between the 
Bethe-chain and SBM phases.
Our best guess is that this phase has period-2 VBS order,
although we do not see a clear signature in the dimer correlations.
Our DMRG states in this region show very weak (if any) dimerization 
of the first-neighbor bonds, which may explain the lack of clear 
order in these dimer correlations.
On the other hand, we see a sizable period-2 dimerization of the 
third-neighbor bonds $\la \vec{S}(x) \cdot \vec{S}(x+3) \ra$, 
but have not measured the corresponding bond-bond correlations to 
confirm long-range order.  
We have not explored possible theoretical routes to understand
the origin of this phase yet and leave our discussion of this 
region as is.

\begin{figure}
\centerline{\includegraphics[width=\columnwidth]{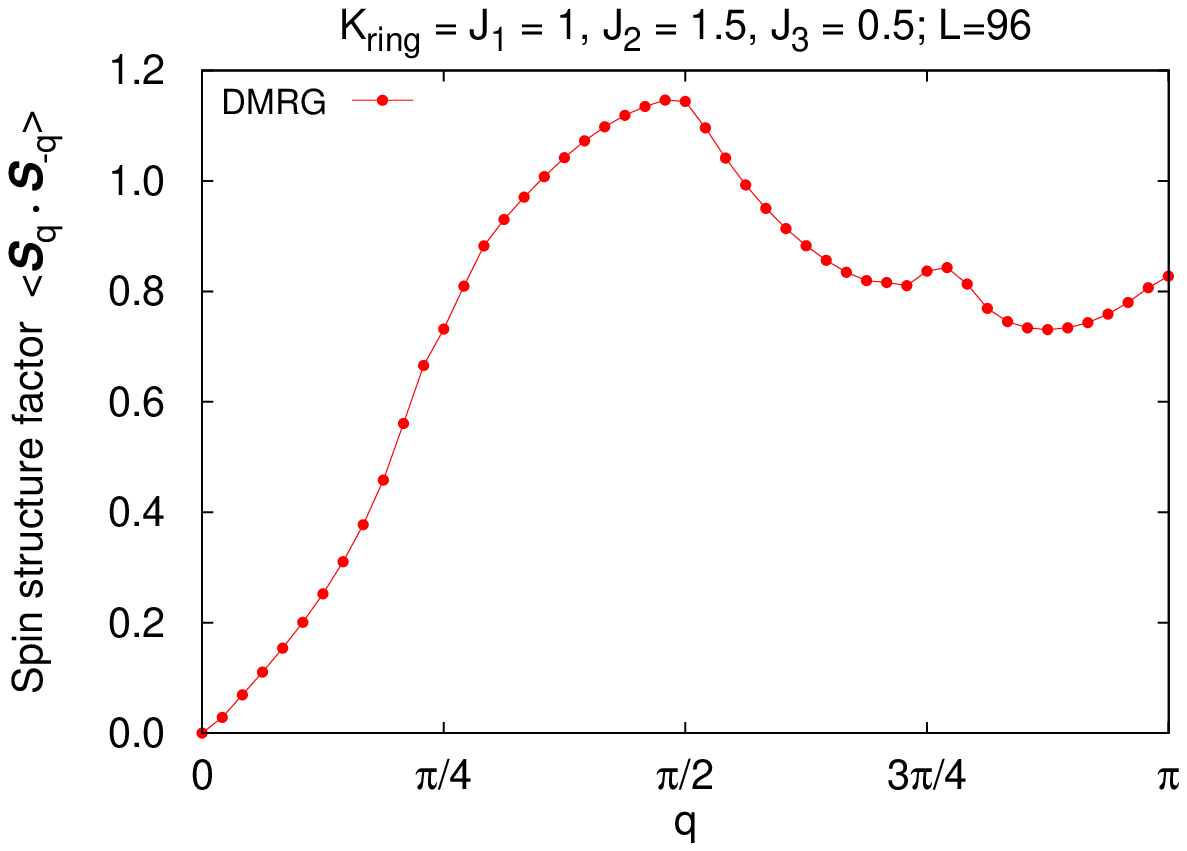}}
\centerline{\includegraphics[width=\columnwidth]{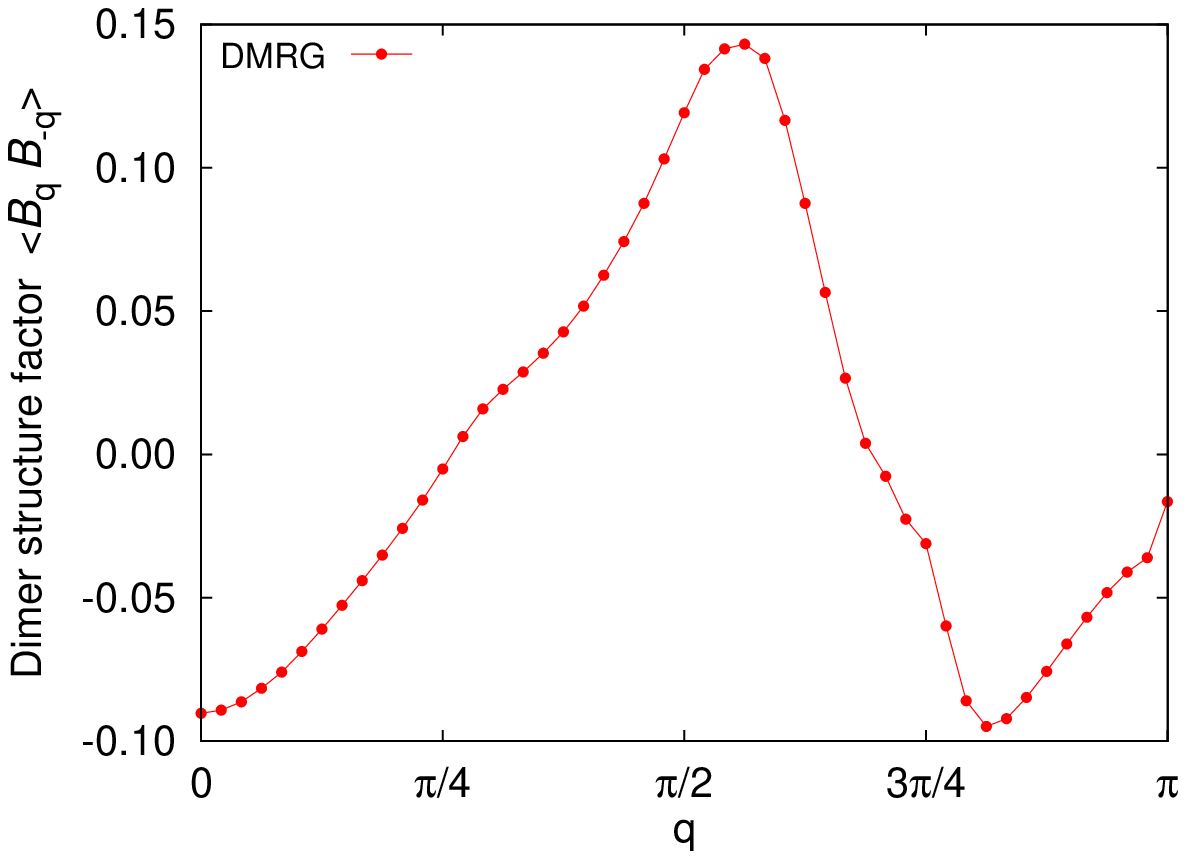}}
\centerline{\includegraphics[width=\columnwidth]{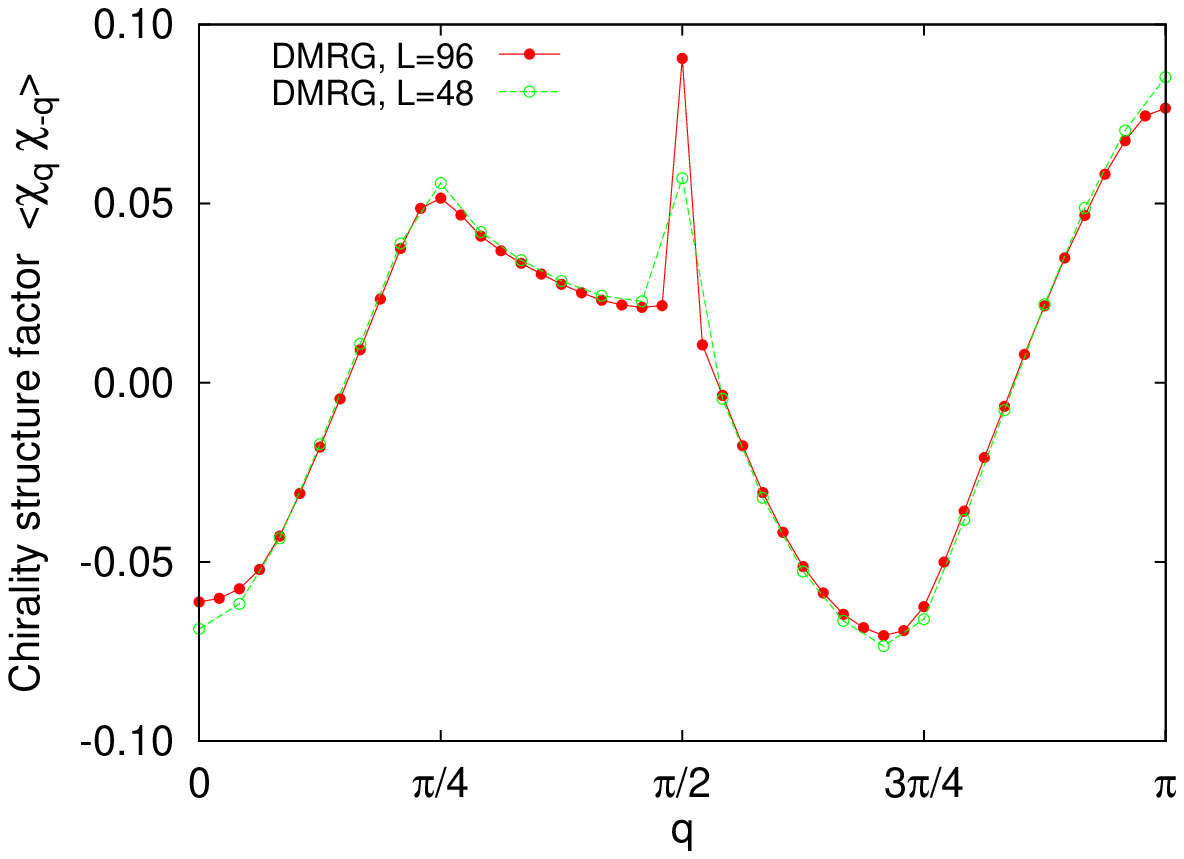}}
\caption{
(Color online)
Spin, dimer, and chirality structure factors at a point in the tentative 
Chirality-4 phase, $K_{\rm ring} = J_1 = 1$, $J_2 = 1.5, J_3 = 0.5$, 
measured in the DMRG for system size $L = 96$. 
Note the absence of sharp features in the spin correlations, which 
suggests a spin gap, while the dimer correlations have only a feature at 
$\pi$ corresponding to period-2 modulation.  
On the other hand, the chirality structure factor shows a Bragg peak at 
$\pi/2$ that grows with increasing system size.
The pattern of chirality correlations in real space is consistent 
with the order shown in Fig.~\ref{fig:chirality4}.
} 
\label{fig:J2_1p5_J3_0p5}
\end{figure}

We now turn to one more new phase found in the $J_3 = 0.5$ model.
In a narrow region inside the SBM phase not far from the left end
of the VBS-3, we again find a spin-gapped phase.  We identify this 
as having period-4 order in the chirality (Chirality-4 phase).
Figure~\ref{fig:J2_1p5_J3_0p5} presents a point $J_2 = 1.5, J_3 = 0.5$.
Looking at the singular wavevectors in Fig.~\ref{fig:qsing_J3_0p5},
we see that this point is roughly where $q_{\rm low} = 2k_{F2}$ 
passes $\pi/4$.  Analysis in Sec.~\ref{subsec:commens_other} suggests an
instability gapping out all modes and leading to some period-4 structure. 
Indeed, the DMRG spin and dimer structure factors show only some 
remnants of features near $2k_{F1}, 2k_{F2}$, while the chirality shows 
a sharp peak at $\pi/2$.
Looking at real space correlations, our $L = 96$ DMRG state breaks 
translational symmetry.  The pattern of chirality correlations
is consistent with the period-4 order shown in Fig.~\ref{fig:chirality4}.
The pattern of dimer correlations is also consistent with this picture
and shows modulation with period 2, which can be seen as a feature at 
$\pi$ in the dimer structure factor in Fig.~\ref{fig:J2_1p5_J3_0p5}.

To summarize, with the help of modest $J_3 = 0.5$ we have stabilized the 
spin-singlet SBM states between the Bethe-chain and VBS-3 phases.
By suppressing potential weak ferromagnetism, we have uncovered 
the Chirality-4 phase, which can be understood as arising from
the instability of the SBM at the special commensuration
discussed in Sec.~\ref{subsec:commens_other}.

%%%%%%%%%%%%%%%%%%%%%%%%%%%%%%%%%%%%%%%%%%%%%%%%%%%%%%%%%%%%%%%%%%%%%%%
%%%%%%%%%%%%%%%%%%%%%%%%%%%%%%%%%%%%%%%%%%%%%%%%%%%%%%%%%%%%%%%%%%%%%%%
%%%%%%%%%%%%%%%%%%%%%%%%%%%%%%%%%%%%%%%%%%%%%%%%%%%%%%%%%%%%%%%%%%%%%%%
\section{An attempt to bring out partially magnetized Spin Bose-Metal 
by ferromagnetic third neighbor coupling $J_3 = -0.5$}
\label{sec:ferroJ3}

In the original model Eq.~(\ref{Hring}), we do not have a clear 
understanding of the ``partial FM'' states to the left of the 
VBS-3 phase in Fig.~\ref{fig:phased_J3_0p0}.  
As discussed in Sec.~\ref{subsec:qsing}, we suspect that there is a 
tendency to weak ferromagnetism in the second Fermi sea.  However, 
the magnetizations that we measure are not large: e.g., they are 
significantly smaller than if we were to fully polarize the second 
Fermi sea, and it is difficult to analyze such states.
Here we seek better control over the spin polarization by adding 
modest ferromagnetic $J_3 = -0.5$ in hopes of stabilizing states
with a full spontaneous polarization of the second Fermi sea, 
which is easier to analyze.  

We do not have as detailed phase diagram as for the $J_3 = 0$ and 
$J_3 = 0.5$ cases.  We expect it to look crudely similar to 
Figs.~\ref{fig:qsing_J3_0p0} and \ref{fig:qsing_J3_0p5},
with a narrower (if any) VBS-3 region, and with a wider
partially polarized SBM region.
We indeed find stronger ferromagnetic tendencies in the range
$0 \leq J_2 \leq 2$.  
However, we cannot claim achieving robust full polarization of the 
second Fermi sea and understanding all behavior.
The largest magnetization and properties closest to our expectations 
are found in the middle region near $J_2 \simeq 1$. 
With this cautionary note warranting more work, we now present results 
for $J_2 \simeq 1$ to illustrate our thinking about such states.

The DMRG study proceeds as follows.  We start as before working in the 
$S^z_{\rm tot} = 0$ sector.  We measure the spin structure factor and 
calculate the total spin $S_{\rm tot}$ from Eq.~(\ref{Stot}), which can 
give a first indication of a non-zero magnetization.  
However, for the larger system sizes, the DMRG finds it difficult to 
converge to integer-valued $S_{\rm tot}$ due to a mixing of states
with different total spins, and this leads to significant uncertainty.  
To check the value of the ground state spin, we run the DMRG in sectors 
with different $S^z_{\rm tot}$, expecting the ground state energy to be 
the same for $S^z_{\rm tot} = 0, \dots, S_{\rm tot}$ and then to jump 
to a higher value for $S^z_{\rm tot} > S_{\rm tot}$.

As an example, at a point $J_2 = 1$ for system size $L = 48$
we find that the DMRG energy is the same in the sectors 
$S^z_{\rm tot} = 0, \dots, 5$ and then jumps, so the ground state spin 
is determined as $S_{\rm tot} = 5$.
The DMRG convergence is good and the SU(2)-invariant structure factor 
$\la \vec{S}_q \cdot \vec{S}_{-q} \ra$ is the same measured in the
different sectors $S^z_{\rm tot} \leq 5$, indicating that these states 
indeed belong to the same multiplet.

The situation is less clear for $L = 96$ because of reduced convergence.
At the point $J_2 = 1$ the extensive energies obtained by the DMRG
in the sectors $S^z_{\rm tot} = 0, \dots, 10$ are non-systematic and are 
within $0.1 J_1$ of each other, with the lowest energy found in the 
$S^z_{\rm tot} = 10$ sector. 
Also, the SU(2)-invariant structure factors differ slightly and the 
estimates of $S_{\rm tot}$ vary around $S_{\rm tot} \sim 8 - 10$.
The convergence is best in the highest $S^z_{\rm tot} = 10$ sector,
where the total spin is found to be accurately $S_{\rm tot} = 10$;
the improved convergence is indeed expected since there are fewer 
available low energy excited states to mix with (e.g., spin-wave 
excitations of the ferromagnet are not present in the highest 
$S^z$ sector).
Interestingly, we find a fully converged state in the 
$S^z_{\rm tot} = 11$ sector with $S_{\rm tot} = 11$ whose energy is
only slightly higher, which probably adds to the above convergence 
difficulties. 
More importantly, the energy jumps to a significantly higher value 
in the sector $S^z_{\rm tot} = 12$. 
Our best conclusion is that the total spin of the ground state is 
$S_{\rm tot} = 10$.

Turning to the VMC study, we consider a family of variational 
Gutzwiller states where we allow different spin up and spin down 
populations of the two Fermi seas centered around $k=0$ and $\pi$
(we do not attempt any further improvements on top of such bare 
wavefunctions).  In the model parameter region discussed here, 
we find that the optimal such states have a fully polarized second 
Fermi sea and an unpolarized first Fermi sea.  
For the $L=48$ example quoted above, the optimal VMC polarization 
indeed matches the DMRG $S_{\rm tot} = 5$, while for the $L=96$ case
the optimal VMC state has $S_{\rm tot} = 11$ and a state with 
$S_{\rm tot} = 10$ is very close in energy.
Appendix~\ref{subapp:Gutzw_ferroFS2} provides more details on the 
properties of such Gutzwiller states, while here we simply compare the 
VMC and DMRG measurements.

\begin{figure}
\centerline{\includegraphics[width=\columnwidth]{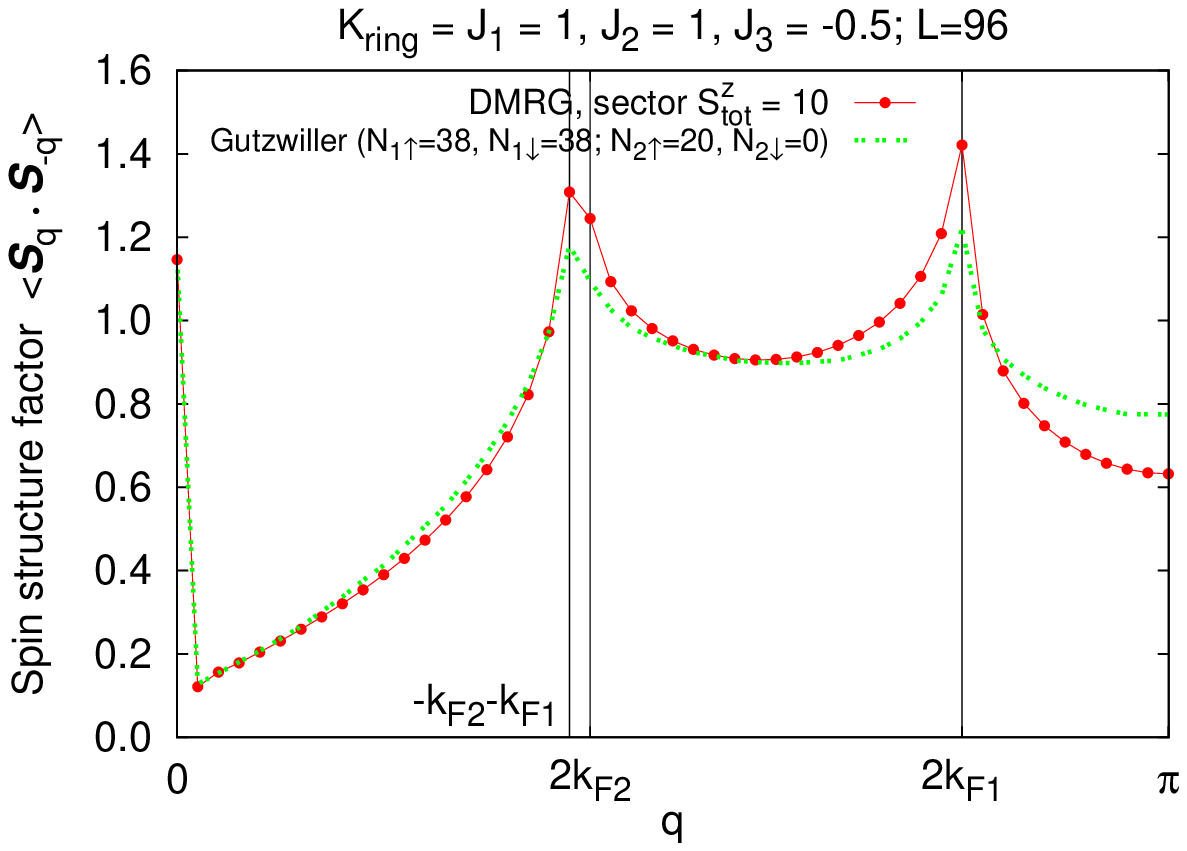}}
\centerline{\includegraphics[width=\columnwidth]{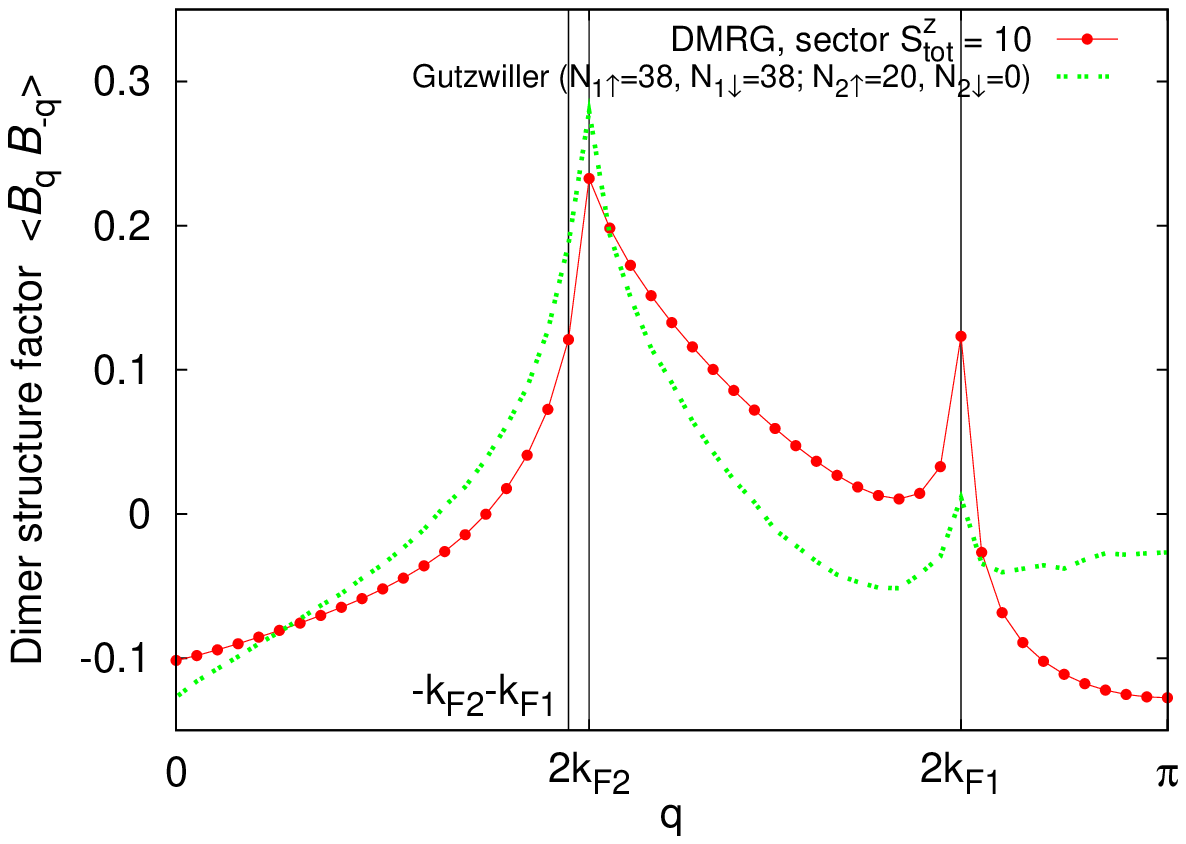}}
\caption{
(Color online)
Spin and dimer structure factors at a tentative point with partial 
ferromagnetism, $K_{\rm ring} = J_1 = 1$, $J_2 = 1$, $J_3 = -0.5$, 
measured in the DMRG for system size $L = 96$.  
The calculations are done in the sector $S^z_{\rm tot} = 10$ where the 
DMRG is well-converged and gives $S_{\rm tot} = 10$, which we think is the 
true ground state spin.
The VMC state has the second Fermi sea fully polarized with 
$N_{2\up} = 20, N_{2\dn} = 0$, while the first Fermi sea is unpolarized 
with $N_{1\up} = N_{1\dn} = 38$. 
Vertical lines label important wavevectors $2k_{F1}$, $2k_{F2}$, and 
$-k_{F2}-k_{F1}$.
}
\label{fig:J2_1p0_J3_m0p5}
\end{figure}

The DMRG structure factors for the $J_2 = 1, J_3 = -0.5$, $L=96$ system 
are shown in Fig.~\ref{fig:J2_1p0_J3_m0p5}, together with the VMC results
for the Gutzwiller wavefunction with $S_{\rm tot} = 10$.
A notable difference from the singlet SBM states of 
Sec.~\ref{sec:HringDMRG} is that the characteristic peaks are no longer 
located symmetrically about $\pi/2$.  
For example, in the trial state we have prominent wavevectors $2k_{F2}$ 
and $2k_{F1}$ that satisfy $2k_{F2} + 2 \times 2k_{F1} = 2\pi$;
also, we have a wavevector $-k_{F2} - k_{F1}$, which is now 
different from $\pi/2$.
It is not easy to discern all wavevectors in 
Fig.~\ref{fig:J2_1p0_J3_m0p5} because the $2k_{F2}$ happens to
be close with the $-k_{F2} - k_{F1}$.
Nevertheless, the overall match between the DMRG and VMC suggests that 
the trial wavefunction captures reasonably the nature of the 
ground state.

\begin{figure}
\centerline{\includegraphics[width=\columnwidth]{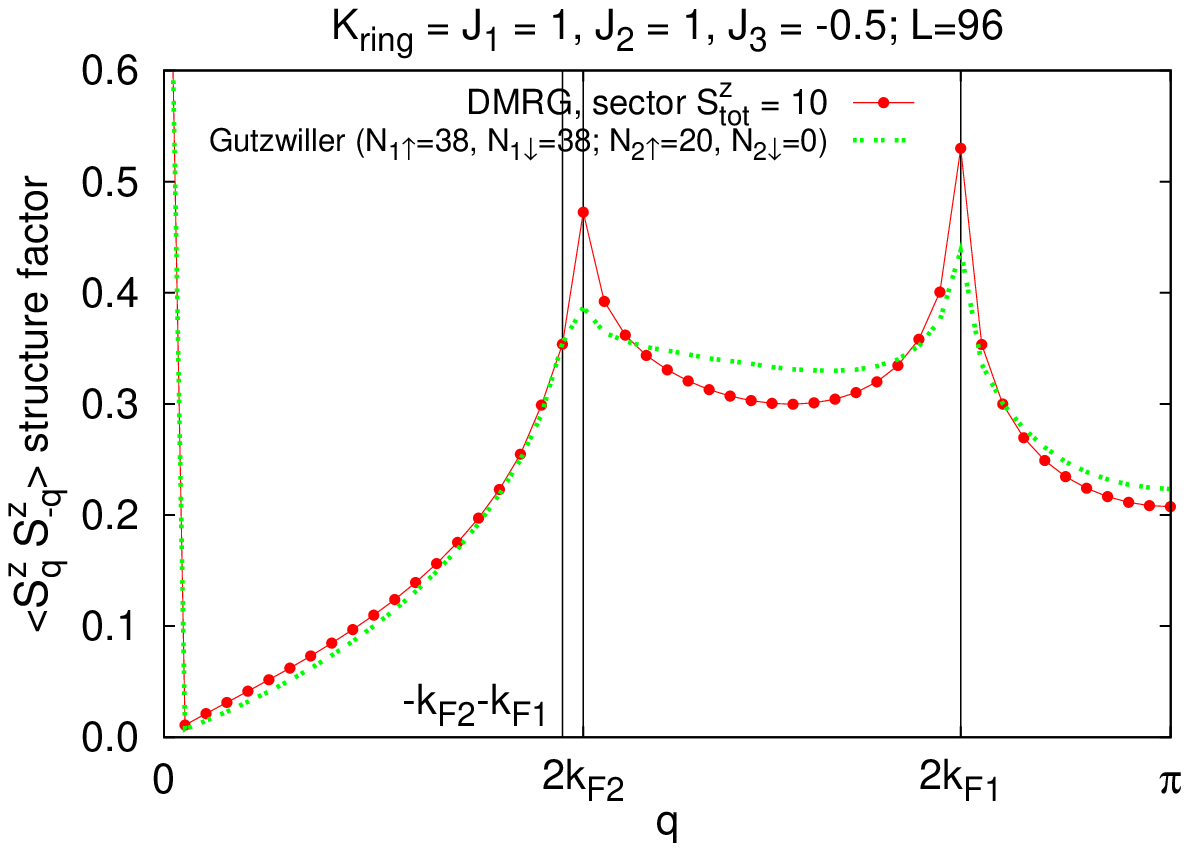}}
\centerline{\includegraphics[width=\columnwidth]{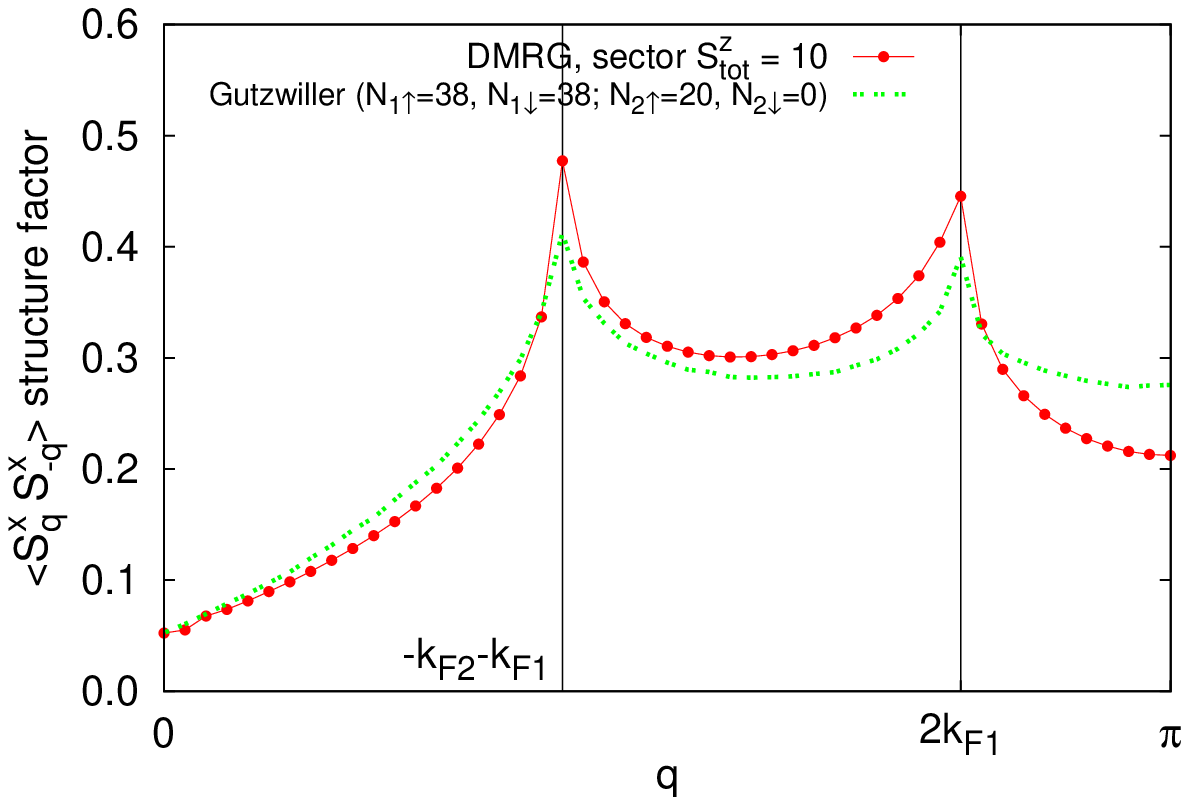}}
\caption{
(Color online)
Separate $\la S^z_q S^z_{-q} \ra$ and $\la S^x_q S^x_{-q} \ra$ 
structure factors for the same system as in Fig.~\ref{fig:J2_1p0_J3_m0p5}.
}
\label{fig:SzSzq_SxSxq_J2_1p0_J3_m0p5_Sztot10}
\end{figure}

Working in the sector $S^z_{\rm tot} = S_{\rm tot}$ also allows more 
detailed comparison between the DMRG and VMC.
In this case, there is a sharp distinction between the 
$\la S^z_q S^z_{-q} \ra$ and $\la S^x_q S^x_{-q} \ra$ structure factors.
The former has singular wavevectors $2k_{F1}$, $2k_{F2}$, and 
$-k_{F2} - k_{F1}$, while the latter is lacking the wavevector $2k_{F2}$ 
since there is no spin-flip process across the second Fermi sea.
Our measurements are shown in 
Fig.~\ref{fig:SzSzq_SxSxq_J2_1p0_J3_m0p5_Sztot10}.
The wavevectors $2k_{F2}$ and $-k_{F2} - k_{F1}$ are too close to make
a more clear-cut distinction; nevertheless, the VMC reproduces
all details quite well.

In analogy with the SBM theory, we expect the $2k_{F1}$ and $2k_{F2}$ 
singularities to become stronger compared with the bare Gutzwiller
and the $-k_{F2} - k_{F1}$ to become weaker; this is roughly consistent
with what we see in the DMRG structure factors.
As discussed in Appendix~\ref{subapp:Gutzw_ferroFS2}, however, 
we do not have a complete description of such a partially polarized 
SBM phase that must incorporate ferromagnetic spin waves as well as the 
low energy SBM modes.  This is left for future work.  
We also mention that the closeness of the $2k_{F2}$ and 
$-k_{F2} - k_{F1}$ warns us that the system is near a commensuration 
point with $k_{F1} = 2\pi/5$ where it can be further unstable, 
which requires more study.

To summarize, by adding modest ferromagnetic $J_3 = -0.5$ we have 
realized the SBM state with fully polarized second Fermi sea,
confirming our intuition about the origin of the weak ferromagnetic
tendencies in the original model discussed in Sec.~\ref{subsec:qsing}.
A more thorough exploration of the phase diagram in the model with 
$J_3 = -0.5$ as well as in the original model in the partial FM region 
is clearly warranted to develop better understanding of such 
partially ferromagnetic states.

%%%%%%%%%%%%%%%%%%%%%%%%%%%%%%%%%%%%%%%%%%%%%%%%%%%%%%%%%%%%%%%%%%%%%
%%%%%%%%%%%%%%%%%%%%%%%%%%%%%%%%%%%%%%%%%%%%%%%%%%%%%%%%%%%%%%%%%%%%%
%%%%%%%%%%%%%%%%%%%%%%%%%%%%%%%%%%%%%%%%%%%%%%%%%%%%%%%%%%%%%%%%%%%%%
\section{Conclusions and future directions}
\label{sec:concl}

We have summarized the main results and presented much discussion
particularly in Sec.~\ref{sec:intro}.
Perhaps one point we would like to reiterate is the remarkable 
coincidence between the sign structure present in the DMRG wavefunctions
for the spin model SBM phase and the sign structure in the spin sector 
of free fermions on the ladder (e.g., metallic electrons).
This sign structure is encoded in the singular wavevectors
(``Bose surfaces''), and indeed the Gutzwiller-projected wavefunction 
with just one variational parameter is sufficient to reproduce the 
locations of all the singularities throughout the observed SBM phase.

We conclude by mentioning some standing questions and future directions.
First, in the ring model, we have focused on the Spin Bose-Metal and 
dealt with other phases only as needed to sketch the rich phase diagram.
For example, we have not studied carefully the VBS-2 region, 
which might harbor additional phases.
We have not studied adequately (numerically or analytically) the 
spin-gap region between the Bethe-chain and SBM phases in the 
$J_3 = 0.5$ model, Fig.~\ref{fig:qsing_J3_0p5}.
One question to ask here is whether there is a generic instability 
when we start populating the second Fermi sea, or whether we can go 
directly from the Bethe-chain phase to the SBM.  More generally, 
we have not studied various phase transitions in the system.

Next, while we understand the long-wavelength SBM theory with its 
single Luttinger parameter $g$, we have found that the Gutzwiller 
wavefunctions represent only the special case $g=1$ and cannot capture 
the general situation $g < 1$.  It would clearly be desirable to 
construct spin-singlet wavefunctions appropriate for the general case.
Even thinking about the Gutzwiller wavefunctions, it could be interesting
to understand the observed $g=1$ analytically and ask if they may be
exact ground states of some Hamiltonians 
(in the spirit of the Haldane-Shastry model\cite{Haldane88,Shastry88}).

On a separate front, we have encountered an interesting possibility 
of the Spin Bose-Metal with partial ferromagnetism occurring in one of
the subbands, but more work is needed to fully understand the numerical 
observations and develop analytical theory.  
Even without a spontaneous moment in the ground state, our observations 
suggest that in the regime between the Bethe-chain and VBS-3 phases the 
second band is narrow in energy.
Some ferromagnetic instability or possibility of spin-incoherent regime 
in one of the bands can lead to anomalous transport properties in such a 
quantum wire, a topic of much current interest.
\cite{Meyer08, Matveev_prl, Matveev_prb, Fiete}

Looking into future, it will be interesting to consider electron 
Hubbard-like models on the two-leg triangular strip and look for 
possible SBM phase.
The Hubbard model has been studied in a number of works,
\cite{Daul, Louis, Capello, Japaridze}
but the focus has been mainly on the conventional insulating phases 
such as the Bethe-chain and VBS-2 states.  This is appropriate in the 
strong Mott insulator limit, $t_1, t_2 \ll U$, where the effective 
spin model is the $J_1 - J_2$ model with $J_1 = 2 t_1^2/U$, 
$J_2 = 2 t_2^2 /U$.  
However, at intermediate coupling just on the insulator side, 
one needs to include multiple-spin exchanges, and the leading new term 
is the ring exchange with $K_{\rm ring} = 20 t_1^2 t_2^2 /U^3$. 
As we have learned, this ring term stabilizes the SBM phase, 
so revisiting the Hubbard model with the insights gained here
is promising.
In Sec.~\ref{subsec:SBMelectron}, we approached the SBM by starting
with a metallic two-band electron system (``C2S2'') and gapping out
only the overall charge mode ``$\rho+$''.  Then it is natural to 
look for the SBM near an extended such C2S2 metallic phase,
and we may need to consider electron models with further-neighbor
repulsion to open wider windows of such phases.

Last but not least, we would like to advance the program of ladder 
studies closer to 2D.  It is prudent to focus on the spin model with 
ring exchanges.  On the exact numerics front, 4 to 6 legs is probably 
at the limit of the DMRG capabilities.
The VMC approach should still be able to capture the critical surfaces
if they are present, since they are dictated by short-distance physics;
on the other hand, the bare Gutzwiller will likely fail even more
in reproducing correct long-distance behavior.
We do not know how far the present Bosonization approach can sensibly 
hold going to more legs.  
These are challenging but worthwhile endeavors given the experimental
importance of understanding weak Mott insulators.

%%%%%%%%%%%%%%%%%%%%%%%%%%%%%%%%%%%%%%%%%%%%%%%%%%%%%%%%%%%%%%%%%%%%%
%%%%%%%%%%%%%%%%%%%%%%%%%%%%%%%%%%%%%%%%%%%%%%%%%%%%%%%%%%%%%%%%%%%%%
%%%%%%%%%%%%%%%%%%%%%%%%%%%%%%%%%%%%%%%%%%%%%%%%%%%%%%%%%%%%%%%%%%%%%
\acknowledgments 

We would like to thank L. Balents, H.-H. Lai, T. Senthil, and S. Trebst
for useful discussions.
This work was supported by DOE grant DE-FG02-06ER46305 (DNS), 
the National Science Foundation through grants 
DMR-0605696 (DNS) and DMR-0529399 (MPAF), 
and the A.~P.~Sloan Foundation (OIM).
DNS also thanks the KITP for support through NSF grant PHY05-51164.

%%%%%%%%%%%%%%%%%%%%%%%%%%%%%%%%%%%%%%%%%%%%%%%%%%%%%%%%%%%%%%%%%%%%%
%%%%%%%%%%%%%%%%%%%%%%%%%%%%%%%%%%%%%%%%%%%%%%%%%%%%%%%%%%%%%%%%%%%%%
%%%%%%%%%%%%%%%%%%%%%%%%%%%%%%%%%%%%%%%%%%%%%%%%%%%%%%%%%%%%%%%%%%%%%
\appendix

\section{Observables in the SBM phase}
\label{app:SBMprops}

We have defined spin $\vec{S}(x)$, bond energy ${\cal B}(x)$, and 
chirality $\chi(x)$ observables in Sec.~\ref{subsec:SBMfp}
[cf.~Eqs.~(\ref{B_def}-\ref{chi_def})].
Here we find detailed Bosonized forms by a systematic construction 
of observables in the SBM (and will find more observables on the way).

In the gauge theory treatment Sec.~\ref{subsec:SBMspinon}, 
we consider gauge-invariant objects constructed from the spinon fields; 
in the interacting electron picture of Sec.~\ref{subsec:SBMelectron}, 
these are operators that do not change the total charge.

We begin with fermion bilinears and first consider the ones composed of a 
particle and a hole moving in opposite directions.
Such bilinears are expected to be enhanced by gauge fluctuations
since parallel gauge currents experience Amperean attraction.
We organize these bilinears as follows:
\begin{eqnarray}
\vec{S}_{2k_{Fa}} &\equiv&
\frac{1}{2} f^\dagger_{La\alpha} \vec{\sigma}_{\alpha\beta} f_{Ra\beta} 
~,\\
\epsilon_{2k_{Fa}} &\equiv& 
\frac{1}{2} f^\dagger_{La\alpha} f_{Ra\alpha}
~,\\
\vec{S}_{\pi/2} &\equiv& 
\frac{1}{2} 
f^\dagger_{R1\alpha} \vec{\sigma}_{\alpha\beta} f_{L2\beta} + 
\frac{1}{2}
f^\dagger_{R2\alpha} \vec{\sigma}_{\alpha\beta} f_{L1\beta}
~,\\
\epsilon_{\pi/2} &\equiv&
\frac{1}{2} f^\dagger_{R1\alpha} f_{L2\alpha} + 
\frac{1}{2} f^\dagger_{R2\alpha} f_{L1\alpha}
~,\\
\vec{\delta}_{\pi/2} &\equiv& 
\frac{1}{2}
f^\dagger_{R1\alpha} \vec{\sigma}_{\alpha\beta} f_{L2\beta} - 
\frac{1}{2}
f^\dagger_{R2\alpha} \vec{\sigma}_{\alpha\beta} f_{L1\beta}
~,\\
\chi_{\pi/2} &\equiv&
\frac{1}{2} f^\dagger_{R1\alpha} f_{L2\alpha} - 
\frac{1}{2} f^\dagger_{R2\alpha} f_{L1\alpha} 
~,
\end{eqnarray}
with $\vec{S}_{-Q} = \vec{S}_Q^\dagger$, etc.
The microscopic spin operator expanded in terms of the continuum fermion
fields readily gives the listed $\vec{S}_{Q}$.

The bond energy can be approximated as the spinon hopping energy,
\begin{eqnarray}
{\cal B}(x) \sim -t[f_\alpha^\dagger(x) f_\alpha(x+1) + \Hc]
\label{B_via_f}
\end{eqnarray}
(recall that we work in the gauge with zero spatial vector potential).
Expansion in terms of the continuum fields gives, up to a real factor,
\begin{eqnarray}
{\cal B}_Q \sim e^{i Q/2} \epsilon_Q ~.
\label{BQ}
\end{eqnarray}
Such connection between ${\cal B}_Q$ and $\epsilon_Q$ is understood
below for all $Q \neq \pi$.  The objects $\epsilon_Q$ are convenient 
because of their simpler transformation properties under 
lattice inversion, $\epsilon_Q \leftrightarrow \epsilon_{-Q}$.

The physical meaning of the operators $\vec{\delta}_{\pi/2}$ and 
$\chi_{\pi/2}$ can be established on symmetry grounds.
Thus, $\chi$ is the spin chirality defined in Eq.~(\ref{chi_def}).
The expression in terms of bilinears can be also found directly by 
considering the circulation of the gauge charge current around the 
$[x-1, x, x+1]$ loop,
\begin{eqnarray}
\chi(x) &\sim& 
\circlearrowleft \hspace{-1.4em} \sum
i t_{rr'} [f_\alpha^\dagger(r) f_\alpha(r') - \Hc] ~.
\label{chi_via_f}
\end{eqnarray}
This is familiar in slave particle treatments:\cite{LeeNagaosaWen} 
the circulation produces internal gauge flux whose physical meaning is 
the spin chirality.  On the other hand, $\vec{\delta}$ is related to the 
following microscopic operator;
\begin{eqnarray}
\vec{D}(x) = \vec{S}(x) \times \vec{S}(x+1) ~, 
\quad \vec{D}_{\pi/2} = e^{i\pi/4} \vec{\delta}_{\pi/2} ~.
\end{eqnarray}
At $Q = \pi/2$, this enters on par with $\vec{S}$, ${\cal B}$, 
and $\chi$.

The bosonized expressions at the $2k_{Fa}$ are:
\begin{eqnarray}
S^x_{2k_{Fa}} &=& -i \eta_{a\up} \eta_{a\dn} 
e^{i\theta_{\rho+}} e^{\pm i\theta_{\rho-}}
\sin(\sqrt{2} \varphi_{a\sigma}) ~, \\
S^y_{2k_{Fa}} &=& -i \eta_{a\up} \eta_{a\dn}
e^{i\theta_{\rho+}} e^{\pm i\theta_{\rho-}}
\cos(\sqrt{2} \varphi_{a\sigma}) ~, \\
S^z_{2k_{Fa}} &=& - e^{i\theta_{\rho+}} e^{\pm i\theta_{\rho-}} 
\sin(\sqrt{2} \theta_{a\sigma}) ~, \\
\epsilon_{2k_{Fa}} &=& i e^{i\theta_{\rho+}} e^{\pm i \theta_{\rho-}} 
\cos(\sqrt{2} \theta_{a\sigma}) ~,
\label{epsilon_2kF}
\end{eqnarray}
where the upper or lower sign in the exponent corresponds to 
$a=1$ or $2$.  
The pinned value $\theta_{\rho+}$, which is determined by minimizing
Eq.~(\ref{H8bosonized}), is left general at this stage.
It is not so important for the qualitative behavior at the 
$2k_{Fa}$ and $\pi/2$, but is crucial at a wavevector $\pi$ later.

For each $a$, the $(\vec{S}, {\cal B})_{2k_{Fa}}$ structure is similar to
that in a single Bethe chain except for the $\theta_\rho$ exponentials.
As in the Bethe chain, we expect the spin and VBS correlations 
to be closely related -- in particular, they decay with the same
power law.  The corresponding scaling dimension in the fixed-point
theory Eq.~(\ref{LSBM0}) is
\begin{equation}
\Delta[ \vec{S}_{2k_{Fa}} ] = \Delta[ {\cal B}_{2k_{Fa}} ]
= \frac{1}{2} + \frac{g}{4} ~.
\label{Delta_2kF}
\end{equation}

The bosonized expressions at the $\pi/2$ are:
\begin{eqnarray}
S^x_{\pi/2} &=& e^{-i\theta_{\rho+}} \Big[
-i \eta_{1\up} \eta_{2\dn} e^{-i \theta_{\sigma-}}
\sin(\varphi_{\rho-} + \varphi_{\sigma+}) \\
&&~~~~~~~~~
-i \eta_{1\dn} \eta_{2\up} e^{i \theta_{\sigma-}}
\sin(\varphi_{\rho-} - \varphi_{\sigma+}) 
\Big] ~, \\
S^y_{\pi/2} &=& e^{-i\theta_{\rho+}} \Big[
-i \eta_{1\up} \eta_{2\dn} e^{-i \theta_{\sigma-}}
\cos(\varphi_{\rho-} + \varphi_{\sigma+}) \\
&&~~~~~~~~~
+i \eta_{1\dn} \eta_{2\up} e^{i \theta_{\sigma-}}
\cos(\varphi_{\rho-} - \varphi_{\sigma+})
\Big] ~, \\
S^z_{\pi/2} &=& e^{-i\theta_{\rho+}} \Big[ 
-i \eta_{1\up} \eta_{2\up} e^{-i \theta_{\sigma+}}
\sin(\varphi_{\rho-} + \varphi_{\sigma-}) \\
&&~~~~~~~~~
+i \eta_{1\dn} \eta_{2\dn} e^{i \theta_{\sigma+}}
\sin(\varphi_{\rho-} - \varphi_{\sigma-})
\Big] ~, \\
\epsilon_{\pi/2} &=& e^{-i\theta_{\rho+}} \Big[
-i \eta_{1\up} \eta_{2\up} e^{-i\theta_{\sigma+}} 
\sin(\varphi_{\rho-} + \varphi_{\sigma-}) \\
&&~~~~~~~~~
-i \eta_{1\dn} \eta_{2\dn} e^{i\theta_{\sigma+}} 
\sin(\varphi_{\rho-} - \varphi_{\sigma-}) 
\Big] ~, 
\label{epsilon_halfpi} \\
\chi_{\pi/2} &=& e^{-i\theta_{\rho+}} \Big[
- \eta_{1\up} \eta_{2\up} e^{-i\theta_{\sigma+}} 
\cos(\varphi_{\rho-} + \varphi_{\sigma-}) ~~~~~~~~\\
&&~~~~~~~~~
- \eta_{1\dn} \eta_{2\dn} e^{i\theta_{\sigma+}} 
\cos(\varphi_{\rho-} - \varphi_{\sigma-}) 
\Big] ~. ~~~~~~~~
\label{chi_halfpi}
\end{eqnarray}
Expressions for $\vec{\delta}_{\pi/2}$ can be obtained from those for 
$\vec{S}_{\pi/2}$ essentially by interchanging sines and cosines.  
As before, ${\cal B}_{\pi/2}$ is given by Eq.~(\ref{BQ}).
The above details are needed particularly when we discuss phases
arising as instabilities of the SBM, 
Secs.~\ref{subsec:g.gt.1}-\ref{subsec:commens_other},
while in the SBM we immediately see that all scaling dimensions
are equal:
\begin{equation}
\Delta[ \vec{S}_{\pi/2} ] = \Delta[ {\cal B}_{\pi/2} ]
= \Delta[ \vec{D}_{\pi/2} ] = \Delta[ \chi_{\pi/2} ]
= \frac{1}{2} + \frac{1}{4g} ~.
\label{Delta_halfpi}
\end{equation}

This completes the ``enhanced'' bilinears.
We also mention, without giving detailed expressions,
``non-enhanced'' bilinears at wavevectors 
$Q = \pm (k_{F2} - k_{F1})$.  Their scaling dimension is
\begin{equation}
\Delta[ \vec{S}_Q ] = \Delta[ {\cal B}_Q ]
= \Delta[ \vec{D}_Q ] = \Delta[ \chi_Q ]
= \frac{1}{2} + \frac{1}{4g} + \frac{g}{4} ~,
\label{Delta_kF2mkF1}
\end{equation}
which is always larger than the spinon mean field value of $1$.

Finally, we have bilinears carrying zero momentum --  
essentially $J_{Paa}, \vec{J}_{Paa}$ from Eq.~(\ref{JP}).
These give conserved densities and currents and have scaling
dimension $1$.
We specifically mention examples leading to 
Eqs.~(\ref{SzQ0}-\ref{chiQ0}):
\begin{eqnarray}
S^z_{Q=0} \sim J^z_{R11} + J^z_{L11} + J^z_{R22} + J^z_{L22} = 
\frac{1}{\pi} \partial_x \theta_{\sigma+} ~, \\
\epsilon_{Q=0} \sim J_{R11} + J_{L11} - J_{R22} - J_{L22} = 
\frac{2}{\pi} \partial_x \theta_{\rho-} ~, \\
\chi_{Q=0} \sim J_{R11} - J_{L11} - J_{R22} + J_{L22} = 
\frac{2}{\pi} \partial_x \varphi_{\rho-} ~.
\end{eqnarray}
(One way we can make the identifications in the last two lines is by
using physical symmetry arguments.)

So far, we have only considered fermion bilinears.
Since the theory is strongly coupled, we should also study
contributions with more fermion fields.
We now include four-fermion terms focusing on the spin, bond energy, 
and chirality operators that are measured in the DMRG.
First, there appears a new wavevector $4 k_{F1} = -4 k_{F2}$ 
in the bond energy, via,
\begin{eqnarray}
\label{e4kF}
\epsilon_{4k_{F1}}:&& 
f_{L1\up}^\dagger f_{L1\dn}^\dagger f_{R1\up} f_{R1\dn} 
\sim  e^{i 2 \theta_{\rho+}} e^{i 2\theta_{\rho-}} ~, \\
&&
f_{R2\up}^\dagger f_{R2\dn}^\dagger f_{L2\up} f_{L2\dn} 
\sim  e^{-i 2 \theta_{\rho+}} e^{i 2\theta_{\rho-}} ~.
\end{eqnarray}
The two contributions come with independent numerical factors
and can be also generated as $(\epsilon_{2k_{F1}})^2$ and 
$(\epsilon_{-2k_{F2}})^2$.  
Once the $\theta_{\rho+}$ is pinned, there is only one qualitatively
distinct contribution and the scaling dimension is
\begin{eqnarray}
\Delta[ {\cal B}_{4k_{F1}} ] = g ~.
\end{eqnarray}
Note that for sufficiently small $g < 2/3$, the power law decay 
is slower than that of the bilinears ${\cal B}_{2k_{Fa}}$. 
There is no comparable $4k_{F1}$ contribution to the spin operator.

Four-fermion terms bring out another important wavevector, $Q = \pi$.
We list independent dominant such contributions to 
$\vec{S}_\pi$, ${\cal B}_\pi$, and $\chi_\pi$:
\begin{eqnarray}
S^z_\pi :&&
\sin(2\theta_{\sigma+}) \sin(2\theta_{\rho+}) ~, \;\;
\sin(2\theta_{\sigma-}) \sin(2\theta_{\rho+}) ~; ~~~~~ \\
{\cal B}_\pi :&&
[\cos(2\theta_{\sigma+}) + \cos(2\theta_{\sigma-})] \sin(2\theta_{\rho+})
~, 
\label{B_pi} \\
&& [\cos(2\theta_{\sigma+}) + \hat\Gamma\cos(2\varphi_{\sigma-})] 
\sin(2\theta_{\rho+}) ~, \\
&& \hat\Gamma \cos(2\varphi_{\rho-}) \sin(2\theta_{\rho+}) ~; \\
\chi_\pi :&&
\hat\Gamma \sin(2\varphi_{\rho-}) \sin(2\theta_{\rho+}) ~.
\end{eqnarray}
These can be generated by combining the previously exhibited bilinears
as follows. 
$S^z_\pi$: 
$S^z_{2k_{F1}} \epsilon_{2k_{F2}} \pm 
 S^z_{2k_{F2}} \epsilon_{2k_{F1}} + \Hc$;
${\cal B}_\pi$: 
$i \epsilon_{2k_{F1}} \epsilon_{2k_{F2}} + \Hc$,
$i (\epsilon_{\pi/2}^2 - \chi_{\pi/2}^2) + \Hc$,
$i (\epsilon_{\pi/2}^2 + \chi_{\pi/2}^2) + \Hc$;
$\chi_\pi$: 
$\chi_{\pi/2} \epsilon_{\pi/2} + \Hc$
The scaling dimensions are
\begin{eqnarray}
\Delta[ \vec{S}_\pi ] &=& \Delta[ {\cal B}_\pi ] = 1 ~, \\
\Delta[ \chi_\pi ] &=& 1/g ~.
\end{eqnarray}
The above observables are present if $\sin(2\theta_{\rho+}) \neq 0$,
e.g., if the $\theta_{\rho+}$ is pinned as in Eq.~(\ref{u8pos}),
which we argued is natural when the spin model is describing a
Mott insulator phase of a repulsive electron model.
On the other hand, the above contributions would vanish if the pinning 
potential Eq.~(\ref{H8bosonized}) had $v_8 < 0$.
Some other physical observables containing $\cos(2\theta_{\rho+})$ and
having different symmetry properties would be present instead.
We do not write these out since both the DMRG and the trial 
wavefunctions have signatures in the spin, VBS, and chirality
at the wavevector $\pi$, suggesting that the pinning Eq.~(\ref{u8pos})
is realized.
We have identified several more new observables at $\pi$ containing 
$\sin(2\theta_{\rho+})$; we do not spell these out here since our
primary focus is to understand features in the numerics measuring the 
familiar $\vec{S}$, ${\cal B}$, and $\chi$.

We finally mention that four-fermion terms produce still more 
wavevectors, $\pm (3k_{F1} + k_{F2}) = \mp (3k_{F2} + k_{F1})$;
for example, $3k_{F1} + k_{F2}$ can be obtained by combining $2k_{F1}$
and $-\pi/2$.
The scaling dimensions are the same as at $\pm (k_{F2} - k_{F1})$,
Eq.~(\ref{Delta_kF2mkF1}).

For completeness, we have also checked six-fermion and eight-fermion
terms.  The only new wavevectors where the scaling dimension can be
smaller than $2$ are $6k_{Fa} = 2k_{Fa} + 4k_{Fa}$ 
(scaling dimension $1/2 + 9g/4$), and 
$8k_{Fa} = 4k_{Fa} + 4k_{Fa}$ (scaling dimension $4g$).
However, one needs small $g$ for these to become visible and in
any case they always have faster power law decay than at $2k_{Fa}$
and $4k_{Fa}$.
Finally, entries listed as ``subd.'' in Table~\ref{tab:SBMprops} can be
constructed, e.g., as $\chi_Q \sim \chi_0 \epsilon_Q$, which has scaling
dimensions $1 + \Delta[\epsilon_Q]$.

Table~\ref{tab:SBMprops} summarizes our results for the correlations
in the Spin Bose-Metal phase.  In words, we expect dominant spin
and VBS correlations at the wavevectors $\pm 2k_{F1}$, $\pm 2k_{F2}$
decaying as $1/x^{1+g/2}$ and at the wavevectors $\pm \pi/2$
decaying as $1/x^{1+1/(2g)}$.  The former decay is more slow since
stability of the phase requires $g<1$.  
Note that the wavevectors $2k_{F1}$ and $2k_{F2}$ are located 
symmetrically around $\pi/2$.
We also expect a $1/x^2$ power law at the wavevectors $0$ and $\pi$.
Next, at the wavevectors $\pm (k_{F2} - k_{F1})$ and 
$\pm (3 k_{F1} + k_{F2})$, which are also located symmetrically
around $\pm \pi/2$, we expect a still faster power law 
$1/x^{1 + 1/(2g) + g/2}$.  
Furthermore, the bond energy shows a power law $1/x^{2g}$ at 
$\pm 4k_{F1}$.
The spin chirality has similar signatures to the above at 
$\pm \pi/2$, $0$, $\pm (k_{F2} - k_{F1})$, and $\pm (3 k_{F1} + k_{F2})$,
but decays as $1/x^{2/g}$ at $\pi$.
These are the simplest observables that can be used to identify the 
SBM phase in a given system.
Figure~\ref{fig:vmc_corrs} shows measurements in the Gutzwiller 
wavefunction projecting two Fermi seas and nicely illustrates all
singular wavevectors, while it appears that such wavefunctions realize a
special case with $g=1$.
We also remark that in the general SBM the presence of the marginally 
irrelevant interactions Eq.~(\ref{vJRvJL}) will lead to logarithmic 
corrections in correlations.
\cite{Singh_logcorrection, Affleck_logcorrection}

We conclude by describing our treatment of the Klein factors
(see, e.g., Ref.~\onlinecite{Fjaerestad02} for more details).
We need this when determining ``order parameters'' of various phases
obtained as instabilities of the SBM, 
Secs.~\ref{subsec:g.gt.1}-\ref{subsec:commens_other}.
The operator $\hat\Gamma$ from Eq.~(\ref{Gamma}) has eigenvalues $\pm 1$.
For concreteness, we work with the eigenstate corresponding to $+1$:
$\hat\Gamma |+\ra = |+\ra$.  We then find the following relation
\begin{eqnarray}
\la +| \eta_{1\up} \eta_{2\up} |+\ra =
\la +| \eta_{1\dn} \eta_{2\dn} |+\ra = \textrm{pure imaginary} ~.
\end{eqnarray}
This is useful when discussing observables at the $\pm \pi/2$
wavevectors; for example,
\begin{eqnarray}
\label{epsilon_halfpi_fixGamma}
\epsilon_{\pi/2} &\!\!=\!\!& 
- e^{-i\theta_{\rho+}} \la +|\eta_{1\up} \eta_{2\up} |+\ra
\Big[
\cos(\varphi_{\rho-}) \sin(\theta_{\sigma+}) \sin(\varphi_{\sigma-})
\nonumber \\
&&~~~~~~~~~~~~~~~~~ 
+ i \sin(\varphi_{\rho-}) \cos(\theta_{\sigma+}) \cos(\varphi_{\sigma-}) 
\Big] ~, \\
\label{chi_halfpi_fixGamma}
\chi_{\pi/2} &\!\!=\!\!& 
-e^{-i\theta_{\rho+}} \la +|\eta_{1\up} \eta_{2\up} |+\ra
\Big[ 
\cos(\varphi_{\rho-}) \cos(\theta_{\sigma+}) \cos(\varphi_{\sigma-}) 
\nonumber \\
&&~~~~~~~~~~~~~~~~~
+ i \sin(\varphi_{\rho-}) \sin(\theta_{\sigma+}) \sin(\varphi_{\sigma-}) 
\Big] ~.
\end{eqnarray}

%%%%%%%%%%%%%%%%%%%%%%%%%%%%%%%%%%%%%%%%%%%%%%%%%%%%%%%%%%%%%%%%%%%%%
%%%%%%%%%%%%%%%%%%%%%%%%%%%%%%%%%%%%%%%%%%%%%%%%%%%%%%%%%%%%%%%%%%%%%
%%%%%%%%%%%%%%%%%%%%%%%%%%%%%%%%%%%%%%%%%%%%%%%%%%%%%%%%%%%%%%%%%%%%%

\section{Details of the wavefunctions}
\label{app:Gutzw}

%%%%%%%%%%%%%%%%%%%%%%%%%%%%%%%%%%%%%%%%%%%%%%%%%%%%%%%%%%%%%%%%%%%%%
%%%%%%%%%%%%%%%%%%%%%%%%%%%%%%%%%%%%%%%%%%%%%%%%%%%%%%%%%%%%%%%%%%%%%
\subsection{Gutzwiller projection of two Fermi seas}

It is convenient to view the spin wavefunction as that of hard-core 
bosons, where up/down spin corresponds to present/absent boson.
In the general spinon construction, we occupy 
$\{ k^\up_j, j=1, \dots, N_\up \}$ orbitals with spin up and 
$\{ k^\dn_j, j=1, \dots, N_\dn \}$ orbitals with spin down;
$N_\up + N_\dn = L$ is the size of the system.
After the Gutzwiller projection, the boson wavefunction is
\begin{eqnarray}
\Psi_{\rm bos}(\{ R_i, i=1, \dots, N_\up \}) = \det[e^{i k^\up_j R_i}]
\det[e^{i p_j R_i}] ~,
\end{eqnarray}
where the set $\{ p_j, j=1, \dots, N_\up \}$ is a complement to
$\{ -k_j^\dn, j=1, \dots, N_\dn \}$ in the Brilloin Zone (BZ).  
The momentum carried by this wavefunction is
$\sum_{j=1}^{N_\up} (k^\up_j + p_j) = 
\sum_{j=1}^{N_\up} k^\up_j + \sum_{j=1}^{N_\dn} k^\dn_j 
+ \sum_{q \in BZ} q$.  
In particular, we see that the wavefunction remains unchanged 
if we shift all occupied spinon momenta by the same integer multiple of 
$2\pi/L$.

\begin{figure}
\centerline{\includegraphics[width=2in]{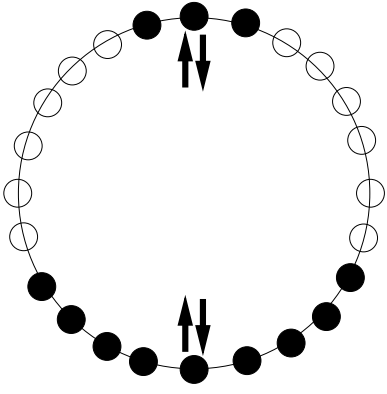}}
\caption{
View of the wavefunction constructed by filling $k$-states of spinons.  
Here momenta $k = 2 \pi n/L$, $n=0, 1, \dots, L-1$, form a closed circle.
Each filled dot is occupied by both spin up and spin down, producing 
spin singlet.  The projected wavefunction remains unchanged if all
momenta are shifted by the same amount.  Only the relative configuration 
matters, and here we show symmetric configuration of the two Fermi seas 
separated by $L/4$ unoccupied $k$-states on either side.  This is our
``bare Gutzwiller'' wavefunction for the Spin Bose-Metal.
}
\label{fig:twoseas}
\end{figure}

We now consider the spin-singlet case when 
$N_\up = N_\dn = L/2$, $\{ k^\up \} = \{ k^\dn \}$.
Here $L$ is even and all momenta are integer multiples of $2\pi/L$,
so $\sum_{q \in BZ} q = \pi$.
For convenience, we assume that $L$ is a multiple of $4$.
Fig.~\ref{fig:twoseas} illustrates filled $k$-points for the 
band in Fig.~\ref{fig:fbands}.  We have two Fermi seas of volume 
$N_1$ and $N_2$ in the symmetric configuration, i.e.,\ separated by 
$L/4$ unoccupied orbitals on each side around the Brilloin Zone.
Relating to the band Fig.~\ref{fig:fbands}, the larger $N_1$ 
corresponds to occupied $k$-points centered around $0$, while $N_2$ 
corresponds to points around $\pi$.  As already noted, 
a solid shift of the occupied states leaves the wavefunction unchanged.
We can then specify such symmetric state as $(N_1, N_2)$, which requires 
only one parameter since $N_1 + N_2 = N_\up = N_\dn = L/2$.
We can readily verify that such $(even, even)$ states carry 
momentum $0$ and are even under site inversion operation
while $(odd, odd)$ states carry momentum $\pi$ and are odd
under inversion.

\begin{figure}
\centerline{\includegraphics[width=\columnwidth]{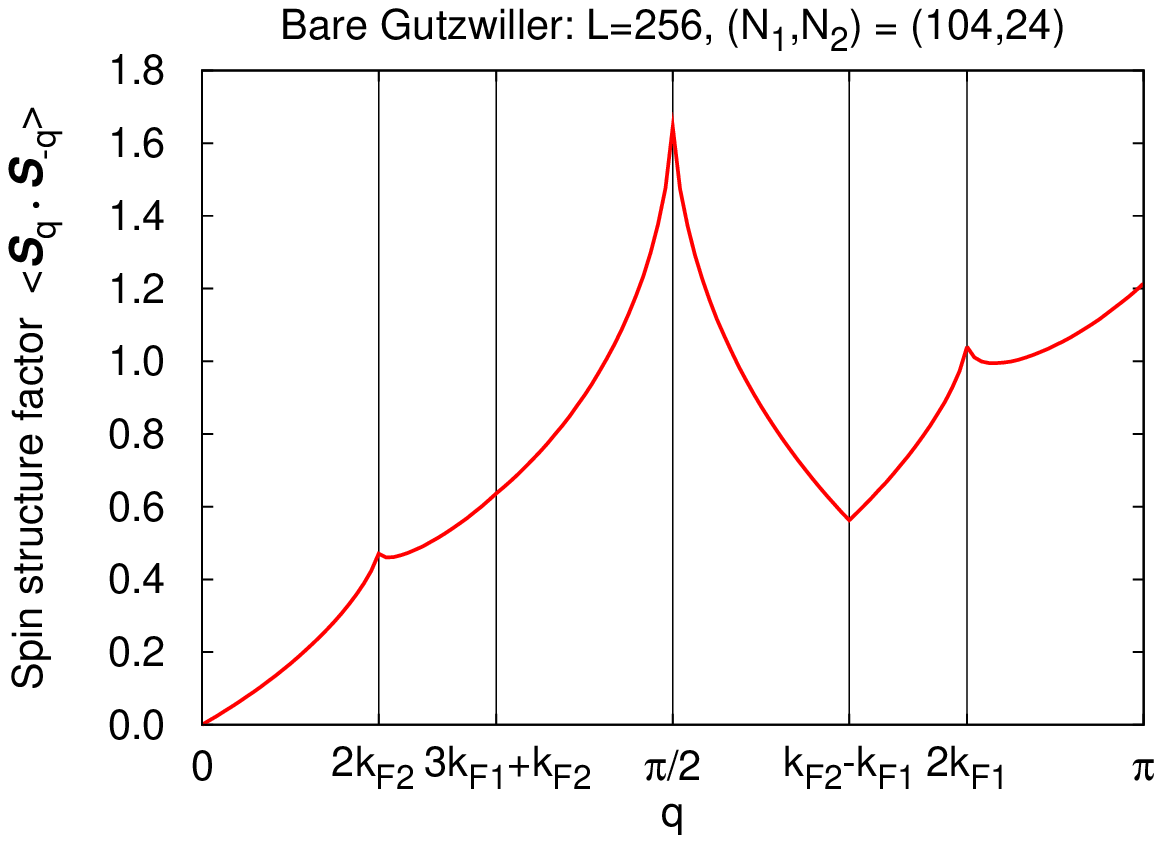}}
\vskip -2mm
\centerline{\includegraphics[width=\columnwidth]{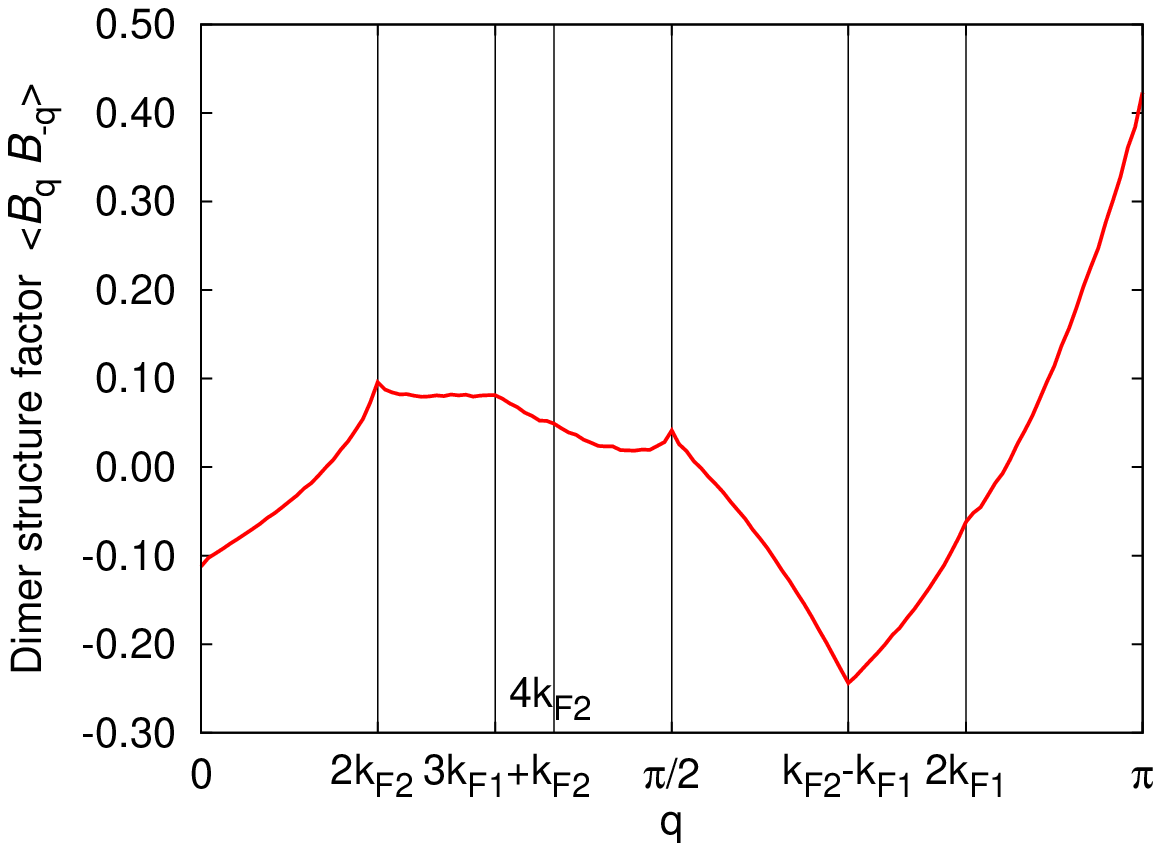}}
\vskip -2mm
\centerline{\includegraphics[width=\columnwidth]{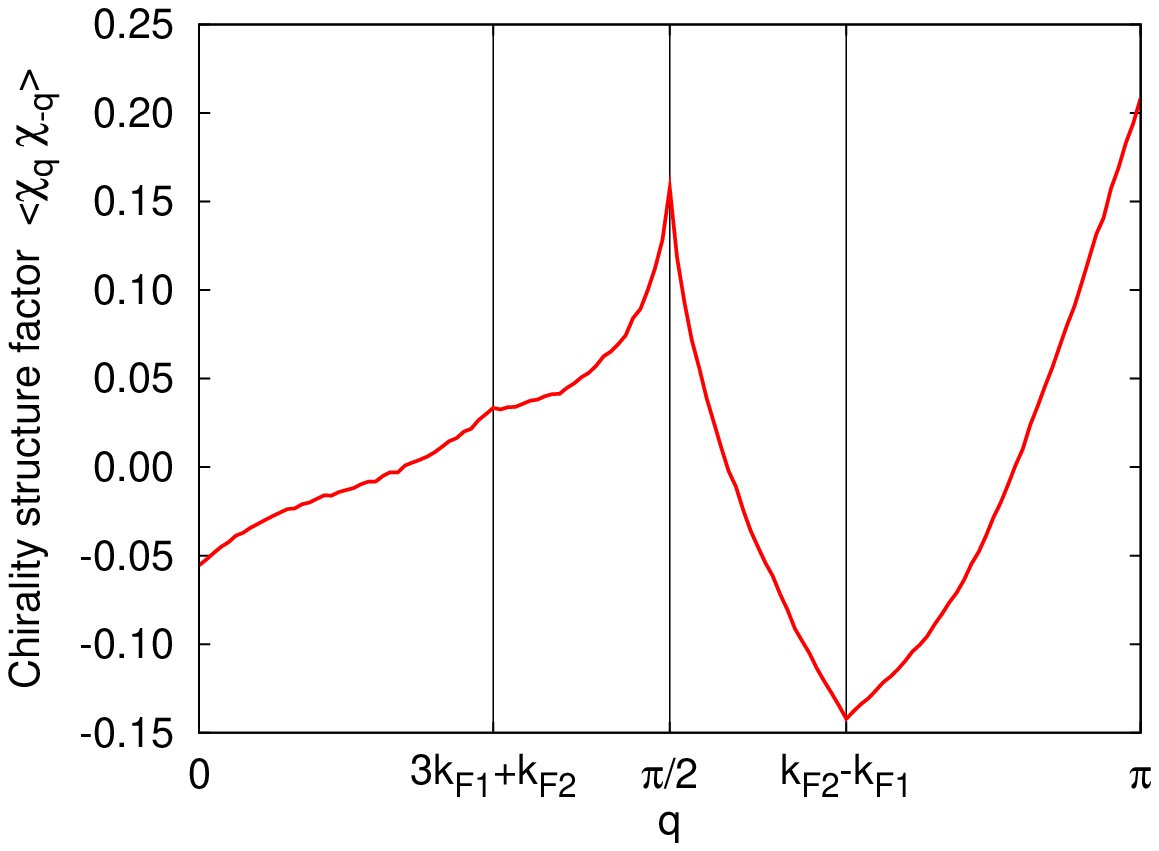}}
\vskip -2mm
\caption{
Spin, bond energy, and chirality structure factors in the bare
Gutzwiller wavefunction with two Fermi seas $(N_1, N_2) = (104, 24)$ 
on the 1D chain of length $L=256$. 
The expected singular wavevectors are marked by vertical lines
(here $k_{F1}$ and $k_{F2}$ are defined as in Fig.~\ref{fig:fbands} 
and all indicated wavevectors are modulo $2\pi$).
The structure factors are symmetric with respect to $q \to -q$,
and we only show $0 \leq q \leq \pi$.
The character of the singularities is consistent with the special
case $g=1$ in the SBM theory of Sec.~\ref{sec:SBMtheory}.
}
\label{fig:vmc_corrs}
\end{figure}

The relative wavevectors connecting the Fermi points are 
gauge-independent and are observed in various quantities, 
see Fig.~\ref{fig:vmc_corrs}.  Specifically, using 
variational Monte Carlo,\cite{Ceperley77, Gros89} we measure the spin 
structure factor and see dominant singularities at the wavevectors 
$\pm 2k_{F1}$, $\pm 2k_{F2}$, and $\pm (k_{F1} + k_{F2}) = \mp \pi/2$, 
which connect Fermi points with opposite group velocities.  
By studying sizes up to $L=512$ and performing scaling analysis at these 
wavevectors, the singularities appear to have the same power law.
This is consistent only with the special case $g=1$ in the SBM theory 
[cf.~Eqs.~(\ref{Delta_2kF}, \ref{Delta_halfpi}) and 
Table~\ref{tab:SBMprops}].
Our direct estimates of the scaling dimensions are also consistent
with the value $\Delta = 3/4$ expected in this case.
Turning to other less singular wavevectors, we clearly see V-shaped
($\sim |\delta q|$) features at $0$ and $\pi$ corresponding to scaling 
dimension $1$, which is expected generally.  
We also see $\pm (k_{F2} - k_{F1})$ with scaling dimension $1$,
which requires $g=1$.

We next consider VBS correlations and see all of the above wavevectors, 
but we can not quantify the singularities as accurately.
The VBS correlations also show singularities at
$\pm (3 k_{F1} + k_{F2})$ and $\pm 4k_{F2} = \mp 4k_{F1}$
(the former is also expected in the spin structure factor but is 
not visible there, probably due to amplitude effect, while
the $\pm 4k_{F2}$ is expected only in the bond energy).

Finally, we measure spin chirality correlations and see dominant 
singularity at $\pm \pi/2$ consistent with $\Delta = 3/4$.
We also see singularities at wavevectors 
$0$, $\pi$, $\pm (k_{F2} - k_{F1})$, and $\pm (3 k_{F1} + k_{F2})$
consistent with $\Delta=1$, again as expected in the special case $g=1$.

The above appears to hold for a range of relative populations of the 
Fermi seas (away from the limiting situations of a single or two equal 
Fermi seas).
We are then led to conjecture that such spin-singlet wavefunctions 
with two Fermi seas have correlations given by the SBM theory 
of Sec.~\ref{sec:SBMtheory} and Appendix~\ref{app:SBMprops} with $g=1$.
This conjecture is natural since in the theory the parameter $g$ 
depends on the ratio of the two Fermi velocities, cf.~Eq.~(\ref{g0}),
while the wavefunction knows only about the occupied/unoccupied states
and does not contain the band energy parameters.
We leave proving this conjecture analytically as an open problem.

Given the preceding discussion, it appears that such bare Gutzwiller 
wavefunctions cannot capture fully the properties of the generic
Spin Bose-Metal as described by the theory of Sec.~\ref{sec:SBMtheory}
with general $g<1$.
It is possible that they are appropriate wavefunctions for some critical 
end-points of the SBM phase where the parameter $g=1$, e.g., 
for the transitions out of the SBM discussed in Sec.~\ref{subsec:g.gt.1}.
While the energetics study with the bare Gutzwiller wavefunctions
gives us first indications for the SBM phase, 
it is desirable to have more accurate trial states.
This is what we turn to next, although only with limited success.

%%%%%%%%%%%%%%%%%%%%%%%%%%%%%%%%%%%%%%%%%%%%%%%%%%%%%%%%%%%%%%%%%%%%%%%
%%%%%%%%%%%%%%%%%%%%%%%%%%%%%%%%%%%%%%%%%%%%%%%%%%%%%%%%%%%%%%%%%%%%%%%
\subsection{SU(2)-invariant improvement of the Gutzwiller wavefunctions}
\label{subapp:gaplessSC}

To allow more variational freedom, we consider mean field with both 
spinon hopping $\xi(k)$, Eq.~(\ref{xi}),
and spinon pairing in the singlet channel with real gap function
$\Delta(k)$ (this way, the wavefunction remains spin rotation and time
reversal invariant).  Generic $\Delta(k)$ would open up gaps, 
while we want the wavefunction to be critical.
One way to maintain gaplessness is to require $\Delta(k)$ to vanish
at the Fermi points.  This can be achieved, for example, by taking
\begin{equation}
\Delta(k) = f(k) \xi(k) ~
\end{equation}
with a smooth $f(k)$.
We have tried several simple functions $f(k)$, e.g., 
expanding in harmonics
\begin{equation}
f(k) = \sum_n f_n \cos(n k) ~,
\end{equation}
with few $f_n$ treated as variational parameters. 
Upon writing out the corresponding Gutzwiller wavefunction, 
one can see that the dispersion $\xi(k)$ enters only through its sign 
$\xi < 0$ or $\xi > 0$, so in the case with two Fermi seas like in 
Fig.~\ref{fig:twoseas} we can use the same label $(N_1, N_2)$ and
expect similar singular wavevectors encoded in the relative positions
of the ``Fermi points.''
We refer to such a state as ``improved Gutzwiller'' and can view
it as a ``gapless superconductor,'' although with caution because
of the non-intuitive effects of the projection.
For example, the SU(2) gauge structure of the projective construction 
implies that $f_n = A \delta_{n,0}$ gives the same state as the bare 
Gutzwiller independent of $A$.
In practice, we often fix $f_0$ and vary $f_1, f_2$.

For the zigzag ring model in the Spin Bose-Metal regime, such approach 
improves the trial energy by about 50-60\% compared to the 
difference between the bare Gutzwiller energy and the exact 
DMRG ground state energy.  The exponents of the power law correlations 
in the improved wavefunctions appear to remain unchanged from the 
bare case, although the numerical amplitudes are redistributed to 
resemble the DMRG correlations better, as can be seen in the examples
in Sec.~\ref{subsec:DMRG:SBM}, Figs.~\ref{fig:J2_0p0_J3_0p0} 
and \ref{fig:J2_3p2_J3_0p0}.
Thus, this approach is only partially successful since we can not 
produce the long-distance behavior expected in the generic SBM and 
tentatively seen in the DMRG.  
Still, the fact that we can significantly improve the trial energy 
while retaining the underlying gapless character gives us more 
confidence in the variational identification of the SBM phase.
  
We also mention that in the Bethe-chain regime where the bare Gutzwiller 
projects one Fermi sea, the ``gapless superconductor'' improvement with 
parameters $f_0, f_2$ works even better, bringing the trial energy 
much closer to the exact DMRG value and better reproducing short-scale 
features in the spin correlations, see Fig.~\ref{fig:J2_m1p0_J3_0p0}.
Such good trial states for the competing Bethe-chain phase give our VMC 
more accuracy in determining where the SBM phase wins 
energetically and more confidence interpreting the DMRG results.

%%%%%%%%%%%%%%%%%%%%%%%%%%%%%%%%%%%%%%%%%%%%%%%%%%%%%%%%%%%%%%%%%%%%%%%
%%%%%%%%%%%%%%%%%%%%%%%%%%%%%%%%%%%%%%%%%%%%%%%%%%%%%%%%%%%%%%%%%%%%%%%
\subsection{States with fully polarized second Fermi sea}
\label{subapp:Gutzw_ferroFS2}

Motivated by the possibility of partial ferromagnetism in some
regimes discovered in the DMRG study of the ring model
(see in particular Sec.~\ref{sec:ferroJ3}),
we have also considered Gutzwiller projection of states with 
unpolarized large Fermi sea and fully polarized small Fermi sea.
The spin correlations here can be understood using a naive 
Bosonization treatment starting with such spinon mean field state
and following the same procedure as for the unpolarized Spin Bose-Metal
in Sec.~\ref{sec:SBMtheory}.  
The naive long-wavelength theory now has two free Boson modes.  
The dominant correlations are expected to be at wavevectors that connect 
Fermi points with opposite group velocities.  
Taking the polarization axis to be $\hat{z}$, the spin structure factor 
$\la S^z_q S^z_{-q} \ra$ has dominant singularities at $2k_{F1}$, 
$2k_{F2}$, and $-k_{F2} - k_{F1}$, while the $\la S^x_q S^x_{-q} \ra$ 
is missing the $2k_{F2}$ since there is no spin-flip process across the 
second (polarized) Fermi sea.
We indeed observe such correlations in the wavefunctions, 
and the dominant power law envelope is consistent with $x^{-4/3}$,
which is what such naive theory would give if we assume equal velocities 
near all Fermi points and ignore all interactions other than 
gapping out the overall ``charge'' mode, Eq.~(\ref{m4rho+}).
We note, however, that to properly describe such a partially polarized
phase in the system with short-range interactions, 
we would need to also account for the ferromagnetic spin wave, 
which is not present in our wavefunctions\cite{Haldane88}
and not treated in the more general (but still naive) Bosonization 
theory outlined above.  We do not pursue this further here.

%%%%%%%%%%%%%%%%%%%%%%%%%%%%%%%%%%%%%%%%%%%%%%%%%%%%%%%%%%%%%%%%%%%%%%%
%%%%%%%%%%%%%%%%%%%%%%%%%%%%%%%%%%%%%%%%%%%%%%%%%%%%%%%%%%%%%%%%%%%%%%%
%%%%%%%%%%%%%%%%%%%%%%%%%%%%%%%%%%%%%%%%%%%%%%%%%%%%%%%%%%%%%%%%%%%%%%%

\section{DMRG results in conventional phases on the zigzag chain}
\label{app:DMRG4nonSBM}

For ease of comparisons, here we show our DMRG measurements in the 
conventional Bethe-chain and VBS-2 phases identified on the zigzag chain 
in earlier works.\cite{White_zigzag, Klironomos}
We take representative points from the same cut $K_{\rm ring} / J_1 = 1$ 
studied in detail in Sec.~\ref{sec:HringDMRG}, since this allows us to 
better relate to the SBM phase at such significant $K_{\rm ring}$ values.

%%%%%%%%%%%%%%%%%%%%%%%%%%%%%%%%%%%%%%%%%%%%%%%%%%%%%%%%%%%%%%%%%%%%%%%
%%%%%%%%%%%%%%%%%%%%%%%%%%%%%%%%%%%%%%%%%%%%%%%%%%%%%%%%%%%%%%%%%%%%%%%
\subsection{Bethe-chain phase}
\label{subapp:DMRG:Bethe}

\begin{figure}
\centerline{\includegraphics[width=\columnwidth]{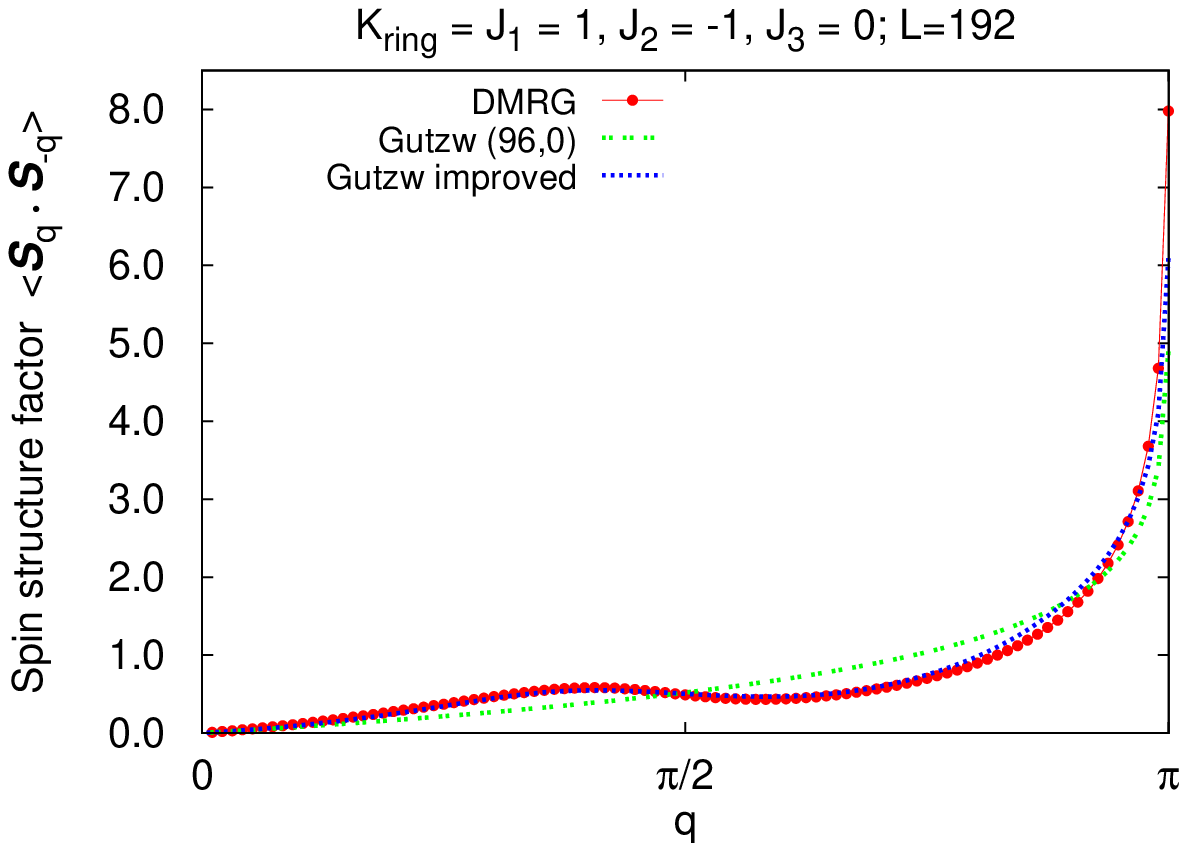}}
\centerline{\includegraphics[width=\columnwidth]{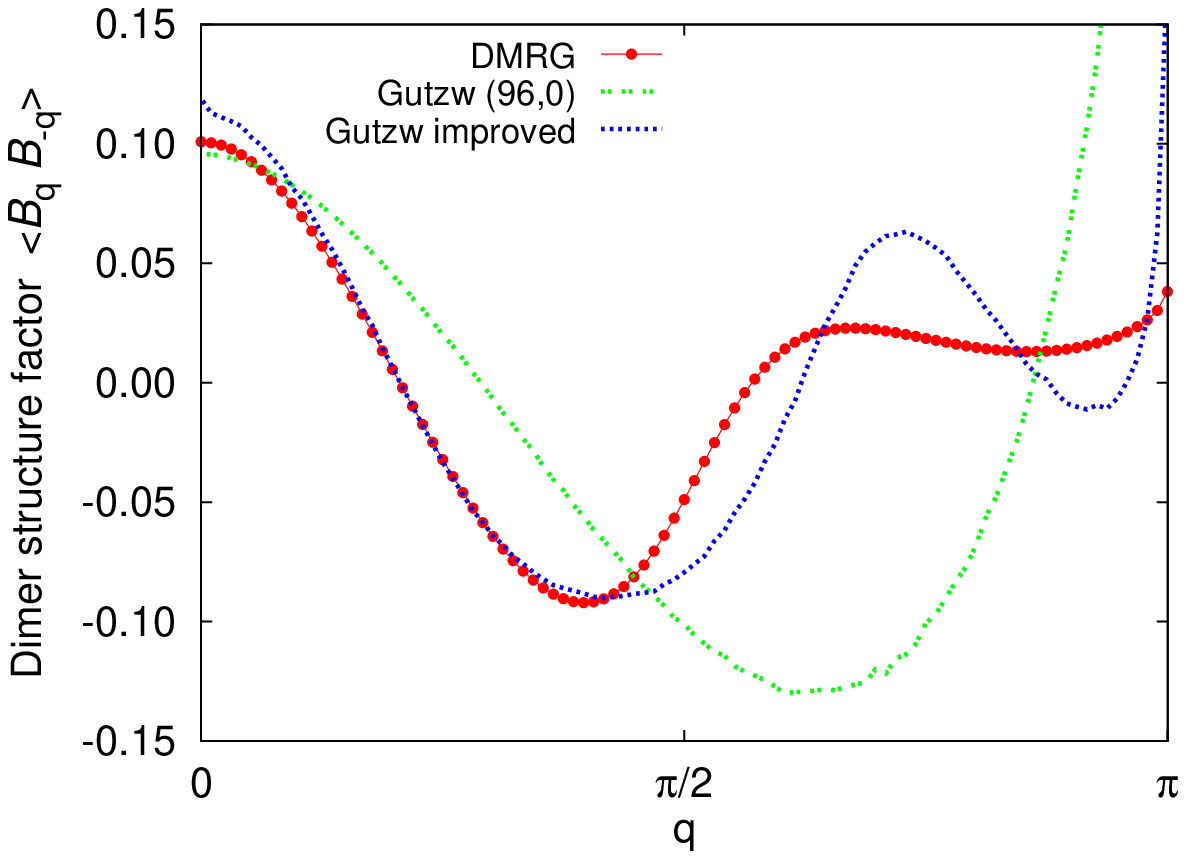}}
\centerline{\includegraphics[width=\columnwidth]{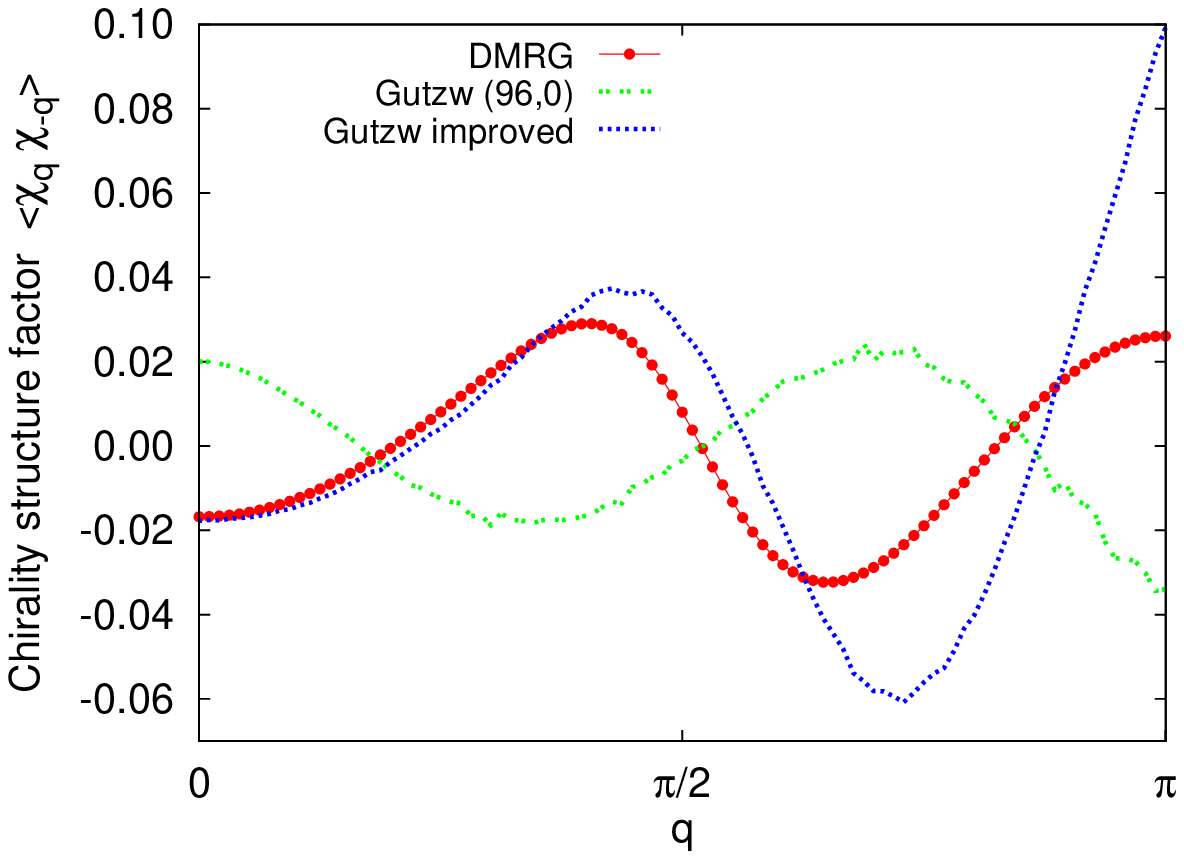}}
\caption{
(Color online)
Spin, dimer, and chirality structure factors at a representative point in
the Bethe-chain phase, $K_{\rm ring} = J_1 = 1$, $J_2 = -1$, measured in 
the DMRG for system size $L=192$.  We also show structure factors in the 
bare Gutzwiller-projected single Fermi sea state, $(N_1, N_2) = (96, 0)$,
and in the improved Gutzwiller wavefunction with parameters 
$f_0 = 1, f_1 = 0, f_2 = -1.4$ 
(see Appendix~\ref{app:Gutzw} for wavefunction details).
} 
\label{fig:J2_m1p0_J3_0p0}
\end{figure}

Figure~\ref{fig:J2_m1p0_J3_0p0} shows spin, dimer, and chirality 
structure factors at $J_2 = -1$, measured in the DMRG for system size 
$L=192$.  The DMRG can still obtain reliable results with $m=3200$ 
states kept in each
block, and this is related to the smaller central charge than in the 
SBM phase (as discussed in Sec.~\ref{subsec:entropy}).
Figure~\ref{fig:J2_m1p0_J3_0p0} also shows the structure factors in the 
bare Gutzwiller-projected single Fermi sea state, $(N_1, N_2) = (96, 0)$,
and in the improved Gutzwiller wavefunction 
(see Appendix~\ref{app:Gutzw});
the latter achieves significantly better trial energy and 
overall match with the DMRG results.

At long distances, we expect both the spin and dimer correlations to 
decay with the same power law:
$\la \vec{S}(x) \cdot \vec{S}(0) \ra \sim (-1)^x/x$,
$\la {\cal B}(x) {\cal B}(0) \ra \sim (-1)^x/x$,
up to logarithmic corrections.
We indeed see roughly such power law in the real space correlations.
Some quantitative aspects are different from the pure Heisenberg chain, 
for which the bare Gutzwiller state is a good approximation. 
Thus, the spin structure factor in Fig.~\ref{fig:J2_m1p0_J3_0p0} 
has a larger amplitude of the $q=\pi$ singularity and also develops a 
hump at wavevectors below $\pi/2$.
Both these features are captured by the improved Gutzwiller wavefunction.
On the other hand, the dimer structure factor has a significantly smaller 
amplitude of the $q=\pi$ singularity than the pure Heisenberg chain
and the bare Gutzwiller; the improved Gutzwiller wavefunction moves 
in the right direction compared to the bare one but still does not 
capture well the amplitude at $\pi$.

For the chirality correlations, we expect 
$\la \chi(x) \chi(0) \ra \sim (-1)^x / x^3 + 1/x^4$,
and we indeed see some fast decay in the real space data comparable
with the power law behavior.  
The corresponding momentum space singularities at $q=\pi$ and $q=0$ are 
very weak.  In agreement with this, we do not see any features in the 
chirality structure factor in Fig.~\ref{fig:J2_m1p0_J3_0p0}.

This Bethe-chain phase example allows to contrast with the SBM phase in
Sec.~\ref{subsec:DMRG:SBM}, 
where we see different singular wavevectors and prominent features 
in all these observables including the chirality.
The experience of being able to improve significantly the short-scale 
features in the trial wavefunctions carries over to the SBM,
although in the Bethe-chain phase we have an advantage that our 
wavefunctions also capture the long-distance power laws correctly.

%%%%%%%%%%%%%%%%%%%%%%%%%%%%%%%%%%%%%%%%%%%%%%%%%%%%%%%%%%%%%%%%%%%%%%
%%%%%%%%%%%%%%%%%%%%%%%%%%%%%%%%%%%%%%%%%%%%%%%%%%%%%%%%%%%%%%%%%%%%%%
\subsection{Valence Bond Solid with period 2}
\label{subapp:DMRG:VBS2}

\begin{figure}
\centerline{\includegraphics[width=\columnwidth]{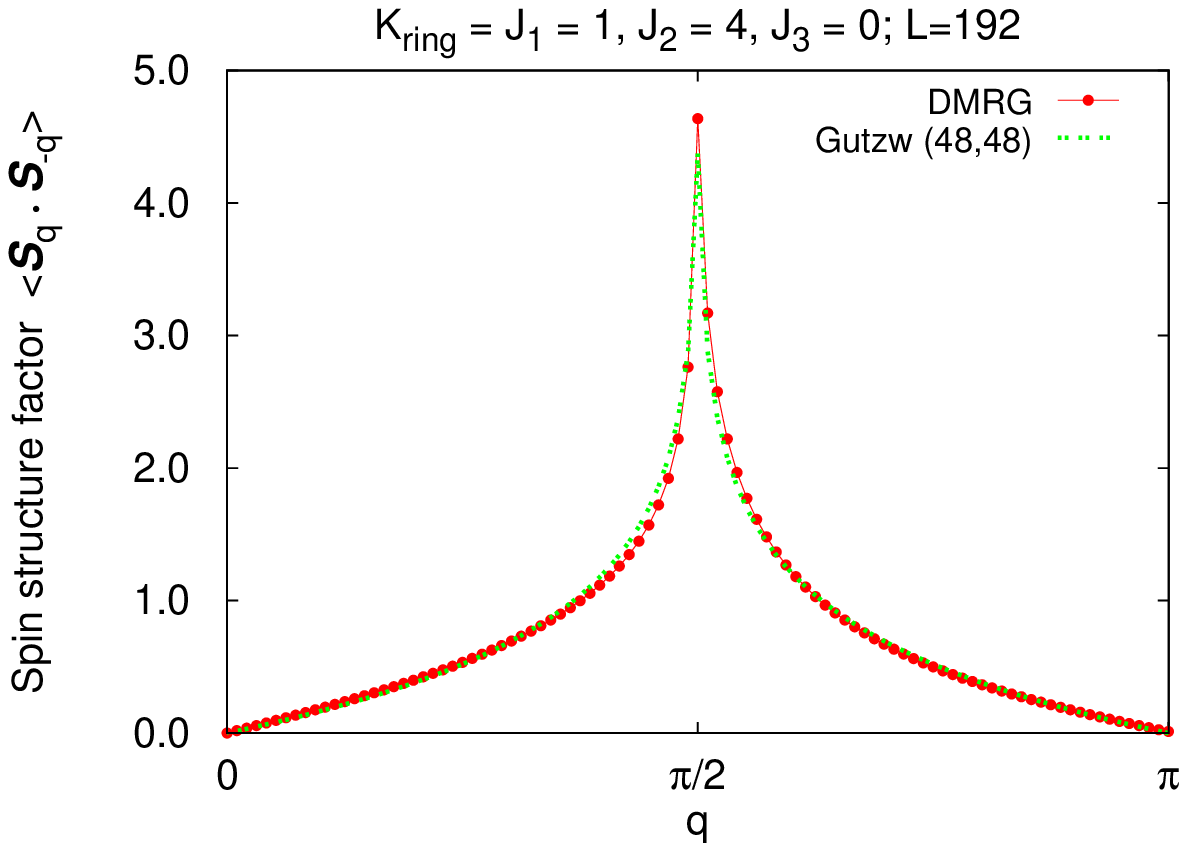}}
\centerline{\includegraphics[width=\columnwidth]{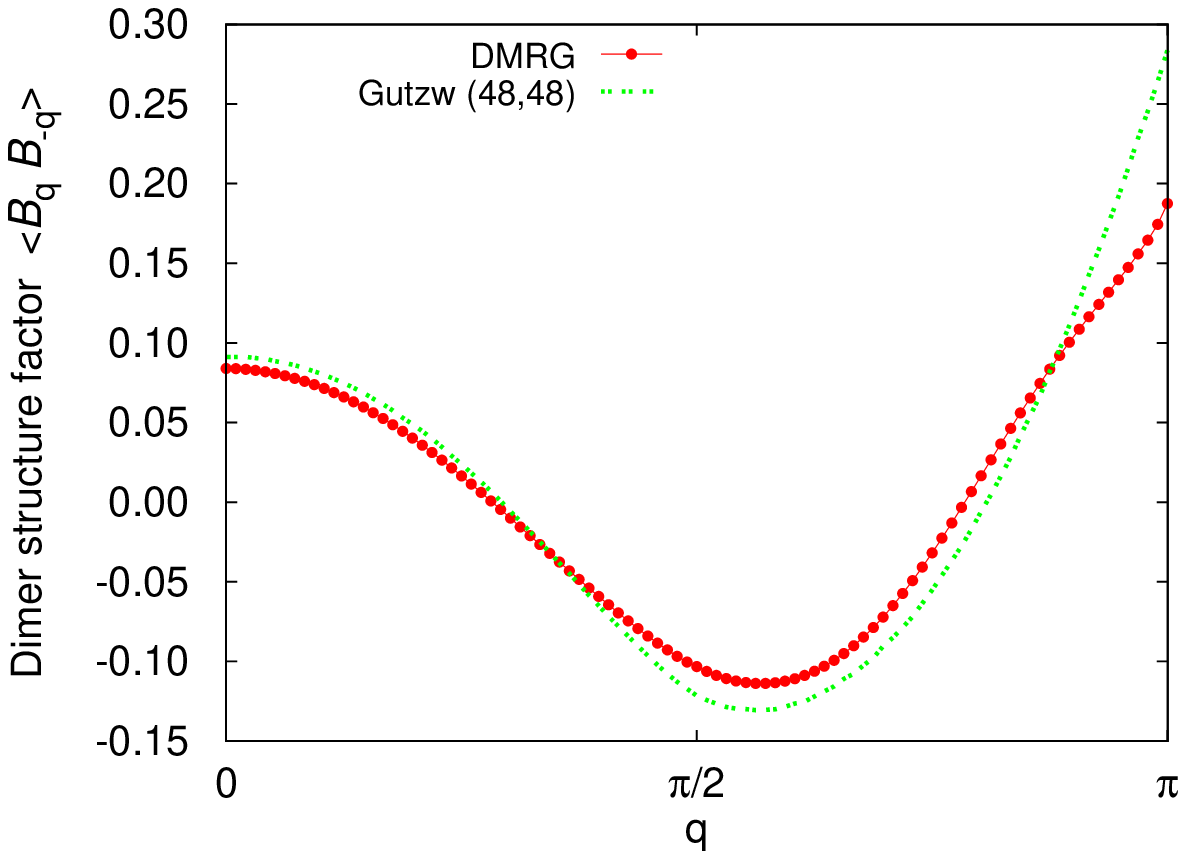}}
\caption{
(Color online)
Spin and dimer structure factors at a point in the VBS-2 phase, 
$K_{\rm ring} = J_1 = 1$, $J_2 = 4$, measured in the DMRG for system 
size $L=192$.  The exhibited trial wavefunction is the Gutzwiller 
projection of two equal Fermi seas, $(N_1, N_2) = (48, 48)$.  
This wavefunction gives decoupled legs expected in the large $J_2$ limit 
and does not have any VBS-2 order; however, it reproduces the DMRG
data quite well, so at this $K_{\rm ring} = J_1$ cut the system is close
to the decoupled legs limit even just outside the SBM.
}
\label{fig:J2_4p0_J3_0p0}
\end{figure}

Consider now the large $J_2$ case.  In the $J_2 \to \infty$ limit, 
we have decoupled legs, and each behaves as a Heisenberg spin chain.
Finite $J_1/J_2$ and $K_{\rm ring}/J_2$ will couple the two legs and 
will likely open a spin gap\cite{White_zigzag, Nersesyan} producing a 
VBS state with period 2 (Fig.~\ref{fig:dimer2}).
Figure~\ref{fig:J2_4p0_J3_0p0} shows our measurements at a 
representative point $J_2 = 4$ from the $K_{\rm ring} = J_1 = 1$ cut.
The spin correlations show a dominant peak at a wavevector $q = \pi/2$ 
and bond correlations have a peak at $q = \pi$. 
We compare with the Gutzwiller projection of two equal Fermi seas
in the 1D zigzag chain language, or, equivalently, decoupled legs in the 
two-leg ladder picture.
This wavefunction is thus strictly appropriate only in the 
$J_2 \to \infty$ limit, but it clearly reproduces the DMRG data 
quite well.

Looking at Fig.~\ref{fig:J2_4p0_J3_0p0}, there is not much direct
evidence for the VBS-2 order in the DMRG data.  
It is safe to say only that upon exiting the SBM phase along this cut, 
we are close to the fixed point of decoupled legs.
One argument for the VBS-2 here could be the continuity to the
strong VBS-2 phase in the broader phase diagram 
Fig.~\ref{fig:phased_J3_0p0}.  
As is known,\cite{White_zigzag} the region $J_2 \sim 0.4 - 2$ 
along the $K_{\rm ring} = 0$ axis has strong VBS-2 order.
However, this does not preclude possibility of more phases in the
model with ring exchanges. 
Thus, along the way at points like $K_{\rm ring} = 0.3, J_2 = 1.5$
and $K_{\rm ring} = 0.2, J_2 = 1.2$ we also see a dimer feature at 
$q = \pi/2$ in addition to a likely Bragg peak at $q = \pi$.  
Since our primary interest is the SBM phase, we do not explore the 
states at large $J_2$ further, loosely referring to all of them 
as VBS-2 in Fig.~\ref{fig:phased_J3_0p0}.

%%%%%%%%%%%%%%%%%%%%%%%%%%%%%%%%%%%%%%%%%%%%%%%%%%%%%%%%%%%%%%%%%%%%%%%
%%%%%%%%%%%%%%%%%%%%%%%%%%%%%%%%%%%%%%%%%%%%%%%%%%%%%%%%%%%%%%%%%%%%%%%
%%%%%%%%%%%%%%%%%%%%%%%%%%%%%%%%%%%%%%%%%%%%%%%%%%%%%%%%%%%%%%%%%%%%%%%

\end{document}